\documentclass[review,sort]{elsart}
\usepackage{amsmath}
\usepackage{amsfonts}
\usepackage{amssymb}
\usepackage{graphicx}
\usepackage{setspace}
\usepackage{caption}
\usepackage{xcolor}

\journal{Journal of Computational Physics}

\begin{document}

\begin{frontmatter}

\title{Mixed-precision for Linear Solvers in Global Geophysical Flows}
\author[1]{Jan Ackmann}\ead{jan.ackmann@physics.ox.ac.uk}\ref{cor}
\author[2]{Peter D. D\"uben}
\author[1]{Tim N. Palmer}
\author[3]{Piotr K. Smolarkiewicz}

\address[1]{University of Oxford, Oxford, OX1 3PU, UK}
\address[2]{European Centre For Medium Range Weather Forecasts, Reading, RG2 9AX, UK}
\address[3]{National Center for Atmospheric Research, Boulder, CO 80026, USA}
\corauth[cor]{Corresponding Author}

\begin{abstract}
Semi-implicit time-stepping schemes for atmosphere and ocean
models require elliptic solvers that work efficiently on modern
supercomputers. This paper reports our study of the potential
computational savings when using mixed precision arithmetic in the
elliptic solvers. The essential components of a representative elliptic
solver are run at precision levels as low as half (16 bits), and 
accompanied with a detailed evaluation of the impact of reduced precision 
on the solver convergence and the solution quality.

A detailed inquiry into reduced precision requires a model configuration 
that is meaningful but cheaper to run and easier to evaluate 
than full atmosphere/ocean models. This study is therefore conducted in the 
context of a novel semi-implicit shallow-water model on the sphere, 
purposely designed to mimic numerical intricacies of modern all-scale 
weather and climate (W\&C) models with the numerical stability independent 
on celerity of all wave motions. The governing 
algorithm of the shallow-water model is based on the non-oscillatory 
MPDATA methods for geophysical flows, whereas the resulting elliptic 
problem employs a strongly preconditioned non-symmetric Krylov-subspace
solver GCR, proven in advanced atmospheric applications. The classical
longitude/latitude grid is deliberately chosen to retain the stiffness
of global W\&C models posed in thin spherical shells as well as to better
understand the performance of reduced-precision arithmetic in the vicinity
of grid singularities. Precision reduction is done on a software level,
using an emulator. The reduced-precision experiments are conducted for
established dynamical-core test-cases, like the Rossby-Haurwitz wave
number 4 and a zonal orographic flow.

The study shows that selected key components of the elliptic solver, most
prominently the preconditioning, can be performed at the level of half
precision. 
\end{abstract}


\end{frontmatter}

\section{Introduction}
In the simulation of complex multi-scale flows arising in weather
and climate (W\&C), one of the biggest challenges is  to satisfy
strict service requirements in terms of time-to-solution and to
satisfy budgetary constraints in terms of energy-to-solution,
without compromising the accuracy and stability of the application
\cite{GMDEscape}. One way to tackle this challenge is to use
reduced-precision arithmetic in W\&C models. The interest with reduced
or mixed precision has already some history in scientific computation
\cite{Goddeke2007,Baboulin2009,Furuichi2011,ChenGJCP2012}.  However,
it is relatively new in W\&C modelling. This effort started with
\cite{Palmer2012} and \cite{DuebenJCP2014} where reduced precision was
motivated by the large degree of uncertainty that is present in W\&C
models due to their nonlinear dynamics and high level of complexity which
makes it difficult to justify high numerical precision. The study of
mixed precision in numerical modelling is timely, since the recent boom
of machine learning methods is pushing developments of supercomputing
hardware towards very efficient dense linear algebra performed at low
precision. This hardware is specialised for the use in deep learning
where half precision arithmetic---or even less---is often sufficient;
see for example the Tensor Processing Unit (TPU) by Google.

A large part of W\&C models is used to represent subgrid scale
processes---e.g., turbulence and cloud physics---that carry a large
degree of inherent uncertainty \cite{Saffin2019}. In principle one can
thus argue that representation of such processes is the prime candidate
for a precision reduction.  The other dominant part of W\&C models is
the dynamical core (a.k.a dycore) which provides discrete solutions to
the Navier-Stokes equations. Although there is little uncertainty about
the equations per se, their numerical solution procedures introduce
model error that add uncertainty to the dynamics. Early work on thorough
analysis to identify the minimal level of numerical precision that can
be used in different parts of dycores has focused on spectral model
formulations \cite{DuebenMWR2014,Chantry2019}. Similar work on grid-point 
model formulations has only just begun with the use of single precision
arithmetic \cite{Nakano2018,TintoPrims2019,Maynard2019} with 32 bits per variable.

In efficient W\&C dycores the semi-implicit (SI) timestepping is typically
used for the atmospheric \cite{Mengaldo2019} as well as the oceanic
\cite{adcroft2004overview,ICON_Korn,Fesom_Wang} component.  Its key
virtue is extended computational stability with respect to acoustic,
buoyant and rotational modes of motion allowing for the use of relatively
long time steps. On the other hand, the approach  results in an intricate
linear problem and complicated elliptic  boundary value problem for
pressure (viz. Schur complement). In particular, the SI approach is at
the heart of the newly-developed Finite-Volume Module
(FVM) of the Integrated Forecasting System (IFS) at the
European Centre for Medium-Range Weather Forecasts (ECMWF)
\cite{Smolarkiewicz2016,SKG2017,Smolarkiewicz2019,Kuhnlein2019}.
The FVM provides an alternative dynamical core to the spectral transform
based IFS. The two essential ingredients of the FVM timestepping is the
Multidimensional Positive Definite Advection Transport Algorithm  (MPDATA)
approach \cite{SmolarSzmelter05,SzmelterSmolar10,KuhnleinSmolar2017} and
the bespoke preconditioned nonsymmetric Krylov-subspace elliptic solver
\cite{Kuhnlein2019} built on the Generalized Conjugated-Residual (GCR)
 approach of \cite{Eisenstat_etal1983}. While MPDATA controls the
explicit part of the timestepping, GCR inverts the Schur complement of
the linear problem.

Undeniably, elliptic solvers are computationally demanding and
 substantial development has been invested into making them 
as efficient as possible for W\&C models on modern supercomputers
\cite{Muller2014,Yang2017,Kuhnlein2019}. To efficiently solve an elliptic
 problem posed in a thin spherical shell (such as the global atmosphere)
  ultimately requires matrix inversion---if not in the main solver than 
at least in its preconditioner. While there is the common opinion that
high precision is required for this purpose, there are theoretical
studies \cite{Goddeke2007,Baboulin2009,Carson2018,Haidar2018,Anzt2019}
 suggesting that mixed-precision approaches could be exploited, and there
are already some applications of mixed-precision elliptic solvers  in
CFD \cite{Idomura2018,Amritkar2015,Furuichi2011} and also in W\&C 
\cite{Maynard2019}. Although these approaches may differ in choice
of algorithms and applications, a common theme emerges that the
preconditioning step may be a good choice for reducing precision,
where some work goes as low as half precision arithmetic (16 bits per variable) representing
each real number with only 16 bits (as a floating point number) and
decimal precision reduced to three digits.

The aim of this paper is to establish the limits of precision for the
 specific combination of characteristics that make SI W\&C models unique,
 while its objective is to perform an in-depth analysis of the behavior
of mixed-precision elliptic solvers in such an environment. The elliptic
 problems encountered in SI W\&C models are non-symmetric, and are typically 
solved using iterative solvers due to the enormous problem size.
We solve a fluid dynamics problem on the sphere, which comes with grid 
irregularities, giving rise to a specific structure of the involved linear
operator. In effect, a standard classification of the problem based solely
on the operator condition number does not give full justice to the 
complexity of the problem; cf. \cite{Smolarkiewicz2019,Kuhnlein2019}. 
Moreover, the elliptic problem is part of an entire SI timestepping
scheme and as such is in a feedback loop with the non-linear dynamics of 
the model. Consequently, we solve much more at each time step than a sole 
elliptic problem.  Different solver solutions, even obtained to the same
accuracy in terms of the residual errors, can lead to vastly different
behavior and model error growth throughout a simulation. For these
reasons, our mixed-precision elliptic solver needs to be tailored to
the specific application at hand and thoroughly tested in this specific
W\&C environment.
 
Herein, we study the impact of mixed-precision in the elliptic solver
for W\&C  models by investigating a representative SI shallow-water
model on the sphere and using the MPDATA approach. With these choices,
we stay conceptionally close to the ideas employed in the IFS-FVM. The
latter is formulated in the latitude-longitude (lat-lon) coordinate
framework, but circumvents polar grid singularity via a quasi-uniform
unstructured mesh discretisation, whereby the essential problem stiffness
comes from the thin vertical dimension. The shallow-water model of this
paper uses a regular lat-lon grid subject to polar singularity \cite{PRUSA2018331},
but retains the aspect of the problem stiffness which is coming from the
longitudinal dimension. Given the stiffness and grid singularity that are
already prone to creating model errors in double precision arithmetic (64 bits per variable),
we carefully study what happens to solutions in polar regions under
reduced precision. We test our mixed-precision solver on test-cases from
the well-known test-suite for shallow-water dycores \cite{Williamson}.
The first test-case is the standard problem of the Rossby-Haurwitz Wave 
with wave number 4. The second test-case paraphrases the standard zonal 
flow over an isolated mountain by replacing the idealised smooth hill 
with the natural orography of the Earth.

In this paper, whenever we refer to mixed-precision, we mean the use of 
different precision levels within the same simulation. This can either refer
to the use of single and double precision as available on modern computers,
or the use of half, single and double precision. While half precision is
available on machine learning accelerators on modern supercomputing hardware,
it is currently not yet availble for use in most high-level computer languages.
An important milestone in this regard is that Fugaku, the current number one of the TOP500 list of supercomputers, allows for the use of half precision arithmetic from Fortran. 
For this paper, half precision will be emulated in software only. 
This enables to easily explore different precision levels without having to purchase
and use specialised hardware that would also require a time-consuming rewrite of the model code. 

The paper is structured as follows. In Section~2, we describe the
continuous as well as discretised model and introduce our elliptic
solver and the preconditioner. Section~3 presents the test-cases, 
the model setup and a description of the reduced-precision experiment 
suite. In section~4 we show and discuss reduced-precision results. 
Section~5 concludes the paper. 

\section{Model Description}
\subsection{The Governing Equations}\label{GovEqs} 
With our choice of model, we aim to stay conceptionally as close as 
possible to the approach used by other MPDATA based fluid models 
\cite{Prusa2008,SzmelterSmolar10,Kurowski2016mwr} and consequently the 
newly developed IFS-FVM atmospheric model
\cite{Smolarkiewicz2016,Smolarkiewicz2019,Kuhnlein2019}. Thus, we chose
to use the shallow-water equations (SWE) on the sphere written in the form
of generalized transport equations \cite{Smolarkiewicz1998,SzmelterSmolar10}:
\begin{equation}
\begin{aligned}\label{SWE1}
\frac{\partial G \Phi}{\partial t}+\nabla \cdot ({\bf v}\Phi) =0~,
\end{aligned}
\end{equation} 
\begin{equation}
\begin{aligned}\label{SWE2}
\frac{\partial G Q_{x}}{\partial t}+ \nabla \cdot({\bf v}Q_{x}) =  G\,R_x~,
\end{aligned}
\end{equation} 
\begin{equation}
\begin{aligned}\label{SWE3}
\frac{\partial G Q_{y}}{\partial t}+ \nabla \cdot({\bf v}Q_{y}) = G\,R_y~,
\end{aligned}
\end{equation} 
where $\Phi$ is the fluid thickness, $\nabla=(\partial/\partial x,\,\partial/\partial y)$,
$Q_{x}=\Phi\,\dot{x}h_x$ and $Q_{y}=\Phi\,\dot{y}h_y$ 
denote the momenta in x (longitudinal) and y (latitudinal) directions, while 
$G=h_{x}h_{y}$ is the Jacobian of the geospherical framework with $h_{x}$, $h_{y}$ being 
the metric coefficients of the general orthogonal coordinates. Here, $h_{x}=a\cos(\phi)$, 
$h_{y}=a$ for a lat-lon grid with Earth's radius $a$, so longitude $\lambda=x$ and 
latitude $\phi=y$. The advective velocity ${\bf v}$ relates to the momentum ${\bf Q}$ via
\begin{equation}\label{vadv}
{\bf v}=G\,(\dot{x},\,\dot{y})=(Q_x h_y,\,Q_y h_x)/\Phi~;
\end{equation}
cf. \cite{Smolarkiewicz1998,SzmelterSmolar10} for discussions. The right-hand sides
of the equations \eqref{SWE2} and \eqref{SWE3} symbolise the forcing terms $R_{x}$
and $R_{y}$ for the momenta, such that 
\begin{equation}
\begin{aligned}\label{RSWE2}
R_x = -\frac{g}{h_{x}}\Phi \frac{\partial(\Phi+H_{0})}{\partial x}
+fQ_{y}+\frac{1}{G\Phi}\left(Q_{y}\frac{\partial h_{y}}{\partial x} - 
Q_{x}\frac{\partial h_{x}}{\partial y}\right)Q_{y}-\alpha(Q_{x}-{\overline{Q}}_{x})~,
\end{aligned}
\end{equation} 
\begin{equation}
\begin{aligned}\label{RSWE3}
R_y = -\frac{g}{h_{y}}\Phi \frac{\partial(\Phi+H_{0})}{\partial y}
-fQ_{x}-\frac{1}{G\Phi}\left(Q_{y}\frac{\partial h_{y}}{\partial x}
-Q_{x}\frac{\partial h_{x}}{\partial y}\right)Q_{x}-\alpha(Q_{y}-{\overline{Q}}_{y})~,
\end{aligned}
\end{equation} 
with specific terms, from left to right, representing the pressure
gradient, the Coriolis force, the metric forces (viz. Christoffel
symbols of the geospherical framework) and the relaxation forcings
attenuating the solution to some hypothetical ``true'' state
${\overline{\bf Q}}$ with the inverse time scale $\alpha$. $H_{0}$ is
the topography, $g$ the gravitational acceleration and $f$ the Coriolis
parameter.

\subsection{The Discretised Model}\label{DiscModel} 
The below-summarised procedure of formulating an implicit linear
problem for dependent model variables, together with its associated
Schur complement (i.e. elliptic Helmholtz equation) is a special 
case of a substantially more intricate procedure documented recently 
for all-scale atmospheric Euler equations in \cite{Smolarkiewicz2019},
whereto the interested reader is referred for further details. 

The governing Equations~\eqref{SWE1}-\eqref{SWE3} are discretised
in a semi-implicit fashion, using a collocated lat-lon grid with its
well-known pole problem \cite{PRUSA2018331}. In a nutshell, momentum
equations are integrated with the trapezoidal rule
\begin{equation}\label{MPDATA}
{\bf Q}^{n+1}_{\bf i}={\mathcal M}_{\bf i}\left({\bf Q}^{n}+0.5\Delta
t {\bf R}^{n},{\bf v}^{n+0.5},G\right) +0.5\Delta t {\bf R}^{n+1}_{\bf
i}~\equiv {\widehat {\bf Q}}_{\bf i}+0.5\Delta t {\bf R}^{n+1}_{\bf i},
\end{equation} 
where $n$ and ${\bf i}$ index discrete points $(t^n,\,{\bf x}_{\bf i})$ 
separated with uniform intervals $\Delta t$ and $(\Delta\lambda, \Delta\phi)$, 
respectively, in time and the two spatial directions. The first term on the 
rhs of (\ref{MPDATA}) denotes a second-order MPDATA operator---with ${\bf
v}^{n+0.5}$ symbolising an ${\mathcal O}(\Delta t^2)$ explicit predictor
of advective velocity at the intermediate $t^n+0.5\Delta t$ time level;
cf. \S{3.4} and \S{4.1} in \cite{Smolarkiewicz1998}---and it forms the explicit
part of the solution. The second term is unknown, as ${\bf R}^{n+1}$
depends (nonlinearly) both on $\Phi^{n+1}$ and ${\bf Q}^{n+1}$. There
are several alternative approaches to assure an easily invertable linear
problem, involving outer iterations or explicit predictors, or both
\cite{Smolarkiewicz2016,SKG2017,Smolarkiewicz2019,Kuhnlein2019}. As our
goal is to assess the role of mixed precision on the elliptic solver
performance, we here select an ad hock approach that complicates (rather
than simplifies) the target elliptic solver.

Specifically, the pressure gradient term is approximated as
\begin{equation}\label{rhslin}
(\Phi \nabla H)^{n+1}  = 0.5[\Phi^{n+1}\nabla(\Phi^\star+H_{0})
+\Phi^\star\nabla(\Phi^{n+1}+H_{0})]~, 
\end{equation} 
where $\Phi^\star$ denotes the explicit second-order-accurate predictor
of $\Phi^{n+1}$, generated by integrating the mass-continuity equation
(\ref{SWE1}) with MPDATA. The Coriolis and relaxation terms are implicit
at $t^{n+1}$, whereas the nonlinear metric forces---generally acting
on a much longer time scale than the scale associated with the propagation
of external gravity waves---are predicted explicitly at $n+1$ by
linear extrapolation from $n$ and $n-1$ time levels.\footnote{The
resulting integral of (\ref{SWE1})-(\ref{SWE3}) is by construction
second-order-accurate in time and space; cf. \cite{Kuhnlein2019} for a pertinent
discussion.} Thanks to the collocated grid, the resulting linear problem
is inverted analytically, forming the closed-form expressions
\begin{equation}\label{Qexpr}
{\bf Q}^{n+1}_{\bf i}={\widehat{\widehat {\bf Q}}}_{\bf i} +{\bf
A}_{\bf i}\nabla\Phi^{n+1}_{\bf i}+{\bf B}_{\bf i}\Phi^{n+1}_{\bf i}~,
\end{equation} 
where ${\widehat{\widehat {\bf Q}}}$, ${\bf A}$ and ${\bf B}$ denote,
respectively, the modified explicit part of the problem, and the
$2\times2$ matrix and vector of known coefficient fields.  Implementing the
expressions \eqref{Qexpr} in the trapezoidal integral of \eqref{SWE1},
leads to the elliptic Helmholtz problem for $\Phi^{n+1}$, which can be
symbolically written as
\begin{equation}\label{Elliptic}
\forall~{(n+1,{\bf i})}~~\mathcal{L}_{\bf i}\left(\Phi^{n+1}\right)
-{\mathcal R}^{n+1}_{\bf_i}=0~;
\end{equation} 
hereinafter, spatial and temporal grid indices are dropped as there is
no ambiguity.  The linear operator $\mathcal{L}$ is negative definite
but not self-adjoint. It takes the form of a generalized Laplacian
\begin{equation}\label{L_Operator}
\mathcal{L}\left(\Phi\right):=\sum^{M}_{I=1}\frac{\partial}{\partial
x^{I}} \left(\sum^{M}_{J=1}A^{IJ}\frac{\partial \Phi}{\partial x^{J}}
+ B^{I}\Phi\right)-C\Phi~, 
\end{equation} 
where the coefficient fields are straightforward modifications
of those appearing in \eqref{Qexpr}, and $M=2$. The solution of
\eqref{Elliptic} provides the updated $\Phi$---upon which ${\bf Q}^{n+1}$
is recovered from (\ref{Qexpr}) and then ${\bf v}^{n+1}$ from 
(\ref{vadv})---thus allowing to complete \eqref{MPDATA}
for calculations with large $\Delta t$, limited only by the advective
CFL condition.

\subsection{The Elliptic Solver}
\begin{figure}[h]
\centering
\includegraphics[width=1.5\textwidth]{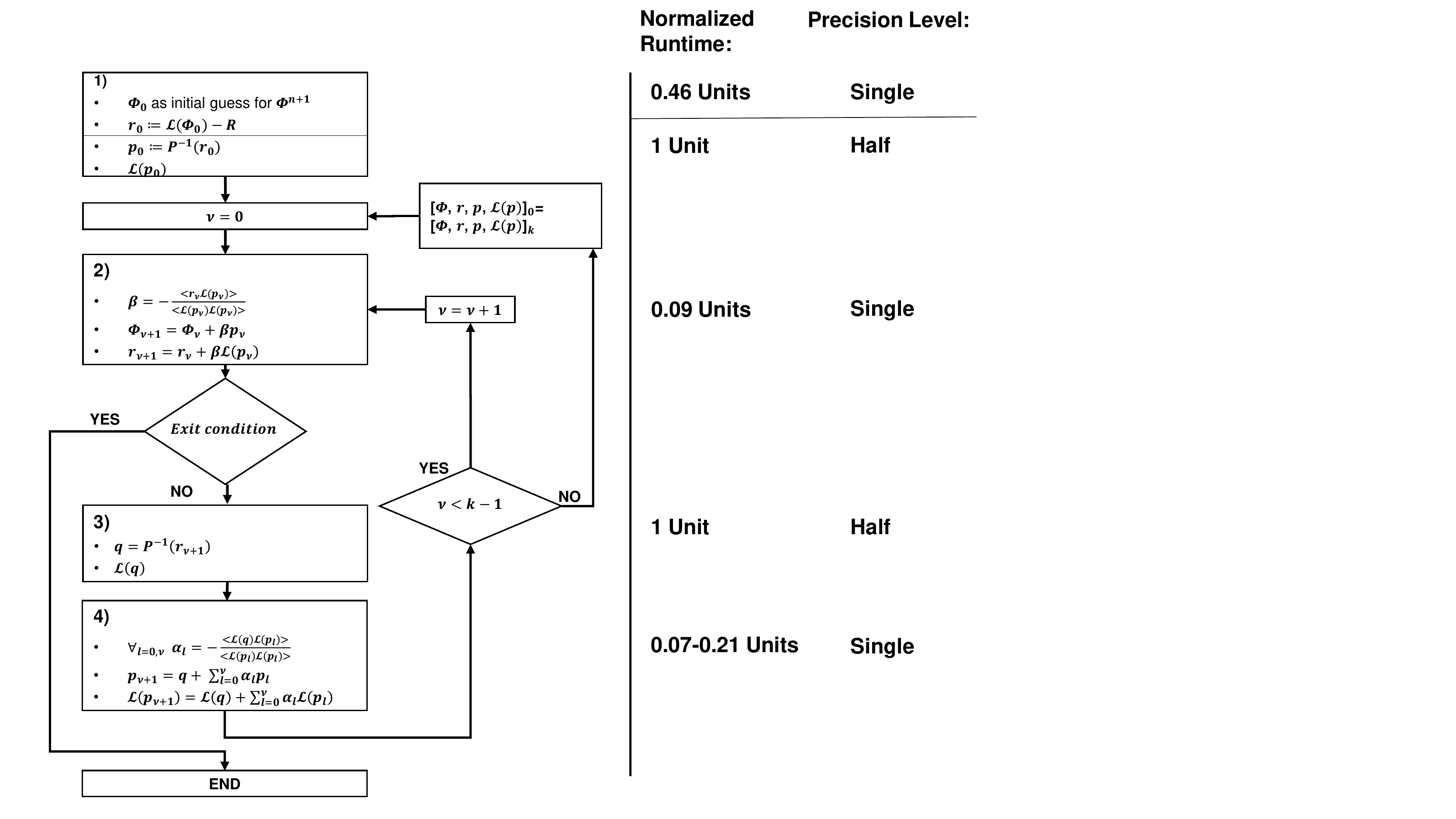}
\caption{From left to right: a flow chart of our preconditioned GCR(k)
implementation; normalized runtimes of the four distinct algorithmic steps for GCR(3) (All performance measurements for this paper have been performend on a single core (CPU: Intel i5, Fortran compiler: GCC 9.3.0.) and they agree with the ratio of the number of occurring floating point operations); and the choice of base precision for each algorithmic step (The choices are described in more detail in \S\ref{MPGCR}).\label{fig:tempevo15}}
\end{figure}

The elliptic problem (\ref{Elliptic}) is solved using the
preconditioned generalized conjugate residual (GCR) approach
\cite{smolarkiewicz2000variational,Eisenstat_etal1983}. GCR is a
Krylov sub-space method that assures monotone convergence of the
solver iterations for a non-self-adjont operator ${\mathcal L}$ (i.e.,
"non-symmetric" in matrix representation) by minimising the $L_2$ norm
$\langle r r\rangle$ (where $\langle ..\rangle$ marks the domain integral)
of the residual error
\begin{equation}\label{Residual}
r_\nu = \mathcal{L}\left(\Phi_\nu\right)-{\mathcal R}~,
\end{equation} 
here $\nu$ numbers the iterations. In essence, GCR cleverly re-optimises
at each iteration all coefficients in the series of the current and past
successive iterations up to some arbitrary specified number, say $k-1$;
cf. \cite{smsz11} for a discussion. When after some number of iterations
$N$, $L_2(r)$ reaches a desired small error tolerance, the solution of
this iterative process $\Phi_{N}$  is set to be $\Phi^{n+1}:=\Phi_{N}$. Figure~\ref{fig:tempevo15} describes our implementation of the
preconditioned GCR algorithm. 

Following the step~2 of the algorithm---the update of the depth $\Phi$ 
and the minimised residual error---there is an exit condition. While a 
variety of exit conditions can be considered \cite{Smolarkiewicz1997}, 
we found that demanding the uniform reduction of the initial residual error, 
\begin{equation}\label{exitcondition}
\parallel r_{\nu+1}\parallel_2 ~\le~ \epsilon\parallel r_0\parallel_2~,
\end{equation}
assures the stability of the semi-implicit model integrator while
maintaining the solution quality throughout a range of spatial
resolutions and physical scenarios. In practice, the choice of
$\epsilon$ is verified experimentally. Based on
our overall experience with W\&C models, in this paper $\epsilon=10^{-5}$
is assumed for all experiments. We
also enforce at least one full GCR(3) cycle regardless of
(\ref{exitcondition}). Further discussion and justification for this 
choice of $\epsilon$ is found in \S\ref{SP_prec_SWM}.

The preconditioner $\mathcal{P}$ in step~3 approximates ${\mathcal
L}$ indirectly, by estimating the solution error $q={\mathcal
P}^{-1}(r)\approx e={\mathcal L}^{-1}(r)$ by means of a stationary
iteration---indexed with $\mu$, such that $q_{\mu=0} = 0$---best
characterised as a semi-implicit Richardson 
scheme~\cite{smolarkiewicz2000variational}. Specifically, the
operator $\mathcal{P}$ is split in two parts.  The first part combines
the second-order zonal derivative term and the Helmholtz term, ${\mathcal
P}^{\mathcal Z} -\mathcal{P}^{\mathcal H}$; whereas the second part,
marked as ${\mathcal P}^{\mathcal M}$ for its meridional predominance, is
the reminder of ${\mathcal L}$ and the first part. In the semi-implicit
Richardson scheme, the first part is taken at the $\mu+1$ iteration
while ${\mathcal P}^{\mathcal M}$ is lagged behind. This results in a
tridiagonal problem
\begin{equation}\label{P_Operator}
\left[{\mathcal I}-\eta\big(\mathcal{P}^{\mathcal Z} 
- \mathcal{P}^{\mathcal H}\big)\right]q_{\mu+1}
 = q_{\mu}+ \eta \left[\mathcal{P}^{\mathcal M}q_{\mu}-r_{\nu+1}\right],
\end{equation} 
where ${\mathcal I}$ denotes the identity operator, and $\eta$ can be
interpreted as a pseudo-timestep, determined from linear stability theory
for the $\mathcal{P}^{\mathcal M}$ operator. For further reference and 
illustration, we implemented four preconditioning options in the
semi-implicit shallow-water model. Options 0 to 2 are elementary,
in that: option 0 corresponds to ${\mathcal P}={\mathcal I}$, viz. no
preconditioning; option 1 is for ${\mathcal P}= {\mathcal D}$,
where ${\mathcal D}$ symbolises the diagonal part of ${\mathcal L}$ in
\eqref{Elliptic}; option 2 is a diagonally preconditioned explicit
Richardson iteration for full ${\mathcal L}$; and option 3 is the
semi-implicit Richardson scheme in (\ref{P_Operator}). Taking the number of
GCR iterations with option 0 as a reference, the subsequent options reduce
the solver's iteration count by about 10, 40 and 98 percent, respectively. 
In the experiments reported in this paper, we use solely option 3 with two 
Richardson iterations that comprise two tridiagonal inversions and one 
evaluation of ${\mathcal P}^{\mathcal M}$. We found that this combination 
yields the best overall performance.

\section{Selected test-cases and experimental setup}
\subsection{Preamble \label{ssec:preamb}}
To assess the impact of mixed precision on the performance of
semi-implicit W\&C models, we exploit two well-established benchmarks
for testing and evaluating shallow-water surrogates of dynamical cores
for global weather and climate, namely the Rossby-Haurwitz wave with
wavenumber four (RHW4) and a zonal flow ($Q_x/\Phi\propto\cos\phi$,
$Q_y=0$) past the Earth orography.  The Rossby-Haurwitz waves are
analytic solutions of the nondivergent nonlinear barotropic vorticity
equation on the sphere, and as such propagate zonally without change
of shape \cite{Haurwitz}. However, Rossby-Haurwitz waves are incompatible
\cite{Temam2006} with shallow-water equations (SWE), and their SWE 
simulations lead to unstable solutions, given sufficiently short
zonal wavelength and sufficient amplitude \cite{Hoskins73}. In
particular when simulated with SWE, the form of RHW4 changes a
little over 24 days, while twice shorter Rossby-Haurwitz waves break
completely within a week \cite{Hoskins73}. Because of this marginal
stability \cite{Thuburn}, RHW4 is a convenient vehicle to test
capability of numerical methods for maintaining subtle nonlinear
balance of wave form solutions over an extended integration time.
The RHW4 simulation with SWE was included in the suite of tests
proposed by Williamson et al.~\cite{Williamson}, and has become a
prominent benchmark in the field.

As the RHW4 problem is particularly smooth, with the analytic solution
confiened to a few spherical harmonics \cite{Thuburn}, we complement it
with a multi-scale variant of the classical problem of a geostrophically
balanced zonal flow past a hill centered in mid latitudes. The original
problem has been studied by Grose and Hoskins \cite{grose}. It is
characterised by small Rossby and Froude numbers---here $R_o=U/Lf$ and
$Fr=U/\sqrt{gH}$, with $U$, $L$ and $H$ denoting the characteristic
flow speed, horizontal scale of the hill and the mean height of the
shallow-water surface, respectively---whereby it is well explained
by the linear theory \cite{grose}. This problem, also proposed by
Williamson et al. \cite{Williamson} for evaluating the efficacy of
numerical methods for global-scale dynamics, has become a benchmark
in the field---see e.g. \cite{smmw01} for its 3D nonhydrostatic
extension. Here we retain the benchmark setup as proposed in
\cite{Williamson}, except for replacing the smooth localised hill
with Earth's orography. This effects in numerical solutions
with broad spectra of scales and, effectively, a stiffer elliptic
problem.

To quantify the impact of the reduced precision on the solver
convergence and solutions quality, we adapt the $L_2$ and $L_\infty$
error norms defined in \cite{Williamson} as normalised deviations from
a ``genuine'' solution. For the RHW4 test-case, the genuine solution
is the analytic solution ${\bf \Psi}_0(x-{\mathcal V} t, y)$,
where ${\bf \Psi}_0 ={\bf \Psi}(x,y)$ represents the initial condition
for the velocity and depth fields---given in eqs.~(143), (144) and (146)
of \cite{Williamson}---and ${\mathcal V}$ is the propagation speed of
the analytic Rossby-Haurwitz wave (eq.~142 in \cite{Williamson}). As
the analytic wave does not satisfy the shallow-water equations
(\ref{SWE1})-(\ref{SWE3}), the deviations of numerical solutions
from ${\bf \Psi}_0(x-{\mathcal V}t, y)$ are expected to grow in time.
However, the comparison provides a meaningful gauge for assessing the impact of 
a reduction in precision. In the orographic flow test-case the genuine solution
is specified as a geostrophically balanced zonal flow, steady in the 
absence of the orography. For both benchmarks, the calculations addressing 
reduced precisions are conducted on the three regular longitude-latitude 
grids, with respective resolutions $NX\times NY = 128\times 64$, 
$256\times 128$ and $512\times 256$.

\subsection{Highlights of RHW4 simulations}
\begin{figure}[!htb]
\centering
\includegraphics[width=0.86\linewidth]{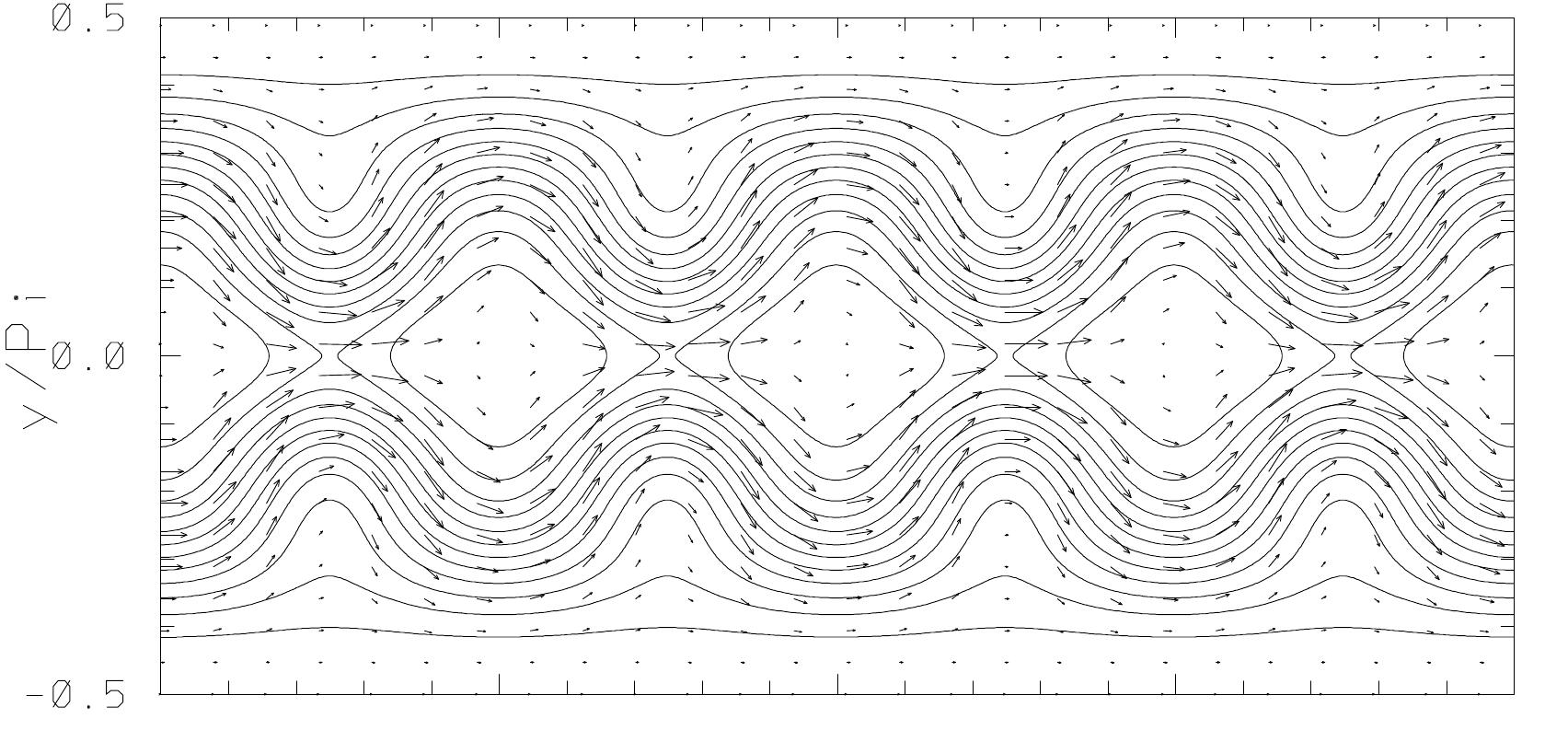}
\includegraphics[width=0.86\linewidth]{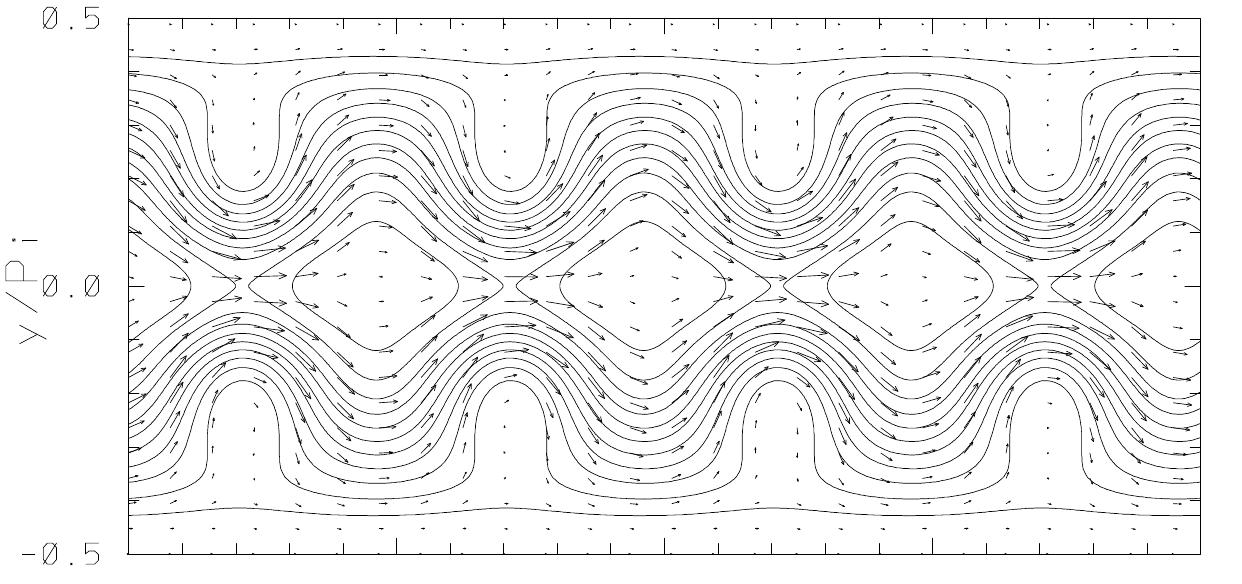}
\includegraphics[width=0.86\linewidth]{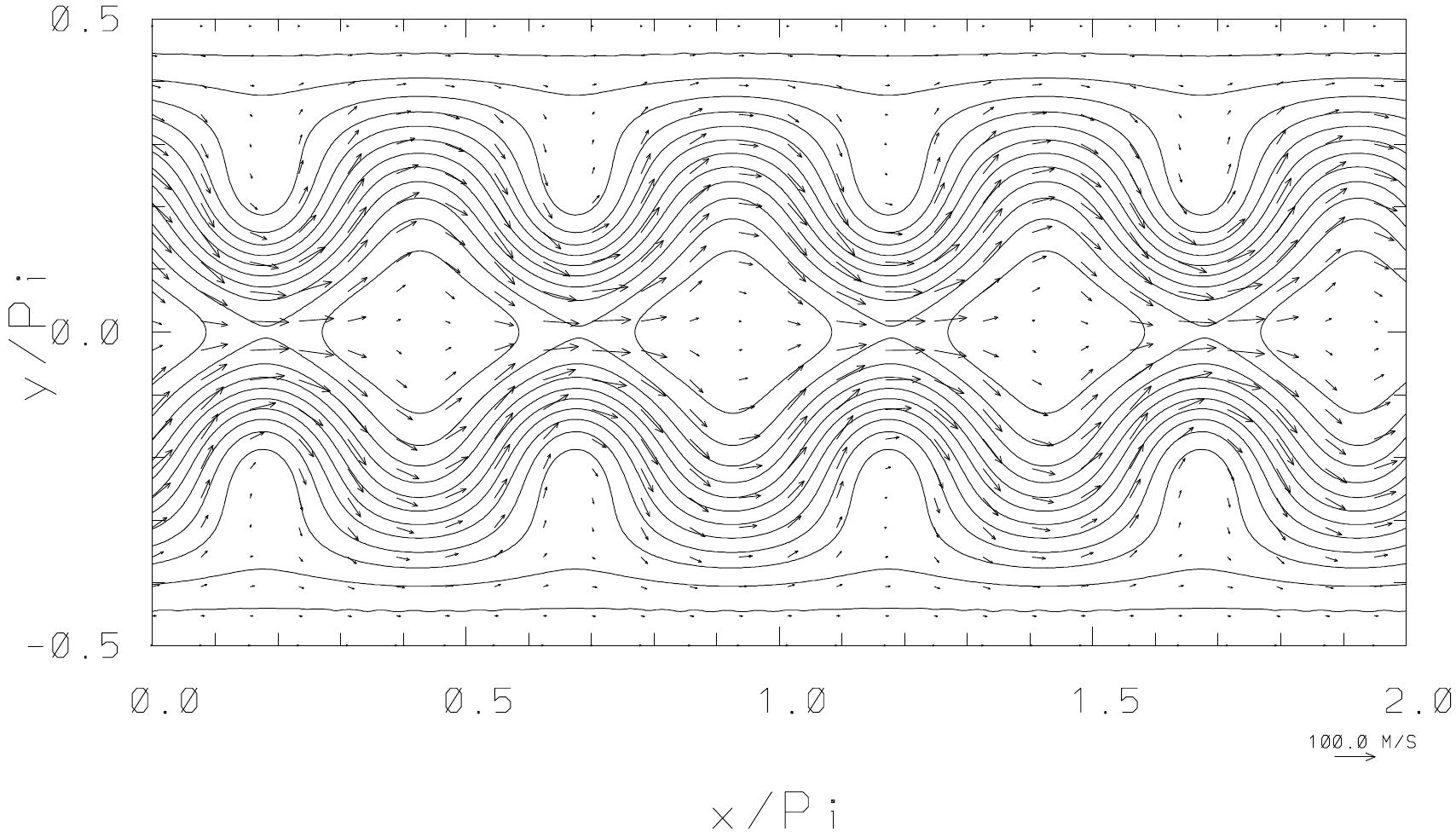}
\caption{RHW4, instantaneous normalised free-surface perturbations,
$(\Phi(x,y,t)-H_o)/H_o$ ($H_o=8~{\rm km}$), with imposed corresponding 
flow vectors.  From the top to bottom, respectively, are the initial 
condition and the numerical results after 7.38 and 14.76 days,
corresponding to one and two periods after which the analytic RHW4
propagates by one and two full wavelengths. The perturbation isolines
are plotted with the contour interval 0.025 (zero contour lines are
not shown), and the reference 100~m/s flow arrow is shown in the lowest
right corner. \label{rhw4rf} } \end{figure}

\begin{figure}[!htb]
\centering
\includegraphics[width=0.86\linewidth]{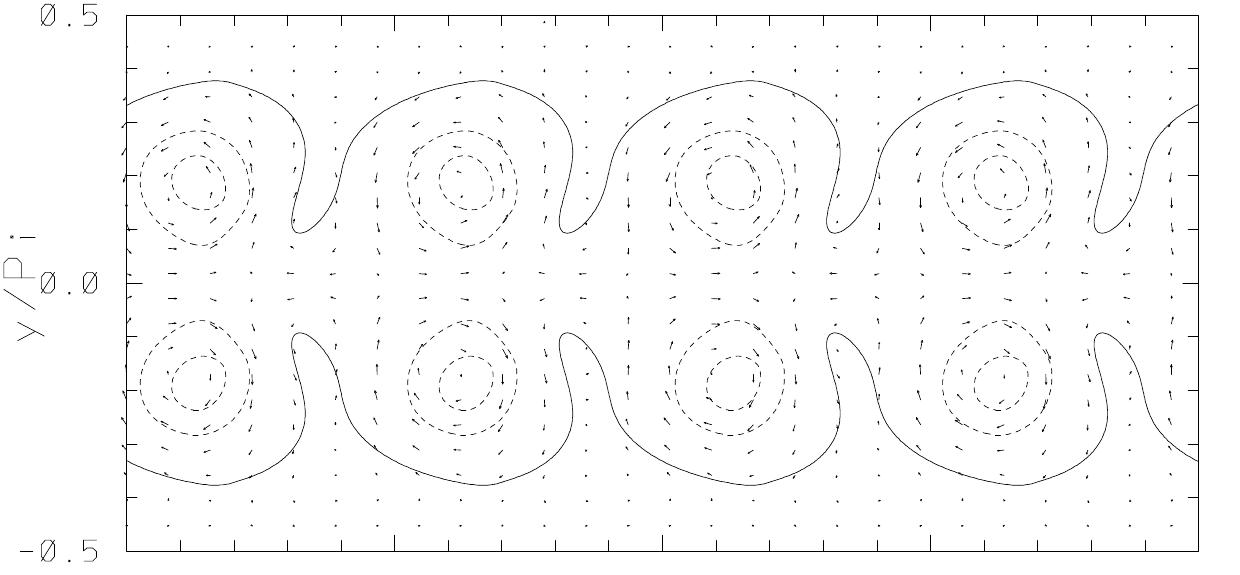}
\includegraphics[width=0.86\linewidth]{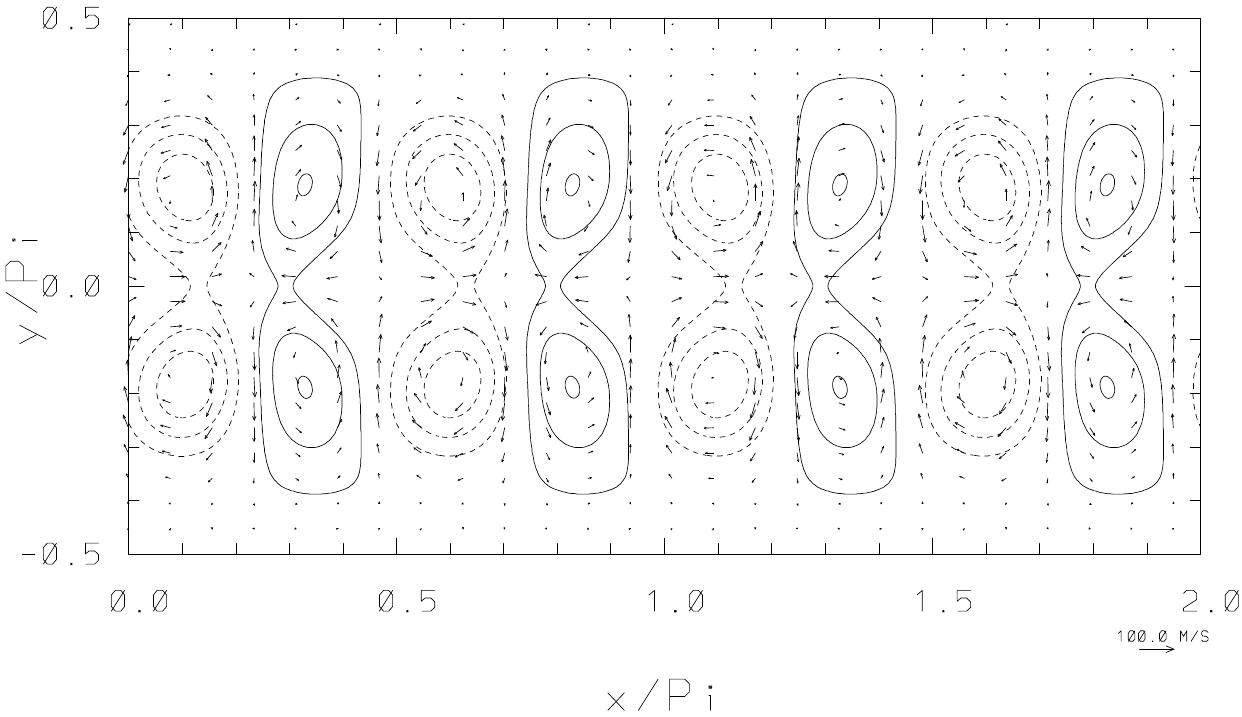}
\caption{RHW4, the normalised free-surface departure from the analytic
result, $(\Phi(x,y,t)-\Phi_0(x-{\mathcal V}t,y))/H_o$, with imposed
corresponding departures of the flow vectors. The top and bottom panels
show the results after $t=7.38$ and $t=14.76$ days, respectively.
The contour interval and the reference flow arrow are the same as in
figure~\ref{rhw4rf}; zero contour lines are not shown and the dashed
contours correspond to negative values. \label{rhw4pr} }
\end{figure}

Figure~\ref{rhw4rf} shows the analytic initial condition and the numerical
results on the $512\times 256$ grid after one and two periods of the
analytic RHW4 when the analytic solution exactly reproduces the initial
condition. The departures from the analytic result are clearly seen. After
one period, the numerical RHW4 is steeper and somewhat retarded compared
to the analytic result. After the two periods, the numerical results
is less steep but the phase retardation is more pronounced. The lower
resolution results (not shown) look similar, which is not surprising as
the RHW4 is well resolved (with 32 grid intervals per wavelength) even
on the coarsest grid; cf. Figure~2 in \cite{Smolarkiewicz1998} that shows
a coarse result after 5 days. As expected (see figure~3 in \cite{Thuburn}),
the current solution is visibly stable, which
however should not be taken for granted, as irregularities of model
discretisation can excite instability of the RHW4 already after 5 days;
cf. Figs.~11 and 12 in \cite{SzmelterSmolar10}. Figure~\ref{rhw4pr}
complements figure~\ref{rhw4rf} with the display of relative departures
of the numerical solutions from the analytic results.
 
To appreciate the strenuosity of the illustrated calculations, consider
that the celerity of the external mode ($\sqrt{gH_o}=280$~m/s) is
comparable to the speed of sound, while the zonal length interval
of the physical grid in the rings of grid points adjacent to the
poles is $\delta x=7675, 1919$ and $480$~m for the three considered
regular grids. The explicit solutions, like those discussed in
\cite{Smolarkiewicz1998,SzmelterSmolar10}, require then temporal intervals
$\Delta t\sim 40, 10$ and 2.5~s, respectively, whereas calculations
reported here were conducted with $\Delta t = 800, 400$, and 200~s,
respectively. The attained linear rather than quadratic reduction of
$\Delta t$ with the reciprocal of $NY$ is due to the specificity of the
flow decaying towards the poles. Although the calculations employed weak 
polar absorbers---with the inverse time scale $\alpha$ increasing linearly 
from zero at the angular distance $\phi \ge 3\pi/64$ away from the poles to
$1/(200\Delta t)$ at the poles---essentially the same solution can be
obtained without any absorbers. However, even weak absorbers can improve
the solver convergence by improving diagonal dominance of the linear
operator (\ref{L_Operator}). For example, the presented results using weak 
polar absorbers $\alpha=1/(200\Delta t)$ have a 30$\%$ reduced Euclidean
norm of the residual $\parallel r_{N}\parallel_2$ on average when exiting 
the GCR algorithm (on average 4 iterations) compared to using no
polar absorbers ($\alpha=0$ everywhere). The calculations illustrated
in Figs.~\ref{rhw4rf} and \ref{rhw4pr}, evinced the maximal Courant
numbers 118 and 0.31, correspondingly with respect to the celerity
of the external mode and the flow speed. 

\subsection{Highlights of orographic flow simulations}
\begin{figure}[!htb]
\centering
\includegraphics[width=0.86\linewidth]{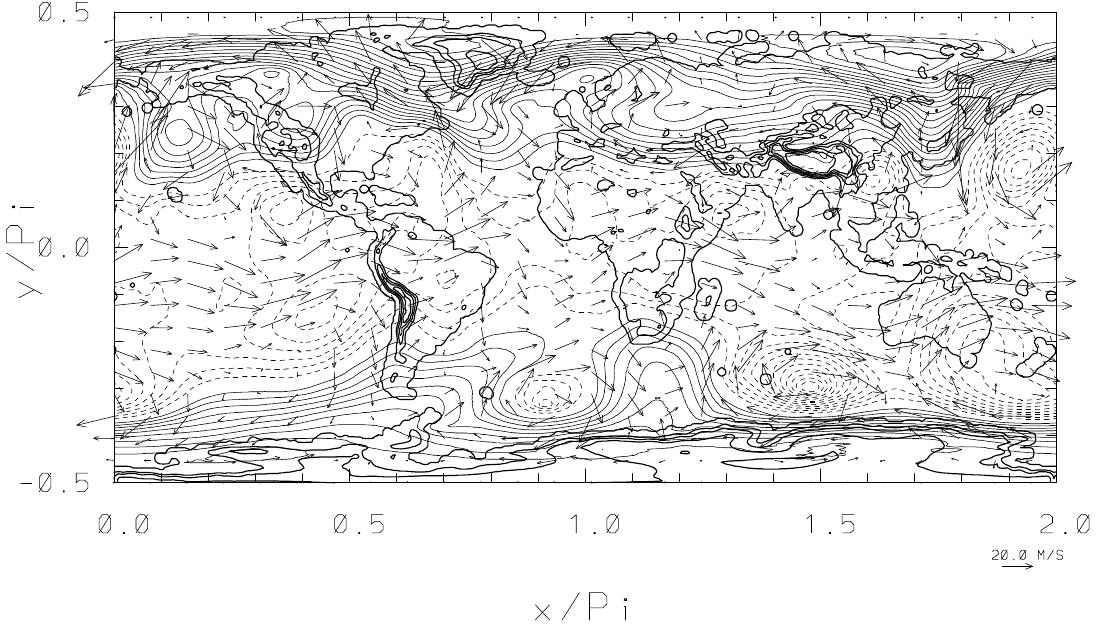}
\caption{Normalised free-surface height perturbations $(H(x,y,t)-H(x,y,0))/H_o$ 
(here $H=\Phi+H_{0}$ and $H_o=8~{\rm km}$), with imposed corresponding 
flow vectors at $t=14.76$ days.  The perturbation isolines are
plotted with the contour interval 0.00625 (dashed lines are for
negative values and zero contour lines are not shown), and the
reference 20~m/s flow arrow is shown in the lowest right corner;
the contours of the orography $H_{0}(x,y)$ span the range 3 to 7003
m with the 1200~m interval. \label{zflet} } \end{figure}

In analogy to the RHW4 case, figure~\ref{zflet} shows the numerical
results on the $512\times 256$ grid after 14.76 days, likewise
simulated with $\Delta t=200$~s. Pattern-wise, the departures from the
initial zonal flows are dramatic, especially in terms of energy and
direction of local flows. Quantitatively, the free-surface height
perturbations (viz. pressure perturbations) are about 3\% and 12\%
in terms of $L_2$ and $L_\infty$ norms; i.e., consistent with real
weather \cite{Smolarkiewicz2019}. For reference, the same measures for the RHW4 simulation are of comparable size, with 3\% and 7\% respectively. In terms of the kinetic energy of the
velocity perturbations, the corresponding $L_2$ and $L_\infty$ norms
are 153\% and 762\% (sic), with the latter reflecting both the direction
(say easterly or northerly) and the high speed of local flows ($\sim~50$
m/s) emerging in the high latitudes of the northern hemisphere. The respective values for the RHW4 case are much smaller in comparison, with 37\% and 35\%. The maximal Courant numbers with respect to the celerity of the external mode is 116, which is about the same as in the RHW4 case. The maximal Courant numbers with respect to flow velocity is 0.71, which is more than doubled compared to the RHW4 case. Nonetheless, the overall 
performance of the elliptic solver was comparable at 4 GCR iterations 
per time step. The essential difference between the orographic flow simulation and the RHW4 
setups is in the importance of the absorbers, its inverse time scale here taken 
as $\alpha=1/(2\Delta t)$ consistently with the theoretical 
considerations \cite{PRUSA2018331}.

\subsection{Reduced-precision Experiments}\label{red_prec_exp}
The precision reduction is achieved in three distinct steps. In the
first step, precision is reduced from double to single precision (23 significand bits) for the entire shallow-water model. The results of these experiments are presented in \S\ref{SP_prec_SWM}. Based on the experience from the tests with the single precision shallow-water model, a mixed-precision shallow-water model is described in \S\ref{MP_SWM}, that behaves similarly to the double precision reference model with respect to the three criteria outlined later in this section. As single precision arithmetic is easily available on modern CPUs, these reduced-precision experiments can be performed on standard hardware. In the third step, a mixed-precision model is presented in \S\ref{MPGCR} that is additionally using half precision for most parts of the elliptic solver. Half precision is emulated with the reduced-precision emulator (rpe v5 library \cite{Dawson2017}). Using the emulator enables us to explore a precision reduction beyond single precision with model code that was developed for standard computer hardware. 

While we emulate the 10 significand bits as in half precision, 
we keep the number of exponent bits at
11---the double precision standard---which means that the dynamical
range of representable numbers remains high. The results should still be representative
for the use of half precision as problems with the dynamical
range can often be mitigated via rescaling of variables, see for instance
\cite{Klower:2019:PAF:3316279.3316281}, especially
since the elliptic problem is linear by design.

The precision levels for the mixed-precision elliptic solver are identified as the
minimum precision required for each step of the algorithm shown in
figure~\ref{fig:tempevo15}. Because the elliptic solver is part of the
timestepping scheme of a non-linear model, it is not obvious how this
could be achieved via analytical means. Thus, we adopt an experimental
approach with the precision being reduced to levels where the elliptic solver
still yields acceptable results. The elliptic solver's performance
is studied for each of the two test-cases defined earlier. This covers a 
wide range of different dynamics, from linear responses up to turbulent 
regimes with a multiplicity of scales. 

We consider three criteria to measure the performance and viability of the elliptic solver when precision is reduced:
i) the solver's convergence rate; ii) whether the required solver accuracy 
was met; and iii) the change in normalised deviations from the respective 
``genuine'' solution in the $L_2$ and $L_\infty$ error norms.
Criterion (i) aims at keeping the solver's convergence rate
unchanged compared to the reference solver. A decrease in the solver's 
convergence rate, and thus an increase in the expected
solver iterations, would be acceptable if each iteration is sufficiently 
cheap when using reduced precision. Unfortunately, since we do
not use real hardware for half precision simulations, we cannot make reliable estimates of the savings. It is thus reasonable to aim for the same
number of solver iterations in the mixed-precision solver. 
Criterion (ii) is straightforward. As soon as the precision reduction is
introduced into parts of the solver, there is a risk that the residual
errors become corrupted with rounding errors. It is therefore important
to show that the elliptic solver is actually capable of converging to the required accuracy. 
Criterion~(iii) is more subtle. The solver accuracy
is chosen based on the additional error in the model solution of a double 
precision reference run, defined as the relative error between the
accuracy measures with respect to genuine solutions specified in the last
paragraph of \S\ref{ssec:preamb}. Given that the residual-error
fields of a double precision and a mixed-precision elliptic solver will
be different---e.g., due to a different spectral composition---even if
the required solver accuracy was reached, the pattern and magnitude of
the additional discretisation error might change as well.

\section{Results}

In this paper, the focus is primarily on the $512\times 256$ resolution, the highest of the three resolutions, as it shows the strongest grid convergence at the poles and is thus expected to be the most challenging test-case for reduced precision.

To evaluate criterion (i) and criterion (ii), the discussed residual error fields are calculated by using definition ~\eqref{Residual}, in double precision arithmetic, applied to the updated fluid thickness field $\Phi_\nu$. We do not use the minimized residual error field $r_\nu$ from step 2 of the GCR algorithm in figure \ref{fig:tempevo15} as the values of $r_\nu$ may already be corrupted by rounding errors.

Given these choices, the convergence plots that are an integral part of the following discussion are here introduced in detail to avoid ambiguity. In the convergence plots, the development of the residual error field in the Euclidean norm is shown over successive solver iterations. We number the solver iterations as follows: iteration 0 denotes the initial residual error field $ \mathcal{L}\left(\Phi_{0}\right)-{\mathcal R}$, iteration 1 to 3 denote the residual error fields $ \mathcal{L}\left(\Phi_{\nu+1}\right)-{\mathcal R}, \; \nu=0,1,2$ of the first cycle of the GCR(3) algorithm. Following this convention, iteration 4 to 6 denote the fields $ \mathcal{L}\left(\Phi_{\nu+1}\right)-{\mathcal R}, \; \nu=0,1,2$ of the second cycle of the GCR(3) algorithm. To highlight the residual error field's evolution over successive solver iterations, values stemming from different iterations of the same call to the elliptic solver are connected by lines.

\subsection{Single Precision Shallow-Water Model}\label{SP_prec_SWM}

 \begin{figure}[!htb] 
\vspace{0.3cm}
\minipage{0.48\textwidth}\centering\textbf{RHW4}\par\medskip
(a) DP
 \includegraphics[width=1.0\linewidth, trim=0cm 0cm 0cm 1.2cm, clip]{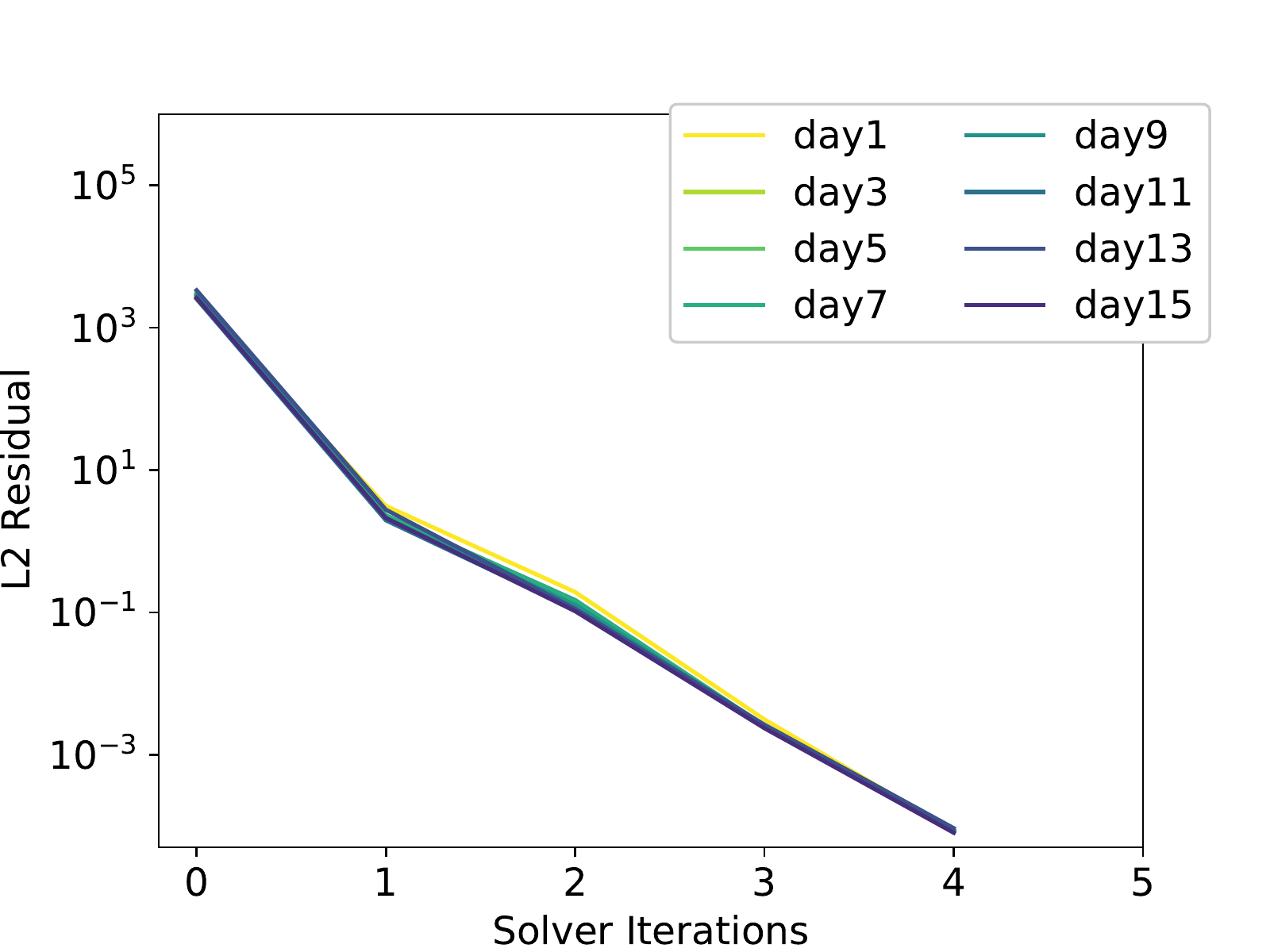} 
\endminipage\hfill
\minipage{0.48\textwidth}\centering\textbf{Orographic Flow}\par\medskip
(b) DP
 \includegraphics[width=1.0\linewidth, trim=0cm 0cm 0 1.2cm, clip]{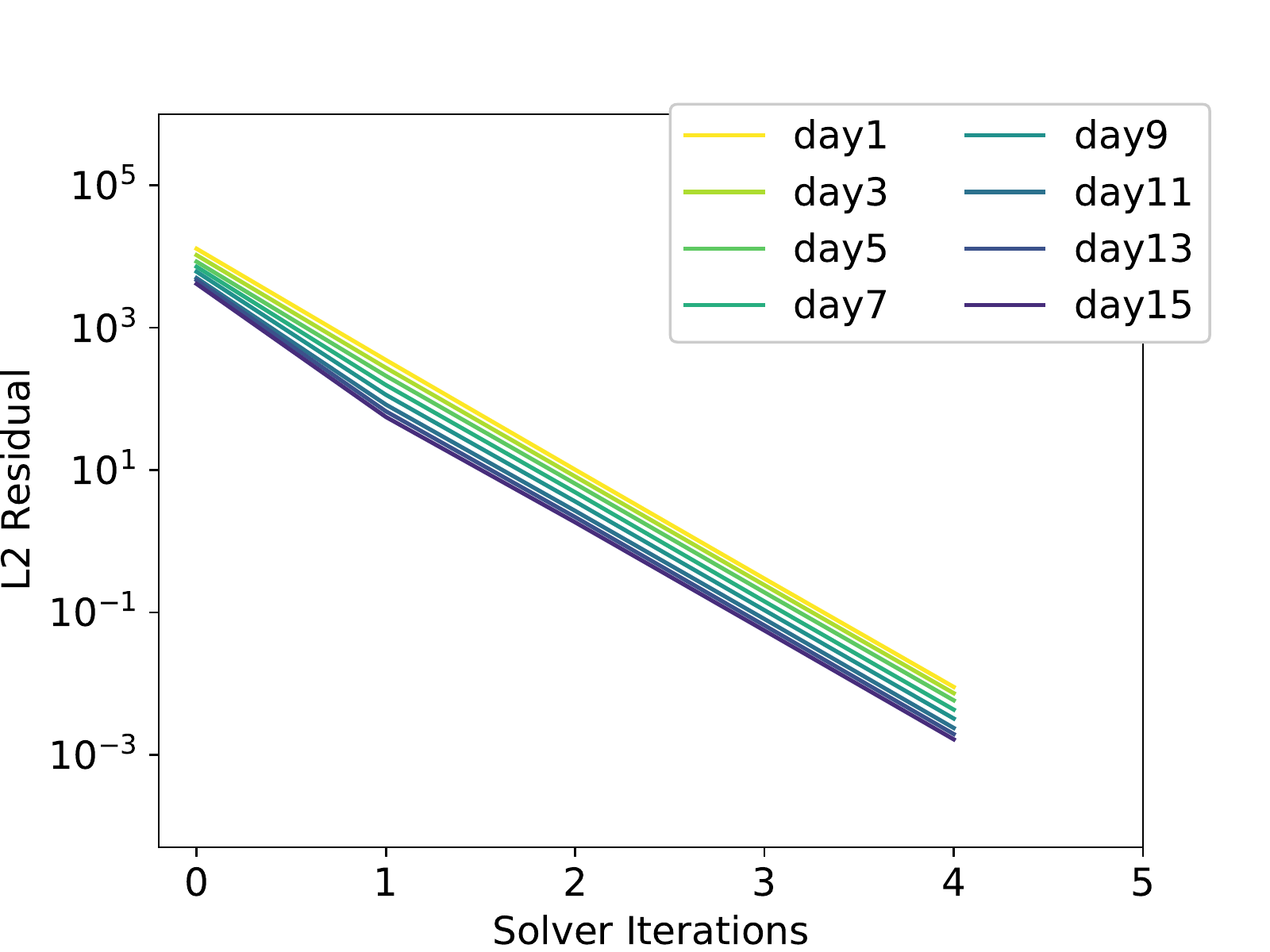}
\endminipage
\vspace{0.3cm}
\minipage{0.48\textwidth}\centering
(c) SP
 \includegraphics[width=1.0\linewidth, trim=0cm 0cm 0 1.2cm, clip]{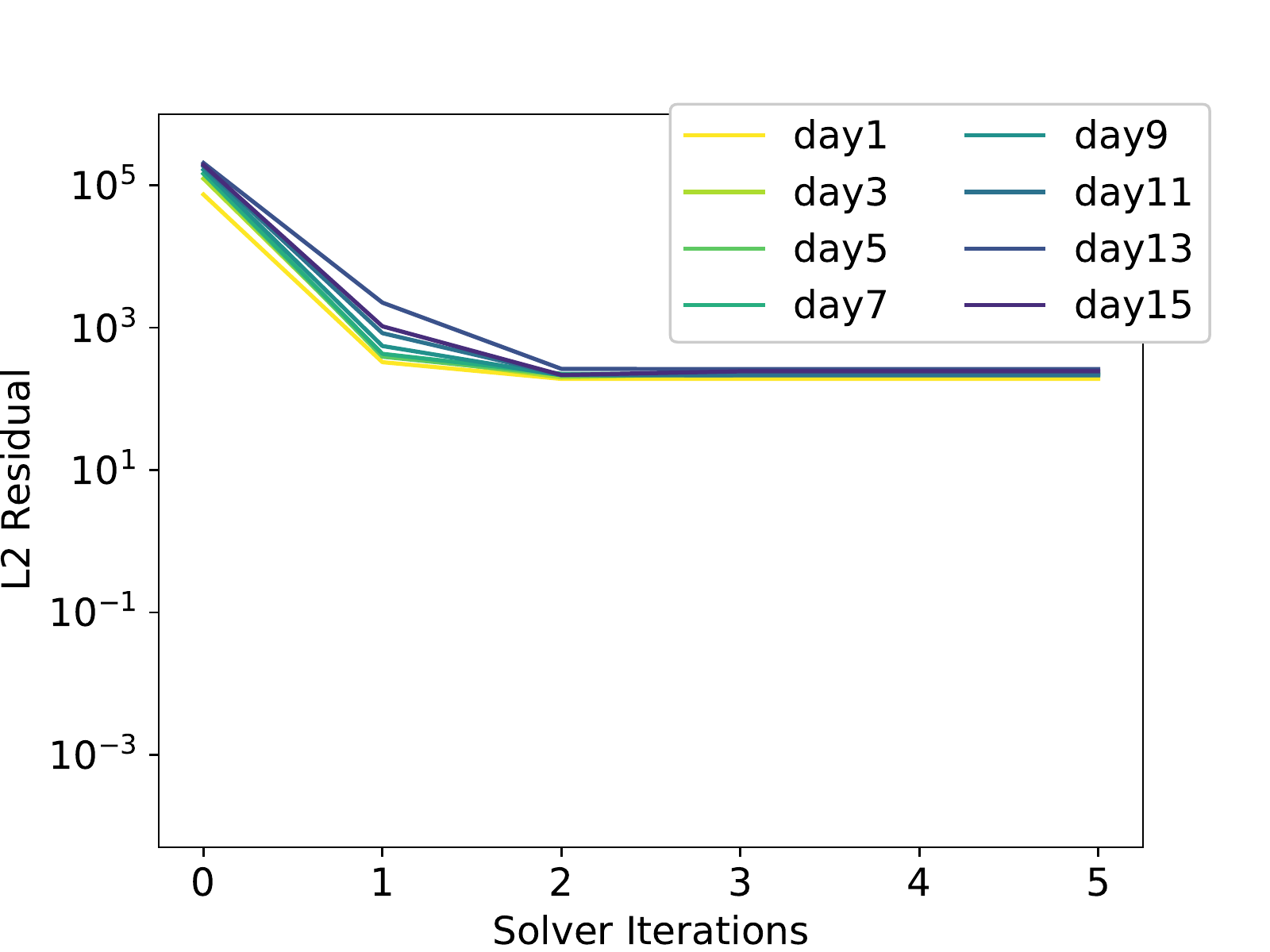} 
\endminipage\hfill
\minipage{0.48\textwidth}\centering
(d) SP
 \includegraphics[width=1.0\linewidth, trim=0cm 0cm 0 1.2cm, clip]{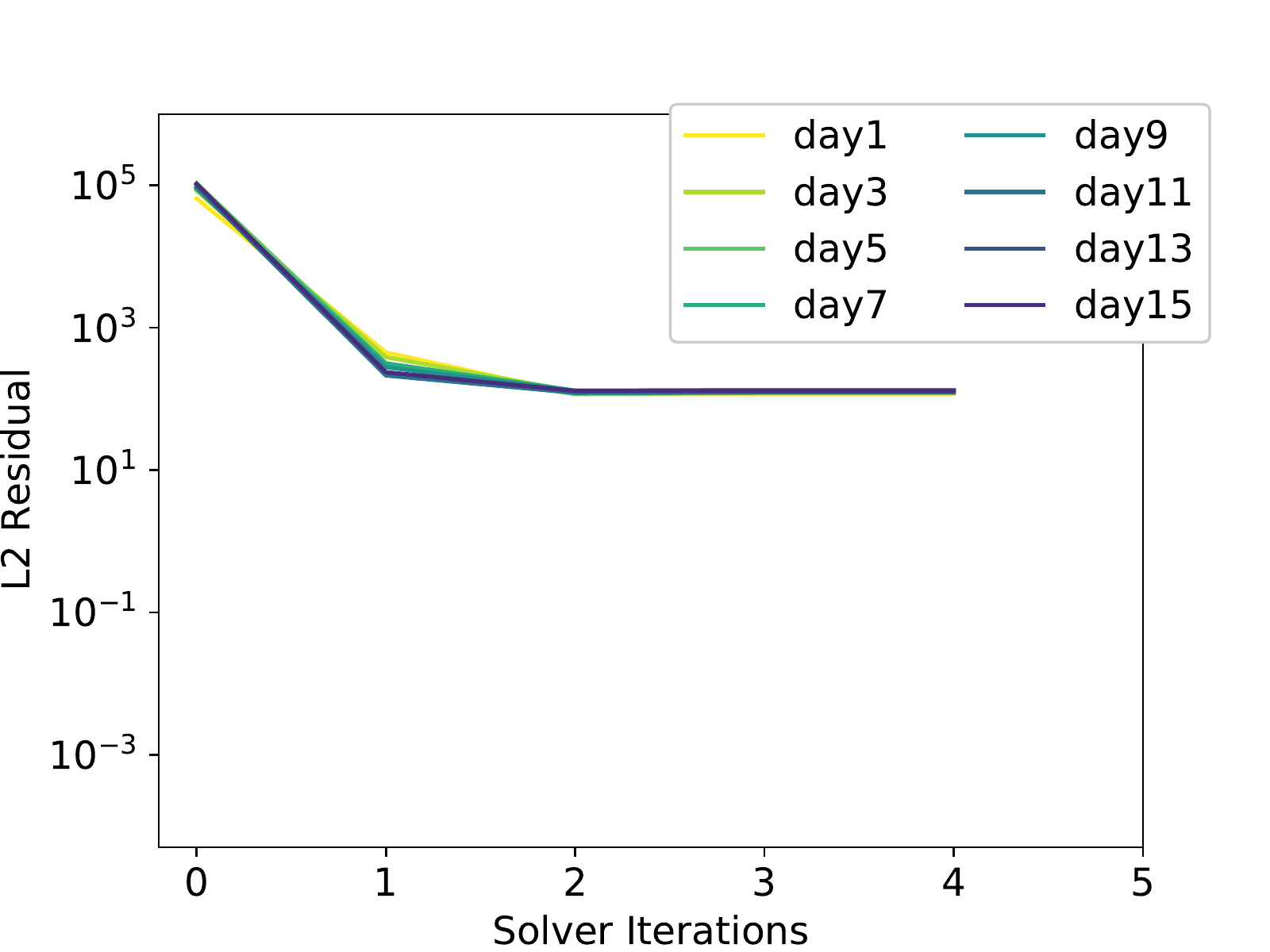}
\endminipage
\vspace{0.3cm}
\minipage{0.48\textwidth}\centering
(e) Mixed-Precision
 \includegraphics[width=1.0\linewidth, trim=0cm 0cm 0 1.2cm, clip]{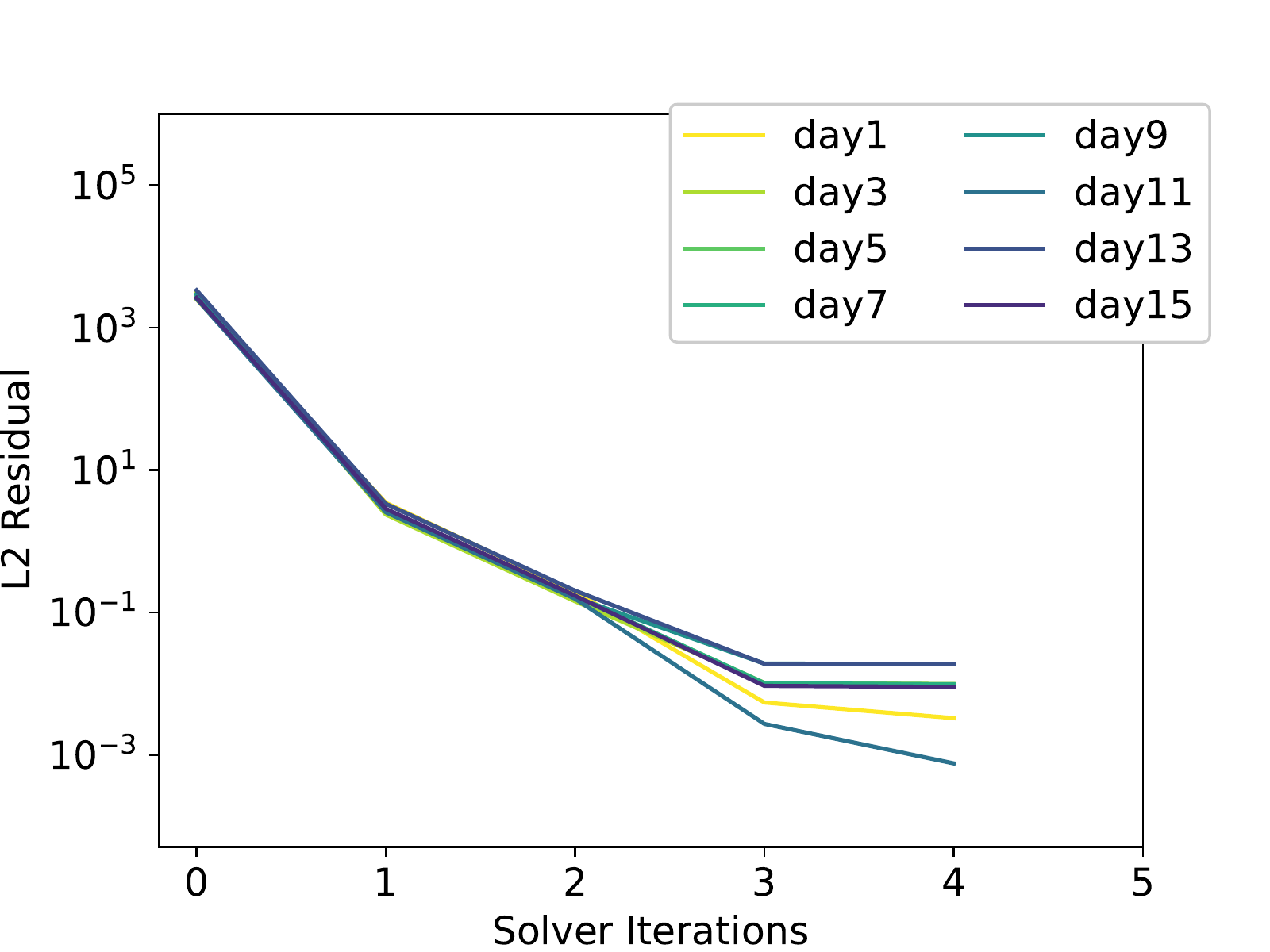} 
\endminipage\hfill
\minipage{0.48\textwidth}\centering
(f) Mixed-Precision
 \includegraphics[width=1.0\linewidth, trim=0cm 0cm 0 1.2cm, clip]{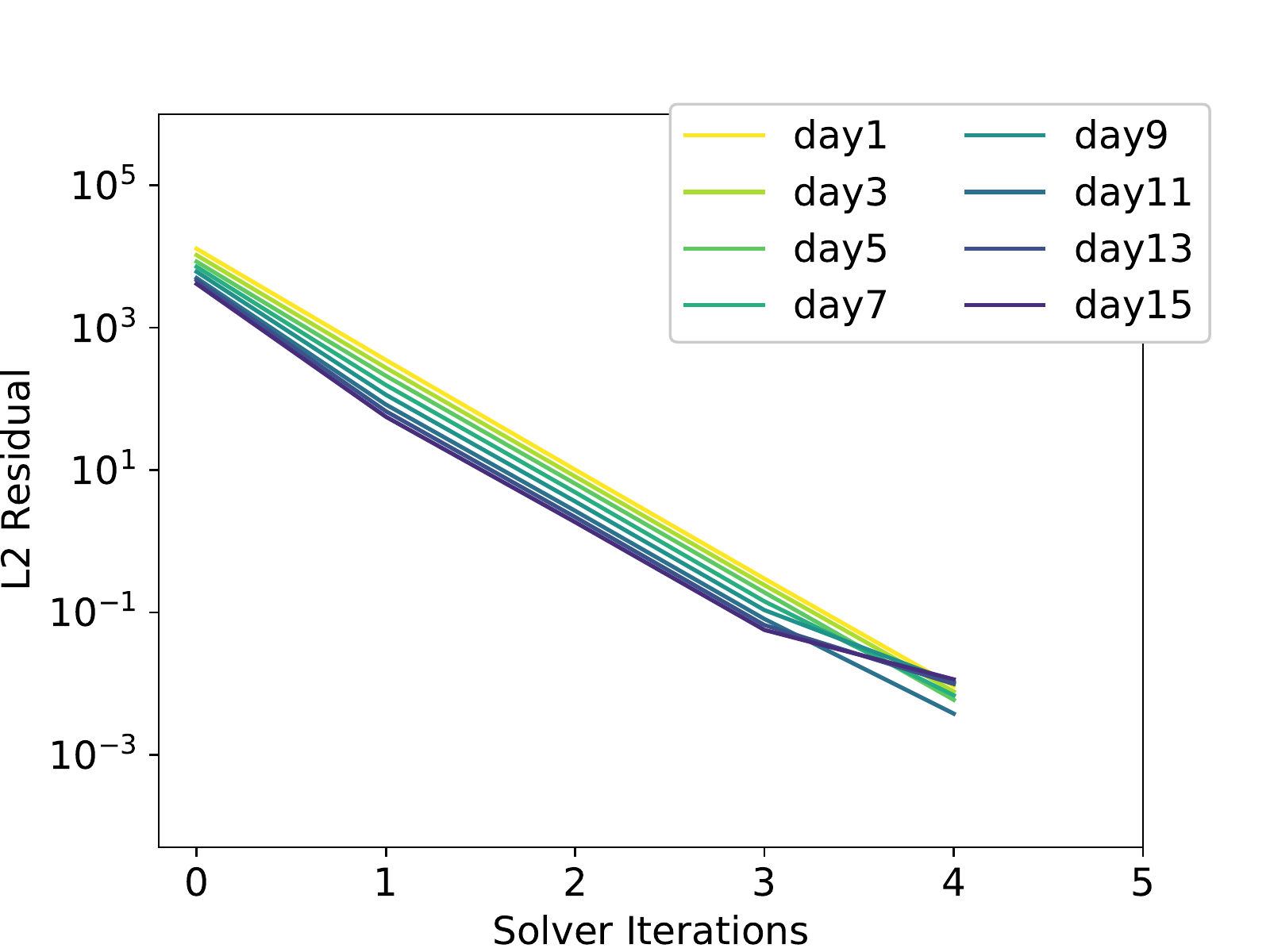}
\endminipage

\caption{Convergence rates of the RHW4 (left) and the orographic flow test-case (right) for three variants of the shallow-water model, a) and b) use double precision arithmetic, c) and d) single precision, e) and f) a mixture of double and single precision. Convergence is shown for 2-day time interval snapshots, and darker colours indicate later simulation time.}
\label{fig:convergence_DPSPMP}
\end{figure}

In double precision arithmetic, the solver is converging steadyly throughout all solver iterations, see figures \ref{fig:convergence_DPSPMP} a) and b). The initial residual is within the range of $10^{3}$ to $10^{4}$ for both test-cases. The solver requires 4 solver iterations for both test-cases which is also the minimal iteration number since we enforce for at least one full cycle of GCR(3). After these 4 iterations, the initial residual is reduced by almost 8 orders of magnitude for the RHW4 and about 6 orders of magnitude for the orographic flow.

\begin{table}
\begin{tabular}{|c|c|c|c|c|}
\hline
Solver iteration & 1 & 2 & 3 & 4 \\
\hline\hline
RHW4 & 
\begin{tabular}{c}
$0.6 m$ \\ $3.5 m$
\end{tabular} 
&  
\begin{tabular}{c}
$5 \cdot 10^{-4}m$ \\ $2 \cdot 10^{-3}m$
\end{tabular} 
& 
\begin{tabular}{c}
$2 \cdot 10^{-5}m$ \\ $3 \cdot 10^{-4}m$
\end{tabular}  
& 
\begin{tabular}{c}
$5 \cdot 10^{-7}m$ \\ $6 \cdot 10^{-6}m$
\end{tabular} 
\\
\hline
Orographic Flow & 
\begin{tabular}{c}
$1.8m$ \\ $48m$
\end{tabular} 
&  
\begin{tabular}{c}
$4 \cdot 10^{-2}m$ \\ $1.1m$
\end{tabular} 
& 
\begin{tabular}{c}
$1 \cdot 10^{-3}m$ \\ $4 \cdot 10^{-2}m$
\end{tabular}  
& 
\begin{tabular}{c}
$3 \cdot 10^{-5}m$ \\ $8 \cdot 10^{-4}m$
\end{tabular} 
\\ \hline

\end{tabular}
\caption{\label{tab:table1} Size of the increments for fluid thickness $\Phi_0$ for the RHW4 and the orographic flow test-case at different iterations of the elliptic solver. The upper values provide the average magnitude and the lower values the maximum value throughout the model run.}
\end{table}

To get a better intuition for the actual size of the corresponding impact to fluid thickness, and how these increments vary over successive solver iterations, statistics of fluid thickness increments are shown in table \ref{tab:table1}. For both test-cases, fluid thickness increments drastically reduce in magnitude over successive solver iterations by 5-6 orders of magnitudes. For the RHW4, this process happens more quickly, consistent with the slightly higher convergence rate for this test-case. 

Given these results, the significance of using $\epsilon=10^{-5}$ should be discussed. For the orographic flow test-case, the fluid thickness increments can still be up to several centimeters in magnitude until the fourth solver iteration. Errors of such magnitude within a single time-step are unacceptable. For the RHW4, the strongly drecreased fluid thickness increments after the first solver iteration suggest that it would be sufficient to only perform one single solver iteration, or at most two. However, experiments with an elliptic solver restricted to only one or two solver iterations respectively lead to model crashes due to exponentially growing instabilities, in the case of a single solver iteration even within the first twelve hours of simulation time.

In comparison to the double precision convergence behaviour, single precision solver convergence rates change significantly, see figures \ref{fig:convergence_DPSPMP} c) and d). Concerning criterion (i) and (ii), for both test-cases the initial residual is two orders of magnitudes larger and only reduced by three orders of magnitude by the elliptic solver. The solver converges more slowly, seemingly plateauing after the first solver iteration. The single precision solver is typically performing five instead of four solver iterations for double precision for RHW4.

While there are differences in the convergence, concerning criterion (iii) the solution quality is still good. The differences in the normalised deviations from the respective ``genuine'' model solution in the $L_2$ and $L_\infty$ norms are insignificant compared to the double precision solution. The largest difference is found for the $L_\infty$-norm of kinetic energy perturbations for the orographic flow with a value of 758\% for the single precision shallow-water model, a relative change of $5 \cdot 10^{-3}$ compared to the double precision solution. 

\begin{figure}[!htb] 
\vspace{0.3cm}
\minipage{0.48\textwidth}\centering\textbf{RHW4}\par\medskip
(a) DP
 \includegraphics[width=1.0\linewidth, trim=0cm 2cm 1.5cm 2.5cm, clip]{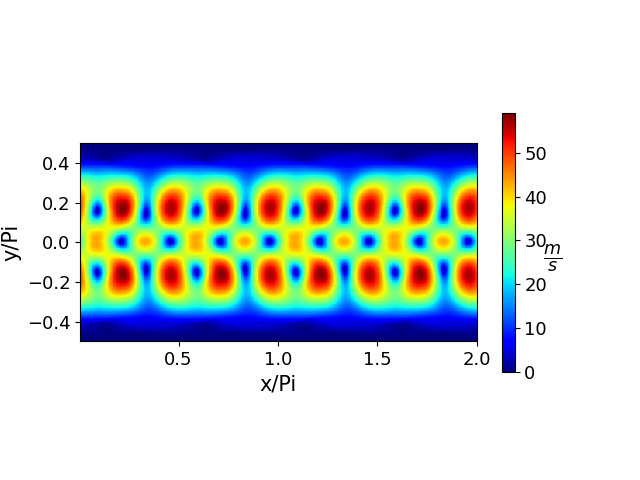} 
\endminipage\hfill
\minipage{0.48\textwidth}\centering\textbf{Orographic Flow}\par\medskip
(b) DP
 \includegraphics[width=1.0\linewidth, trim=0cm 2cm 1.5cm 2.5cm, clip]{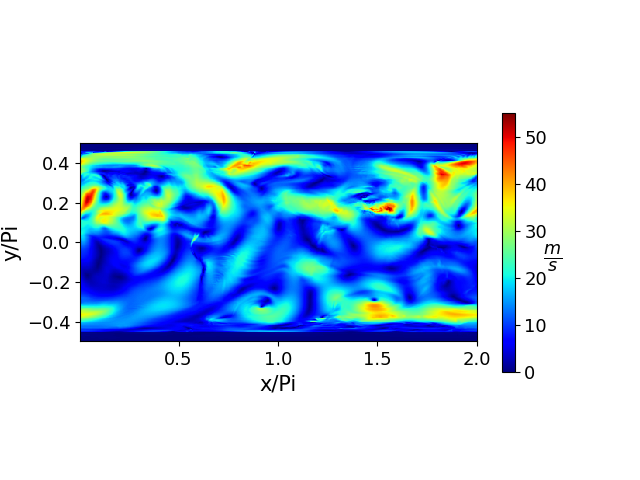}
\endminipage
\vspace{0.3cm}
\minipage{0.48\textwidth}\centering
(c) SP
 \includegraphics[width=1.0\linewidth, trim=0cm 2cm 1.5cm 2.5cm, clip]{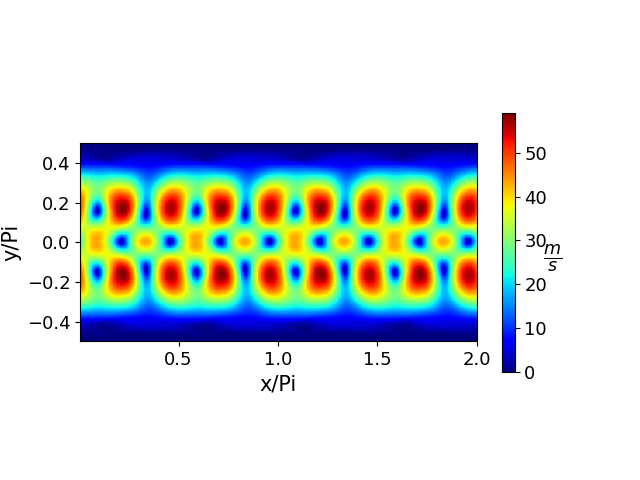} 
\endminipage\hfill
\minipage{0.48\textwidth}\centering
(d) SP
 \includegraphics[width=1.0\linewidth, trim=0cm 2cm 1.5cm 2.5cm, clip]{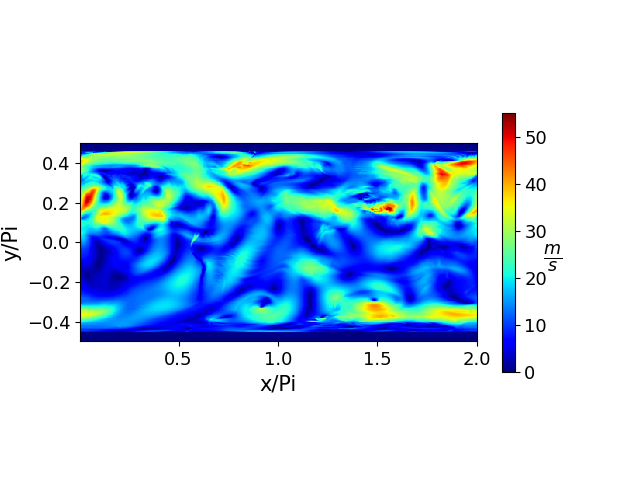}
\endminipage
\vspace{0.3cm}
\minipage{0.48\textwidth}\centering
(e) Difference SP-DP
 \includegraphics[width=1.0\linewidth, trim=0cm 2cm 1.5cm 2.5cm, clip]{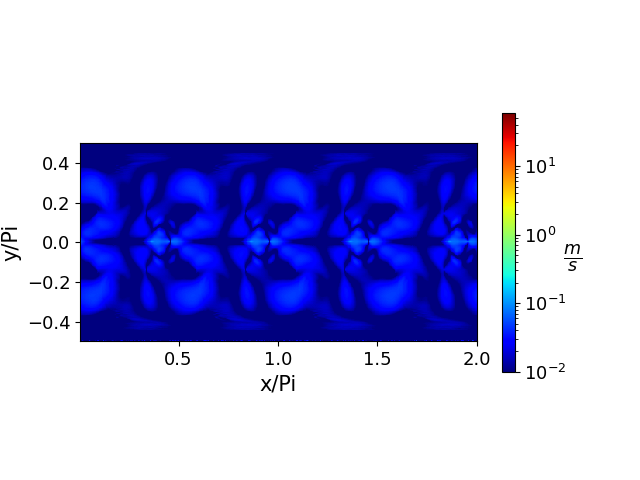} 
\endminipage\hfill
\minipage{0.48\textwidth}\centering
(f) Difference SP-DP
 \includegraphics[width=1.0\linewidth, trim=0cm 2cm 1.5cm 2.5cm, clip]{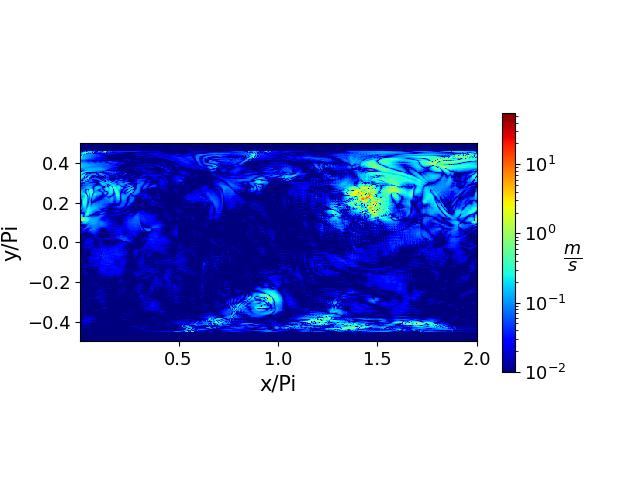}
\endminipage

\caption{Square-root of kinetic energy deviations in $\frac{m}{s}$ from the respective ``genuine'' model solution at $t=14.76$ days for the RHW4 in a) double precision, c) single precision, and the orographic flow test-case in b) double precision, d) single precision. Figures e) and f) show the respective difference in kinetic energy deviations between the double precision and single precision shallow-water model solution in logarithmic scale.}
\label{fig:deviations_DPSP}
\end{figure}

Further elaborating on this point, the above-discussed kinetic energy perturbations as well as their change when going from double to single precision are shown in figures \ref{fig:deviations_DPSP} a)-f). The square-root of the deviations can be up to 50 $\frac{m}{s}$ for both test-cases. For RHW4 the departures are centered around the wave crests and differences are symmetric between both hemispheres. When going from double to single precision, the differences in the deviations are shown over a range of 4 orders of magnitude from $10^{-2}$ to $10^{2}$. The difference is at least two orders of magnitude smaller than the values of the deviations. These differences can be considered insignificant, especially when considering how sensitive the test-case is towards perturbations due to its inherent unstable character. 

For the orographic flow, the deviations from the genuine solution show the locations and magnitude of synoptic-scale eddies, as well as regions with high mountains, such as Himalaya, where fluid thickness is small compared to orography and orography gradients are large making the associated flow exhibit strongly non-linear behaviour. The most significant differences between single and double precision are centered around the Himalaya and its downstream area. The differences are about one order of magnitude smaller than the deviations to the genuine solution.

To further put the change in kinetic energy deviations into perspective, we perform an additional simulation in double precision that is run from randomly perturbed initial conditions for fluid thickness (adding white noise with an amplitude of five centimeters). These perturbations represent the inherent uncertainty of the system and inevitable errors in the initial conditions for simulations of W\&C. It is found that the changes in kinetic energy departures between the perturbed and unperturbed double precision simulations are comparable to the differences between the simulations in single and double precision. Thus, the change in departures when going to single precision will probably be insignificant when compared to the uncertainty range typically encountered in W\&C prediction and can therefore be considered as negligible.

In summary, although the solver of the single precision model variant behaves differently, the model still produces valid solutions for our test-cases.

The fact that the shallow-water model is crashing if only a single solver iteration is performed in the elliptic solver seems to disagree with the result that the single precision model is producing valid solutions. This is because, although the single precision model uses more than one solver iteration, the convergence results in figures 5 c) and d) are showing no significant reduction of the residual error after the first solver iteration. Still, the additional iterations seem to improve the solution sufficiently to prevent model crashes. Therefore, the popular and widespread use of the $L_2$-norm of the residual error field to measure the progress of the elliptic solver seems to have reached its limits for the reduced precision simulation. Further tests with the $L_\infty$-norm as exit condition yield the same result. This clearly indicates that, when using reduced precision in the elliptic solver, extra care needs to be taken when analyzing elliptic solver convergence and also when selecting a suitable---physically-relevant for the problem at hand---exit condition.

To understand why the single and double precision solvers show a different convergence behaviours, we first take a look at the fluid thickness variable $\Phi$. The fluid thickness variable is stored in single precision arithmetic and its magnitude is on the order of $10^{4}$ to $10^{5}$ meters. In comparison, a typical fluid thickness increment at each time-step is on the order of (see again table \ref{tab:table1}) $3 \cdot 10^{-5}m$ to $5 \cdot 10^{-7}m$ for the last solver iteration. However, the relative error of single precision truncation is given by the machine epsilon $\approx 1.19\cdot 10^{-7}$, indicating that thickness increments smaller than $10^{-2}m$ to $10^{-3}m$ cannot reliably be represented in the fluid thickness field $\Phi$. This renders most of the later solver increment field unused, even if the single precision solver were to perform as good as the double precision elliptic solver.

\begin{figure}[!htb] 

\vspace{0.3cm}
\minipage{0.48\textwidth}\centering\textbf{RHW4}\par\medskip
 \includegraphics[width=1.0\linewidth, trim=0cm 0cm 0 1.2cm, clip]{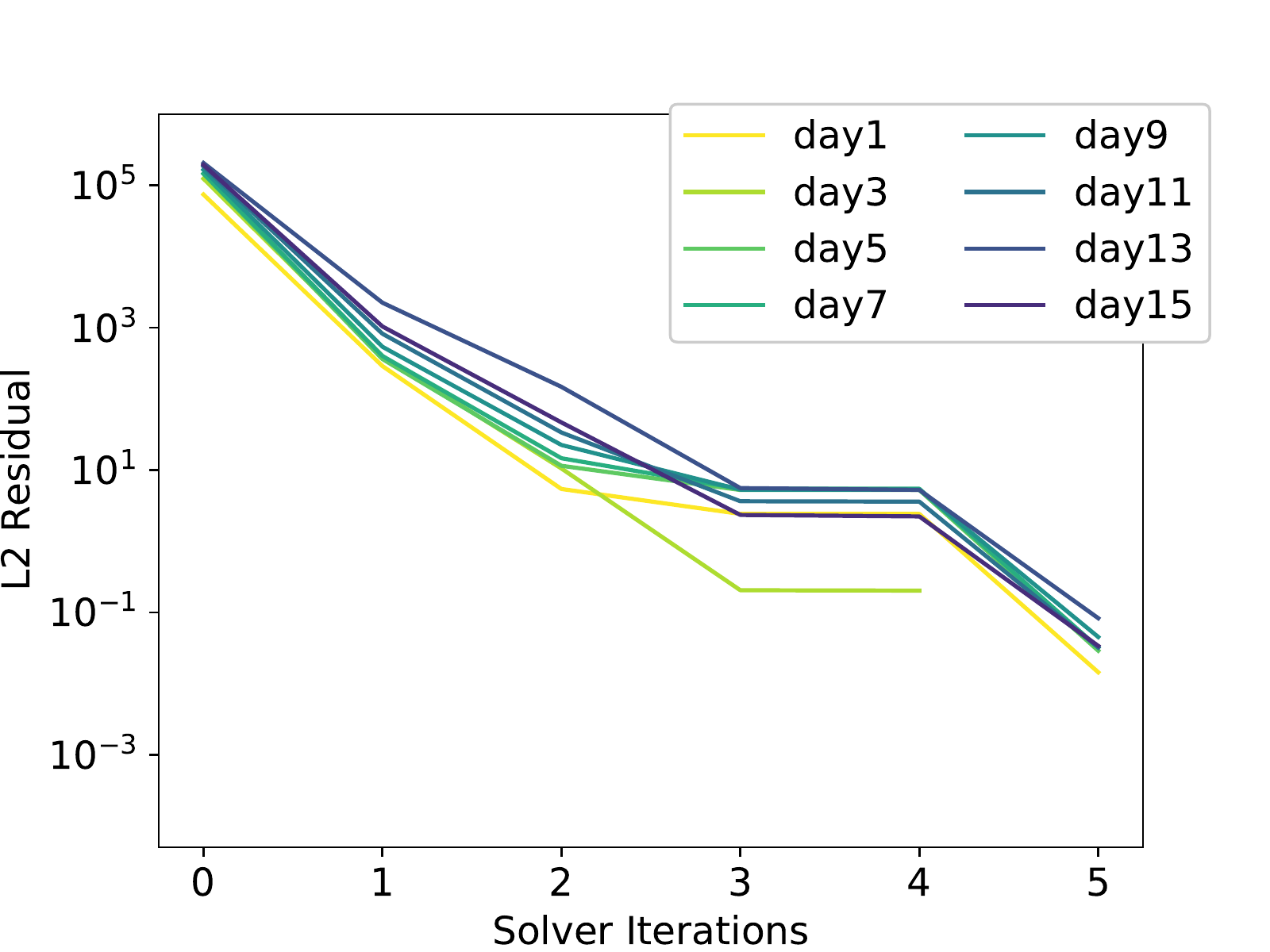} 
\endminipage\hfill 
\minipage{0.48\textwidth}\centering\textbf{Orographic Flow}\par\medskip
 \includegraphics[width=1.0\linewidth, trim=0cm 0cm 0 1.2cm, clip]{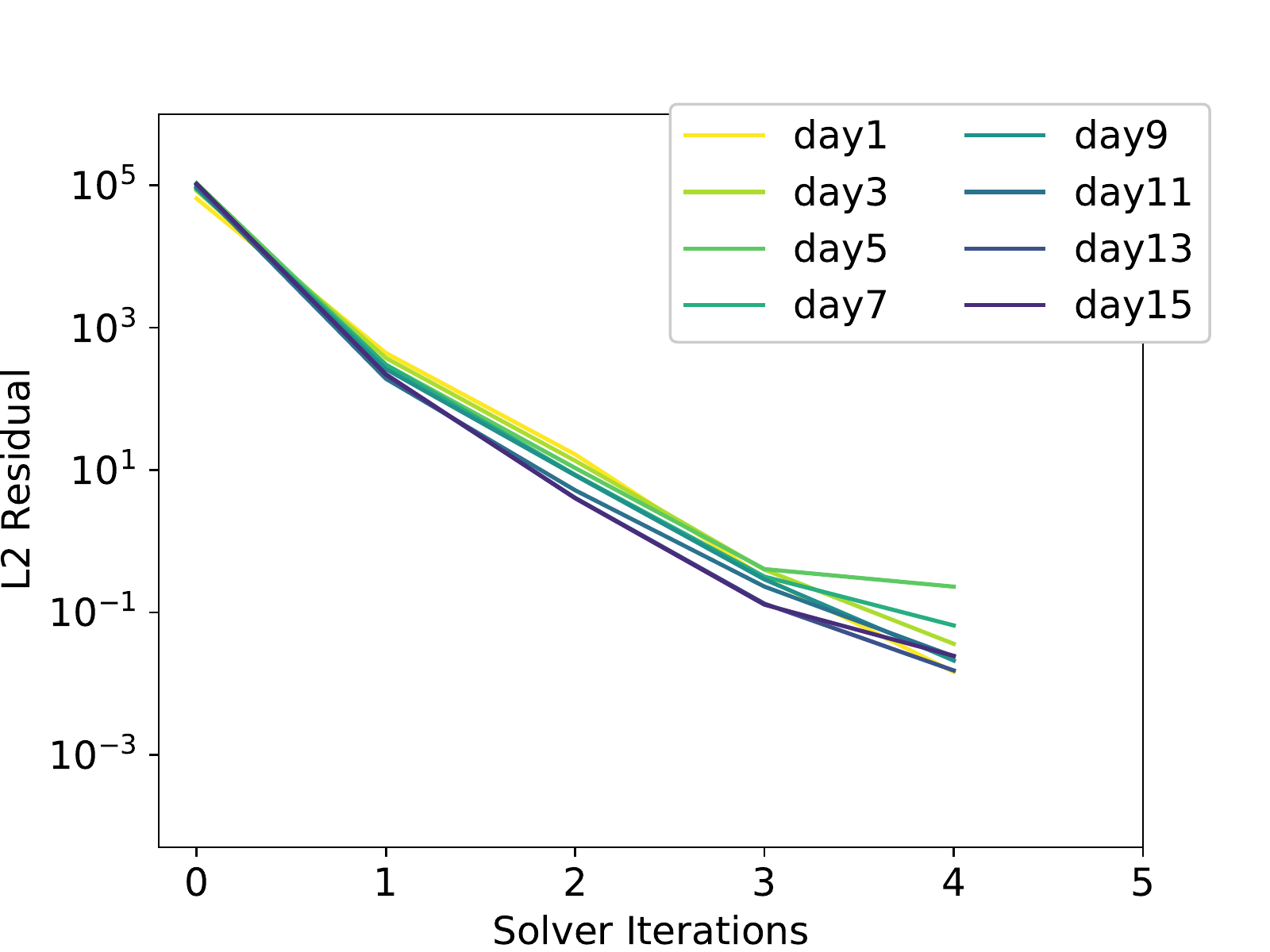}
\endminipage

\caption{Convergence rates of the RHW4 (left) and the orographic flow test-case (right) for the shallow-water model using single precision. In contrast to figures \ref{fig:convergence_DPSPMP} c) and d), the residual error fields are however calculated from a double precision copy of the initial fluid thickness $\Phi_0$ that is updated by the solver increments after each solver iteration.}
\label{fig:convergence_sp_r_nu}
\end{figure}

This gives an indication that the single precision solver variant might actually be performing better than figures \ref{fig:convergence_DPSPMP} c) and d) indicate. This can be investigated by additionally introducing a double precision copy of the initial fluid thickness $\Phi_0$ into the single precision elliptic solver and calculating the residual error fields from this double precision variable. The double precision copy of $\Phi_0$ is continuously kept updated by the single precision solver's fluid thickness increments after each solver iteration. The convergence rates derived in this way, see figure \ref{fig:convergence_sp_r_nu}, look drastically different and much closer to the convergence rates of the double precision variant of the elliptic solver from figures \ref{fig:convergence_DPSPMP} a) and b). The initial residual is still larger when compared to the double precision simulations, but throughout the solver iterations the initial residual error is now reduced by 5 to 6 orders of magnitude. This shows that the single precision solver variant actually converges well even after the first solver iteration, the solver increments are just not reliably being added to the fluid thickness field $\Phi$.

To understand the remaining differences, i.e. the much larger magnitude of initial residual errors and the still reduced convergence rate, the solver's convergence is investigated for each latitude separately. Given the insights from the last paragraph, from now on residual error fields are always calculated from a double precision copy of $\Phi$. The reason being that for the mixed-precision model, which we ultimately aim to design, to behave similarly to the double precision elliptic solver variant, an increase in precision appears to be unavoidable for the fluid thickness variable $\Phi$.

\begin{figure}[!htb] \centering\textbf{Orographic Flow}\par\medskip
\vspace{0.3cm}
\minipage{0.48\textwidth}\centering\textbf{Initial Residual $r_0$}\par\medskip
(a) DP
 \includegraphics[width=1.0\linewidth, trim=0cm 0cm 0cm 0.2cm, clip]{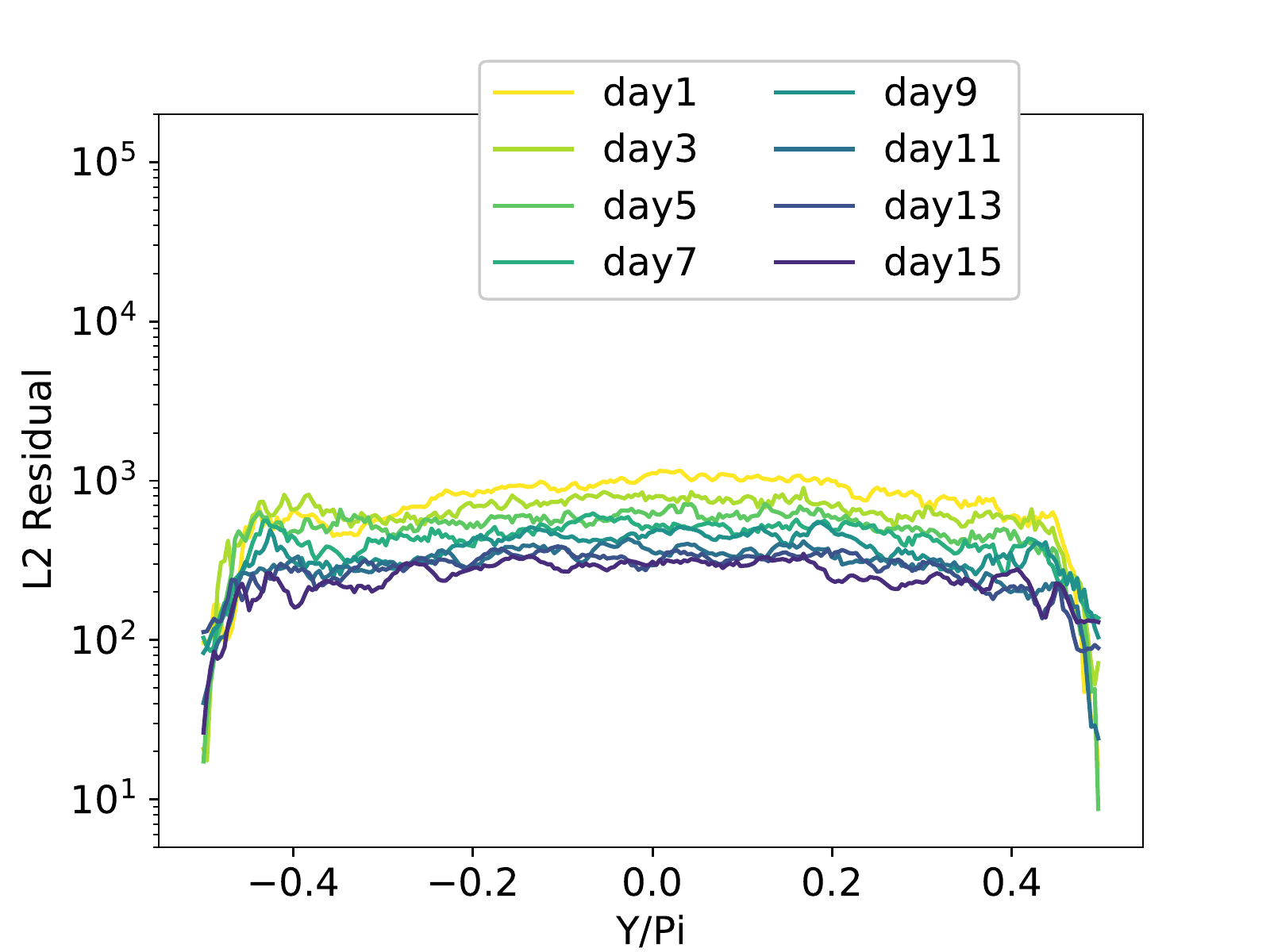} 
\endminipage\hfill
\minipage{0.48\textwidth}\centering\textbf{Final Residual $r_N$}\par\medskip
(b) DP
 \includegraphics[width=1.0\linewidth, trim=0cm 0cm 0 0.2cm, clip]{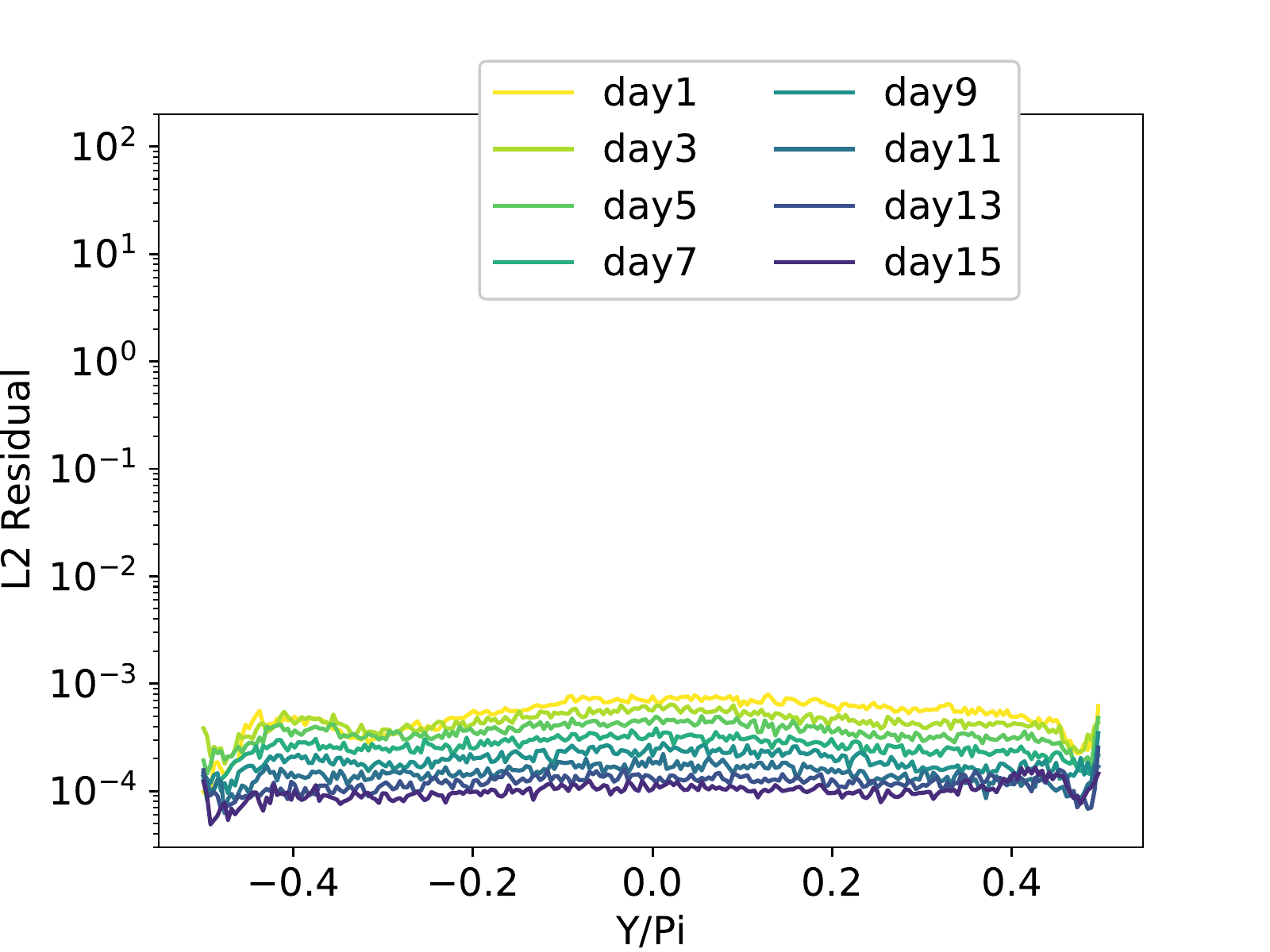}
\endminipage
\vspace{0.3cm}
\minipage{0.48\textwidth}\centering
(c) SP
 \includegraphics[width=1.0\linewidth, trim=0cm 0cm 0 0.2cm, clip]{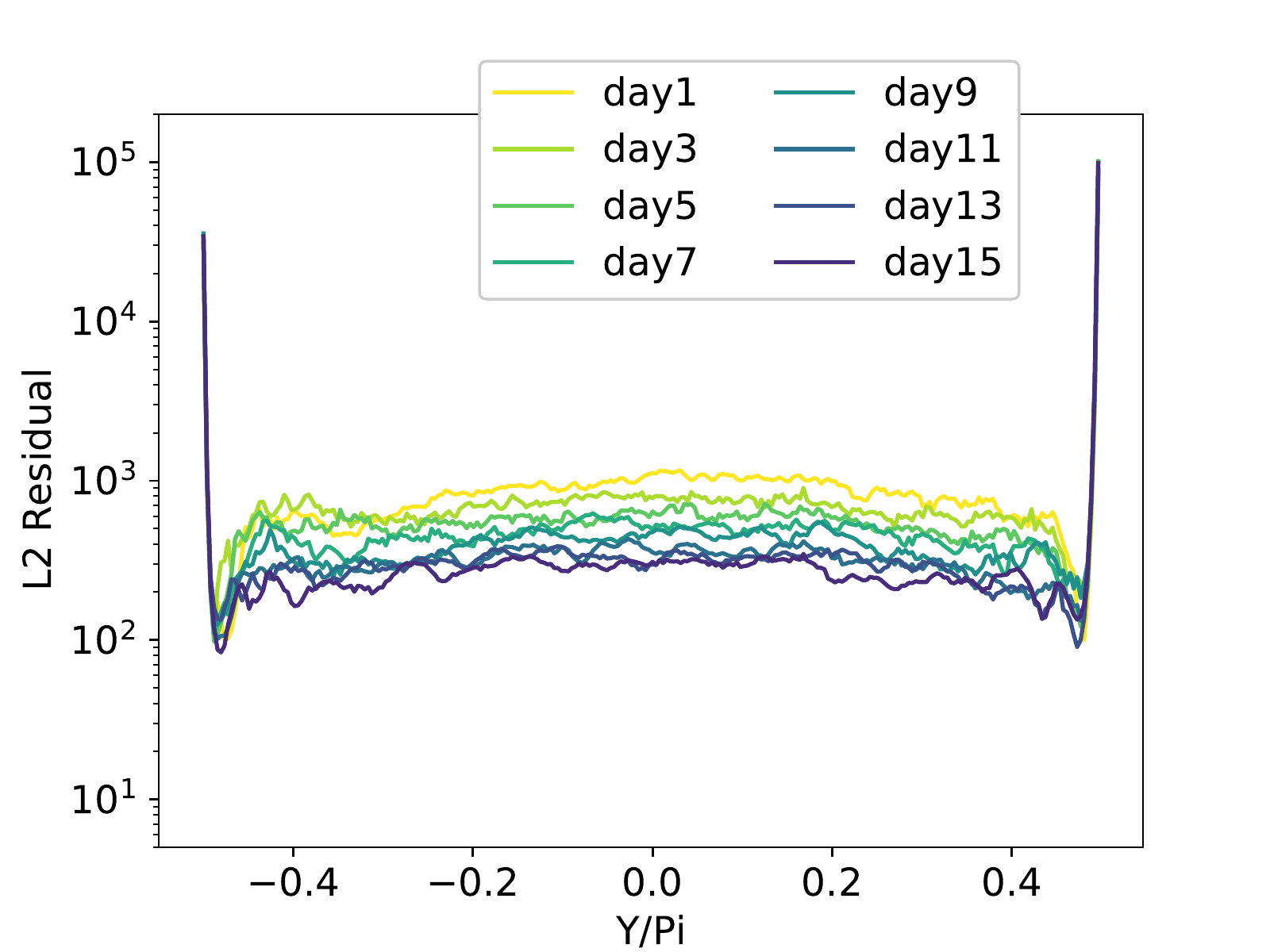} 
\endminipage\hfill
\minipage{0.48\textwidth}\centering
(d) SP
 \includegraphics[width=1.0\linewidth, trim=0cm 0cm 0 0.2cm, clip]{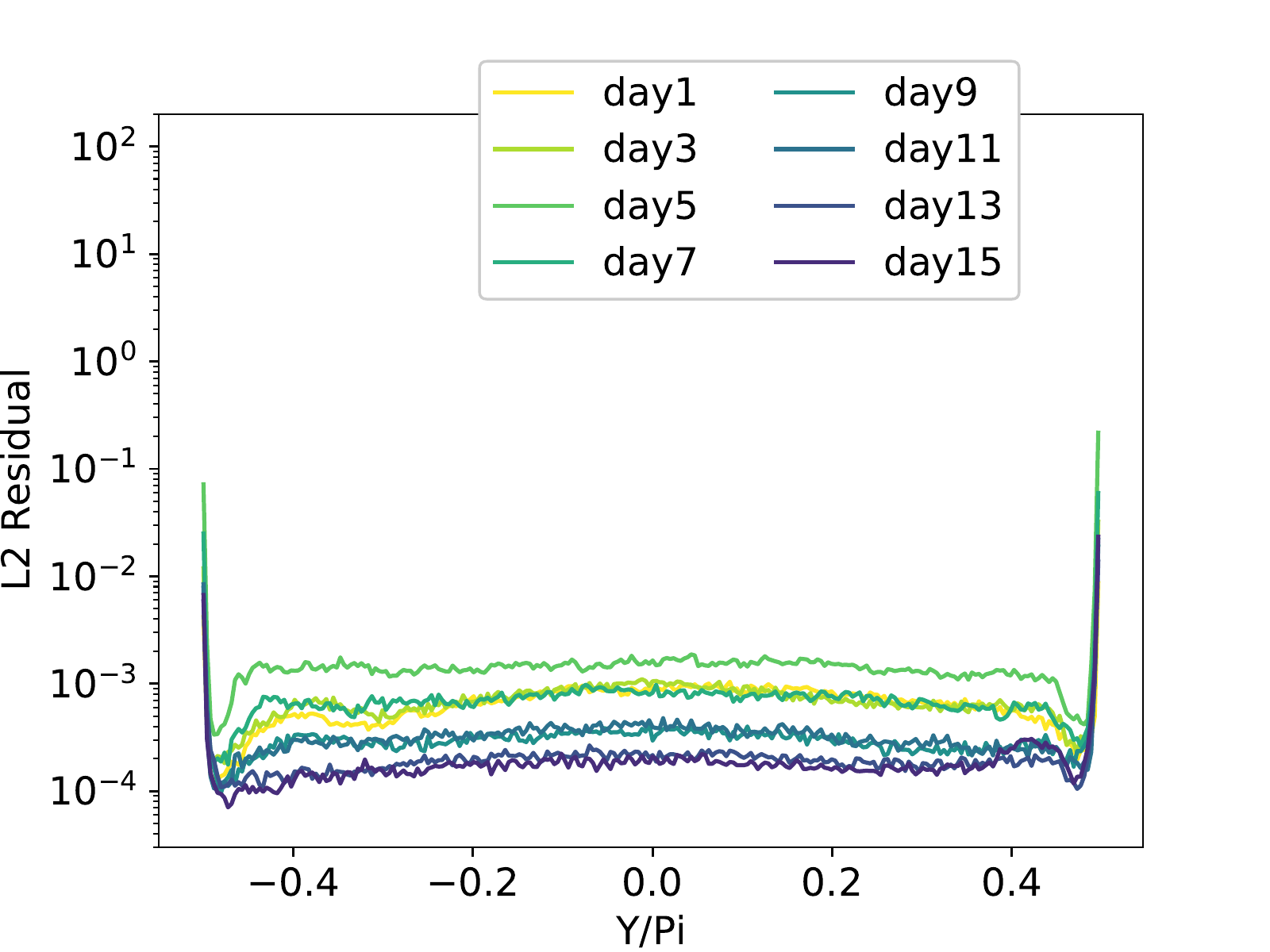}
\endminipage
\vspace{0.3cm}
\minipage{0.48\textwidth}\centering
(e) Mixed-Precision
 \includegraphics[width=1.0\linewidth, trim=0cm 0cm 0 0.2cm, clip]{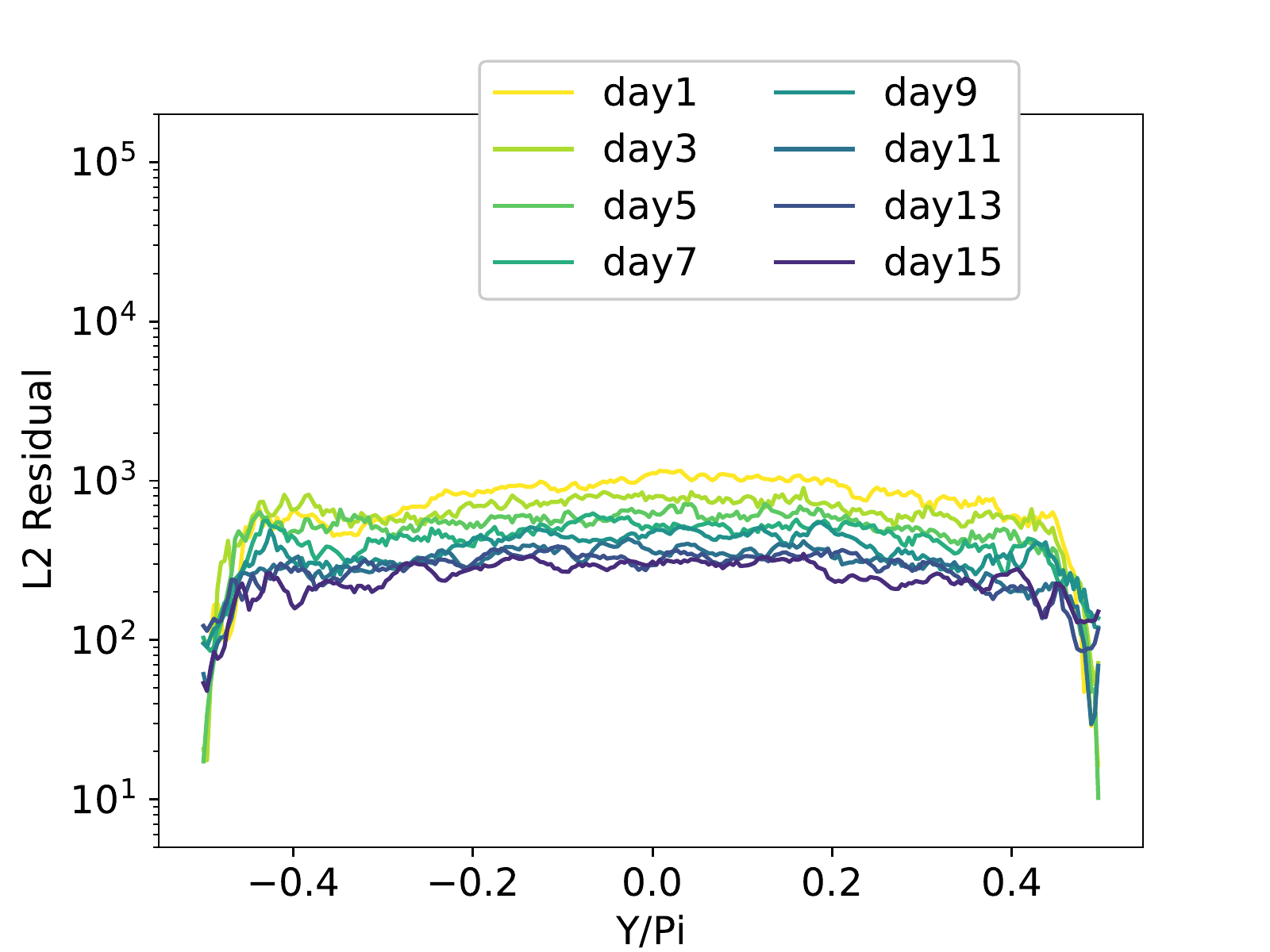} 
\endminipage\hfill
\minipage{0.48\textwidth}\centering
(f) Mixed-Precision
 \includegraphics[width=1.0\linewidth, trim=0cm 0cm 0 0.2cm, clip]{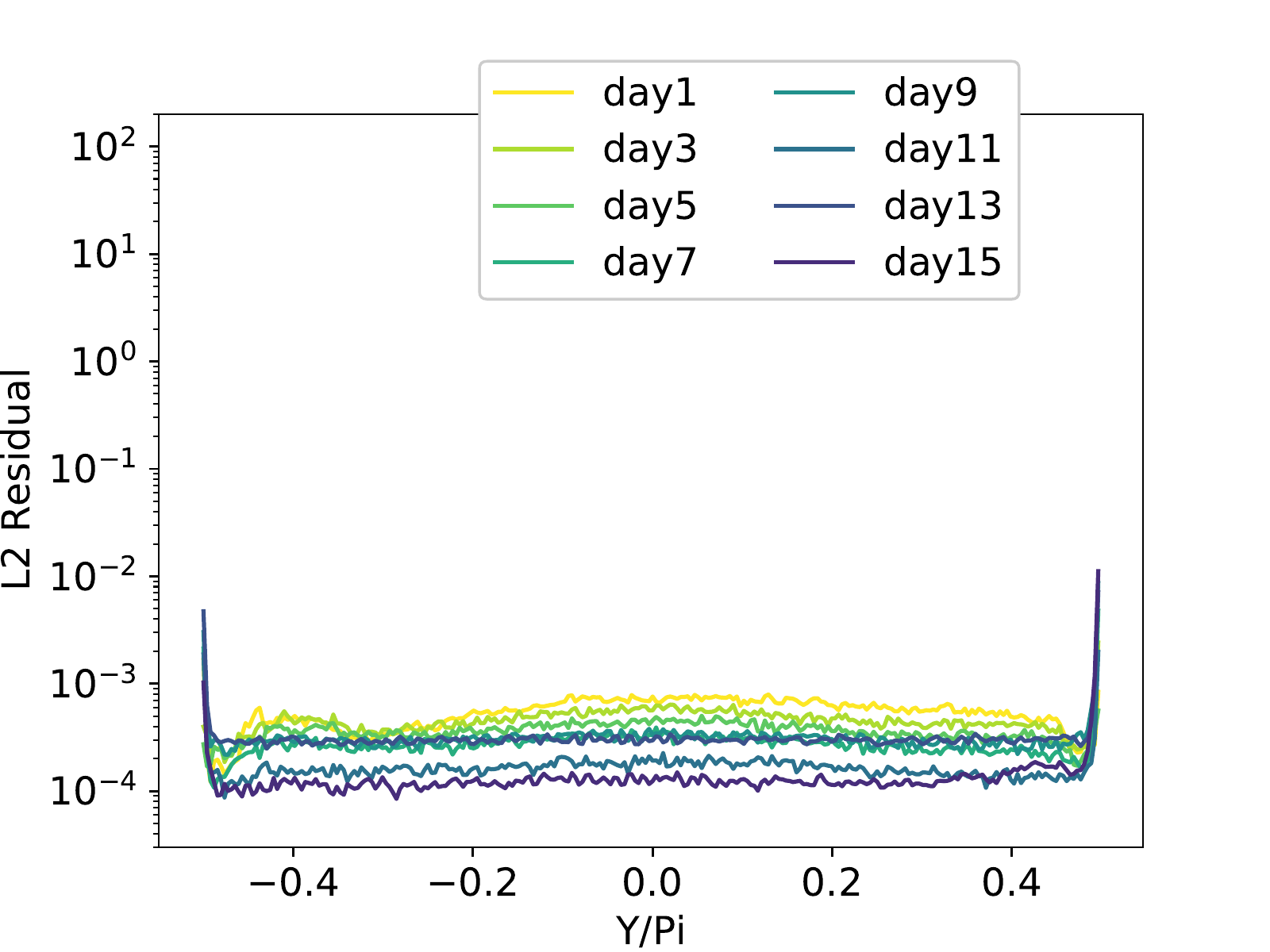}
\endminipage

\caption{Latitude-wise inital and final residual in the $L_2$-norm for the orographic flow test-case for three variants of the shallow-water model, a) and b) use double precision arithmetic, c) and d) single precision, e) and f) a mixture of double and single precision.}
\label{fig:Lat_Convergence_DPSPMP}
\end{figure}

Figures \ref{fig:Lat_Convergence_DPSPMP} a)-d) show the convergence rates for each latitude. Here only results for the orographic flow are shown as the RHW4 test-case behaves qualitatively similar. The differences in the $L_2$-norm of the residual values are only visible near the poles, where the single precision model's values are three to four orders of magnitude larger than for double precision for all solver iterations throughout the simulation. For the other latitudes, the $L_2$-norms of the residual error fields are however indistinguishable by eye for single and double precision. When the solver reaches convergence, see figure \ref{fig:Lat_Convergence_DPSPMP} b) and d), the double precision elliptic solver has managed to consistently reduce the $L_2$-norm of the initial residual values by five orders of magnitude near the poles and six orders of magnitude in the rest of the computational domain. In comparison, in the single precision model variant the initial values are decreased by about six orders of magnitude everywhere.
 
\begin{figure}[!htb] 
\vspace{0.3cm}
\minipage{0.48\textwidth}\centering\textbf{RHW4}\par\medskip
 \includegraphics[width=1.0\linewidth, trim=0cm 0cm 0cm 0cm, clip]{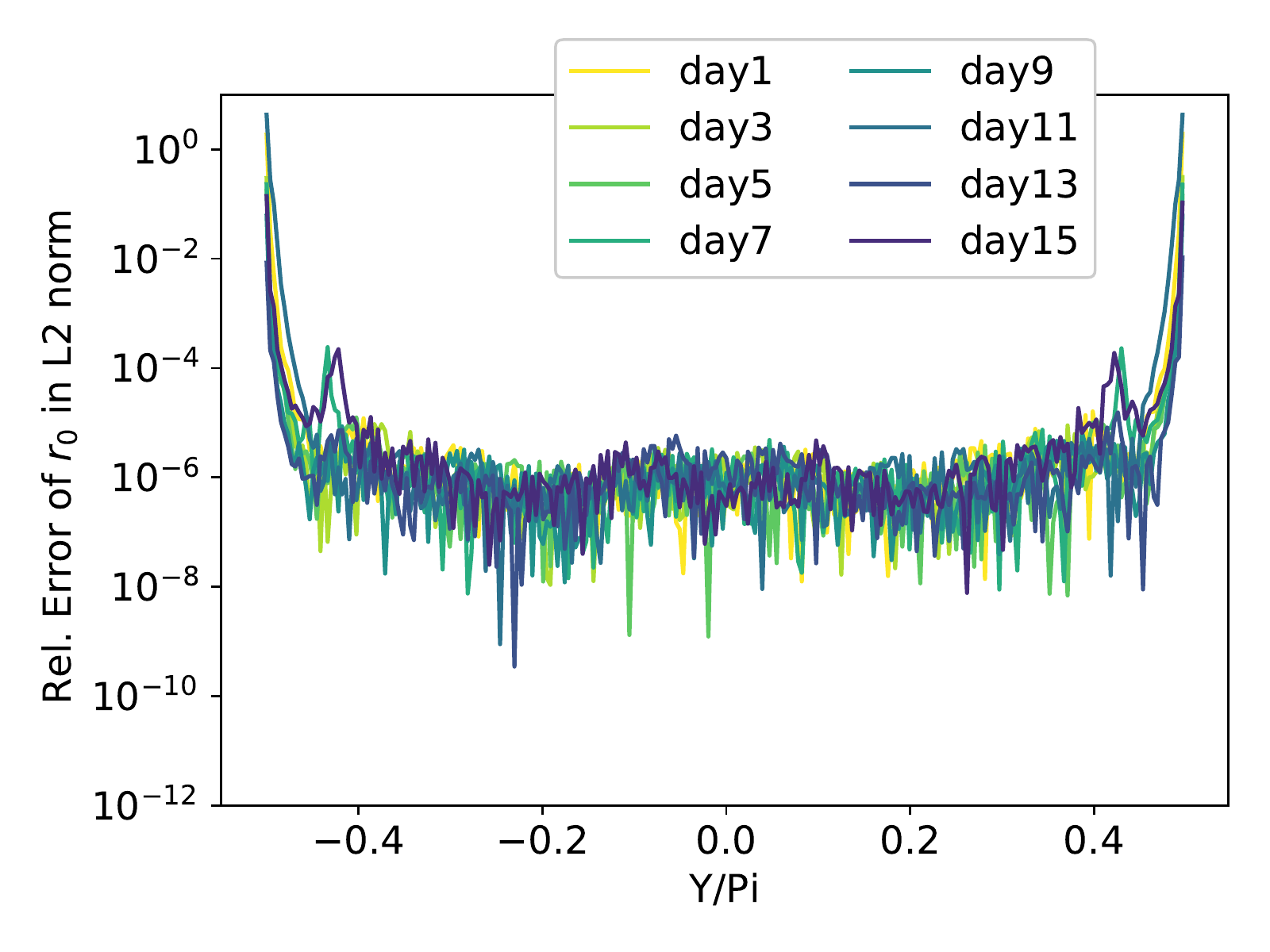} 
\endminipage\hfill
\minipage{0.48\textwidth}\centering\textbf{Orographic Flow}\par\medskip
 \includegraphics[width=1.0\linewidth, trim=0cm 0cm 0 0cm, clip]{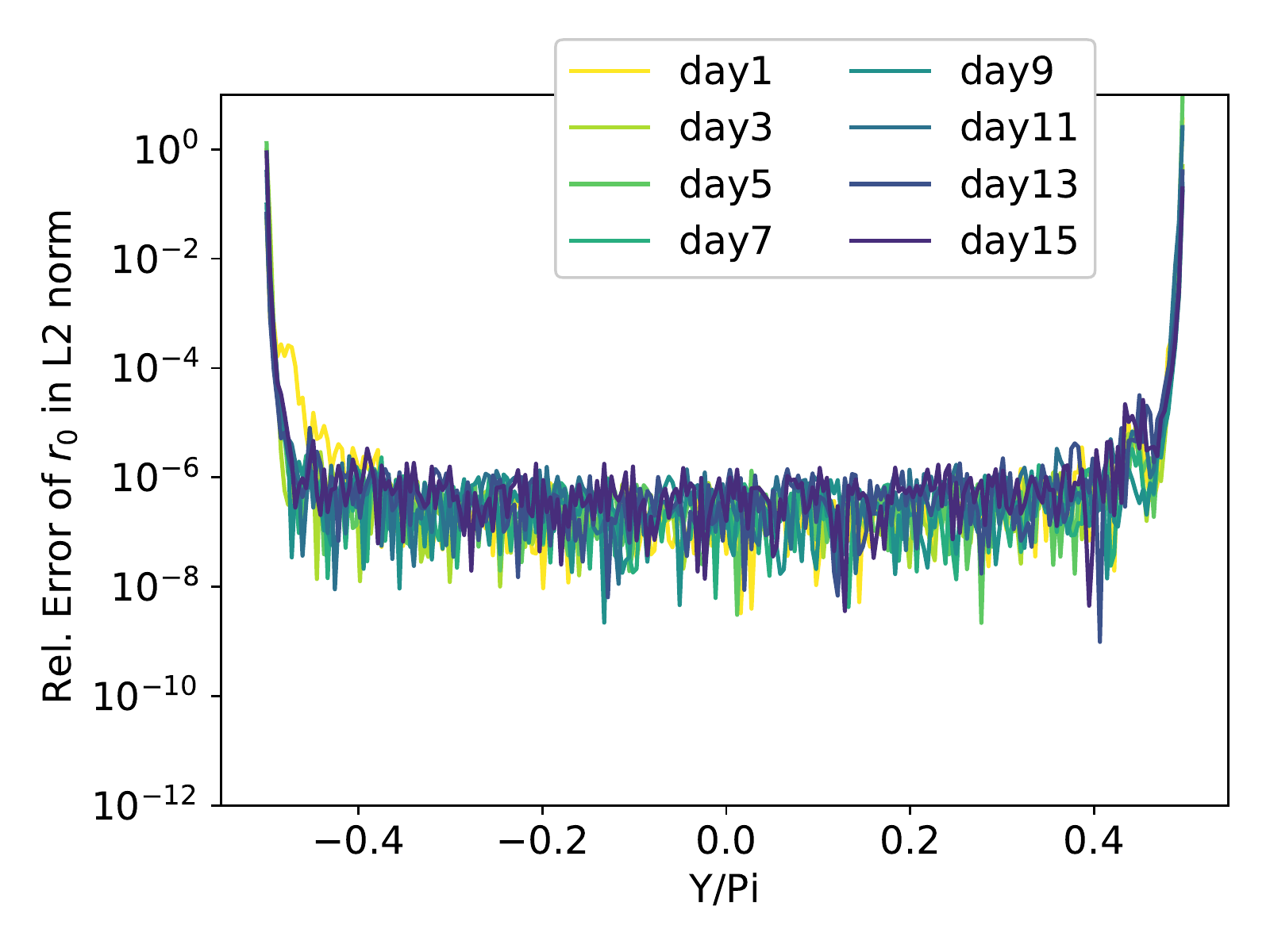}
\endminipage

\caption{Latitude-wise relative numerical error in the $L_2$-norm of the inital residual $r_0$ when the elliptic solver's initial residual is calculated with a fluid thickness variable $\Phi_0$ that was truncated to single precision. Results are shown for the RHW4 (left) and the orographic flow test-case (right) in 2-day time interval snapshots; darker colours indicate later simulation time.}
\label{fig:Lat_Error_MP}
\end{figure}

\begin{figure}[!htb] 

\vspace{0.3cm}
\minipage{0.48\textwidth}\centering\textbf{RHW4}\par\medskip
 \includegraphics[width=1.0\linewidth, trim=0cm 0cm 0 1.2cm, clip]{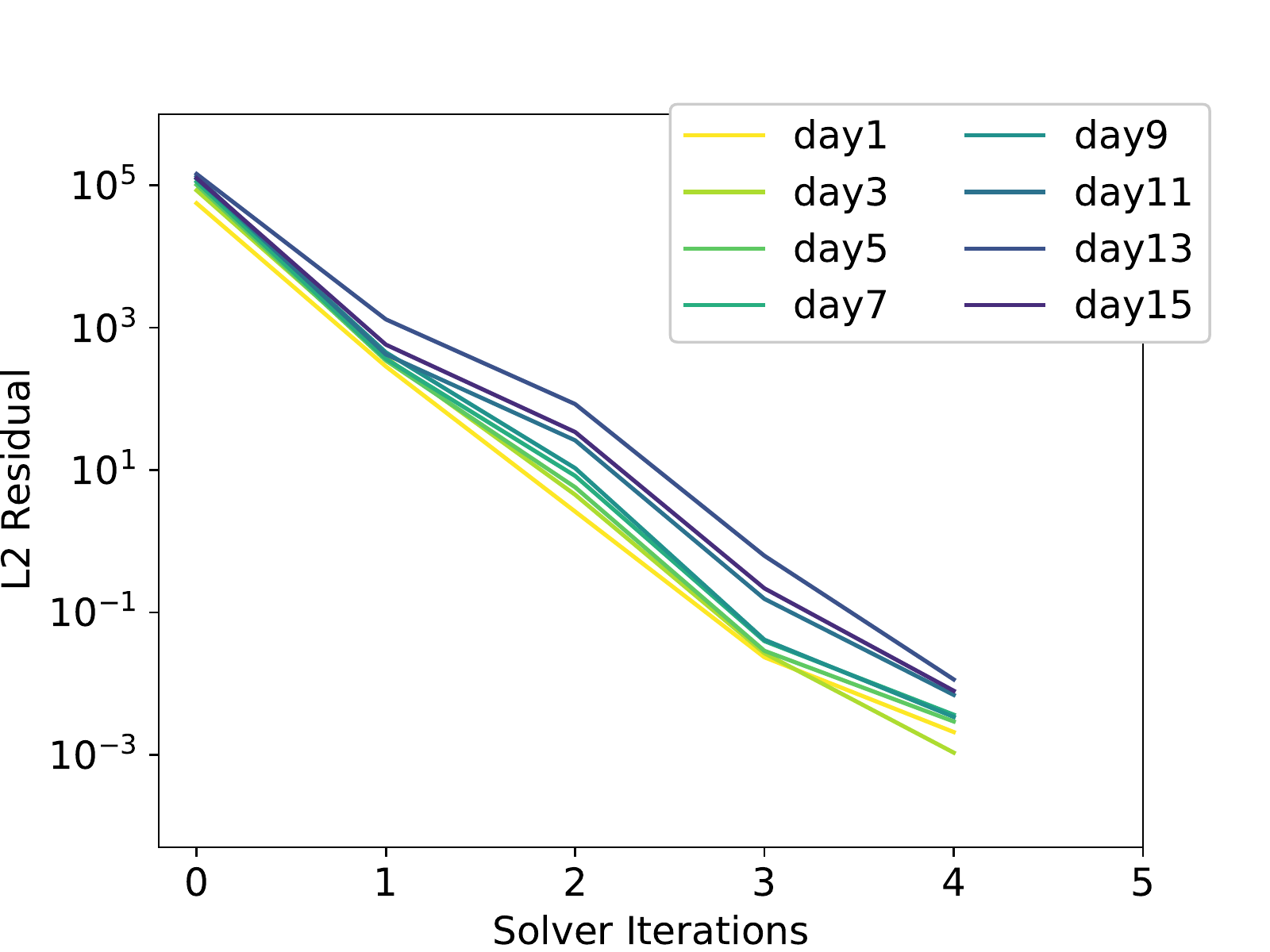} 
\endminipage\hfill 
\minipage{0.48\textwidth}\centering\textbf{Orographic Flow}\par\medskip
 \includegraphics[width=1.0\linewidth, trim=0cm 0cm 0 1.2cm, clip]{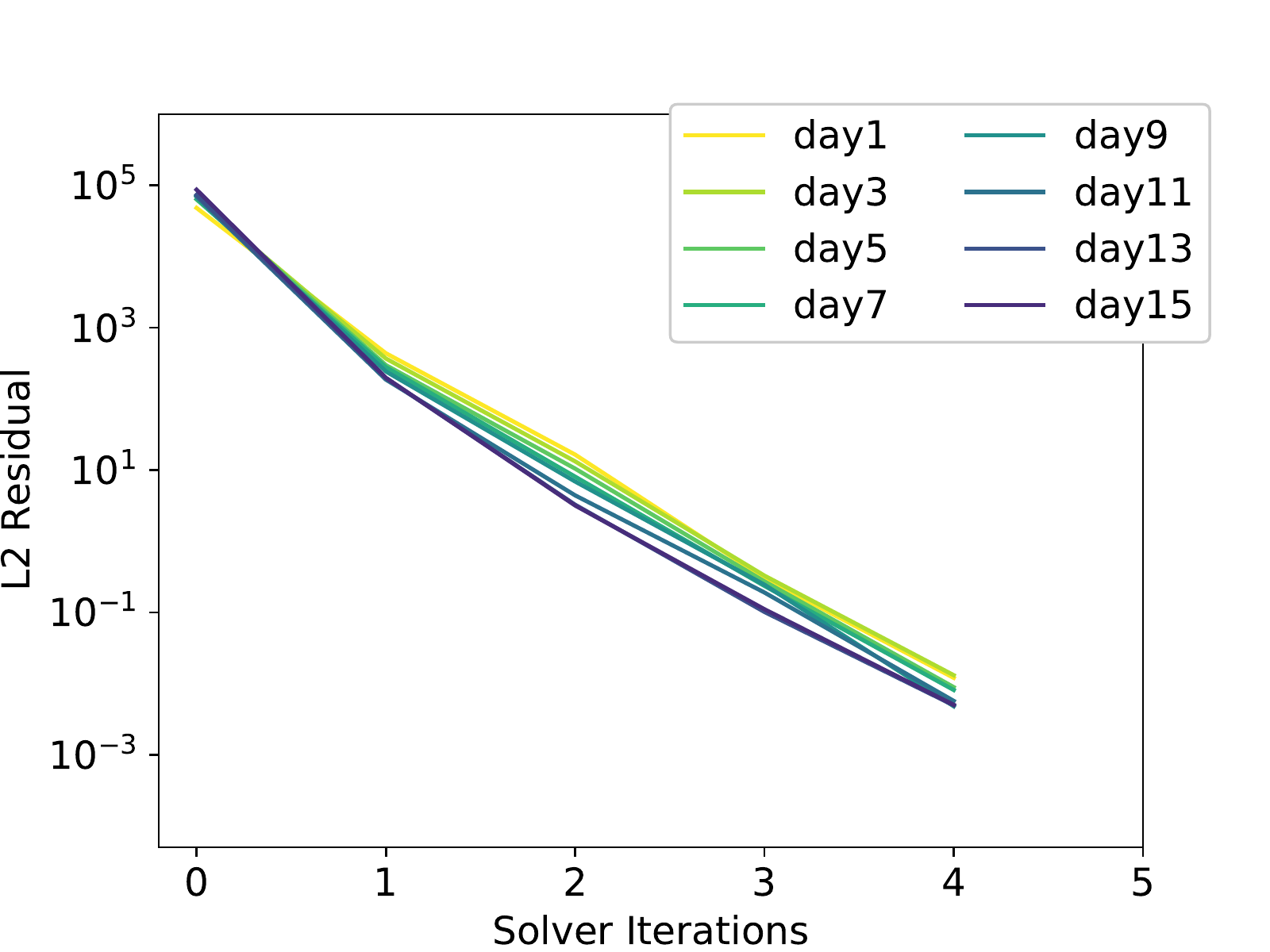}
\endminipage

\caption{Convergence rates of the RHW4 (left) and the orographic flow test-case (right) for the double precision shallow-water model variant where, in contrast to figures \ref{fig:convergence_DPSPMP} a) and b), the elliptic solver's initial residual $r_0$ is calculated with a fluid thickness variable $\Phi_0$ that was truncated to single precision.}
\label{fig:convergence_sp_Phi0}
\end{figure}

The accumulation of numerical errors near the poles of the single precision model can be replicated in the double precision model variant when the elliptic solver's initial residual $r_0$ is calculated with a fluid thickness $\Phi_0$ that was truncated to single precision. Even if all model calculations are performed in double precision arithmetic, the truncation leads to numerical errors in the $L_2$-norm of the initial residual error field $r_0$ on the order of $10^{-2}$ for the RHW4 and $10^{-3}$ for the orographic flow test-case when compared to a double precision solver without precision truncation. The spatial distribution of these numerical errors is illustrated in figure \ref{fig:Lat_Error_MP}. The relative error in the $L_2$-norm is around $10^{-5}$ for the majority of latitudes for both test-cases, while the relative numerical errors in the $L_2$-norm at the poles can easily be as large as $100 \%$. This shows how badly conditioned the linear operator $\mathcal{L}$ is near the poles. The resulting solver convergence of the truncated elliptic solver is illustrated in figures \ref{fig:convergence_sp_Phi0}, showing a clear degradation when compared to the double precision elliptic solver from figures \ref{fig:convergence_DPSPMP} a) and b).
The convergence behaviour is instead similar to the single precision elliptic solver variant that was observed in figure \ref{fig:convergence_sp_r_nu}, especially for the first two solver iterations. The precision of the fluid thickness variable $\Phi_0$ is thus of major importance for the calculation of the initial residual error field $r_0$ and the subsequent convergence behaviour of the elliptic solver. For substantiation, additional experiments are run using stronger polar absorbers. For the RHW4, the single precision model variant is run with $\alpha=1/(2\Delta t)$ but it is found that, although the numerical errors in the $L_2$-norm of the initial residual is reduced by half, solver convergence is not improved.

\subsection{Mixed-precision Shallow-Water Model}\label{MP_SWM}

Here, a mixed-precision model is described that shows a very similar convergence behaviour when compared to the double precision variant while performing as many calculations as possible in single precision.

First, based on our previous analysis, the fluid thickness field $\Phi$ used for the initial residual $r_0$ calculation in algorithmic step 1 of the elliptic solver and the update of the model variable $\Phi$ in step 2 should be kept in double precision. This motivates an approach to keep the prognostic model variables $\Phi$, $Q_{x}$, and $Q_{y}$ in double precision. The variables are kept and updated in double precision while the costly part, the calculation of their respective tendencies within a time-step, are calculated in single precision arithmetic. Updating the fluid thickness field $\Phi_\nu$ directly in algorithmic step 2 is replaced by accumulating the solver increments to fluid thickness in a single precision variable $\Delta \Phi_\nu$ which is then added to the fluid thickness field $\Phi^{n}$ when the elliptic solver is exited. This ensures that summing up the solver increments can still be performed in single precision as the value range of $\Delta \Phi_\nu$ is on the order of $10^{1}m$ which means that adding solver increments of as low as $10^{-7}m$ is possible without excessive rounding error. Second, the initial residual $r_0$ calculation is performed in double precision to ensure that the relative error in $r_0$ is smaller than $\epsilon$. And third, we find that the forces ${\bf R_i}^{n+1}$---the last term in equation ~\eqref{MPDATA}---need to be calculated in double precision to avoid numerical errors at the poles. Again, the fluid thickness field $\Phi$ plays a crucial role as, similarly to the calculation of $r_0$, the pressure gradient term from ${\bf R_i}^{n+1}$ is responsible for a large fraction of the numerical errors at the poles.

The majority of the model is still performed in single precision. This includes the entire second-order MPDATA operator (first term on the rhs of (\ref{MPDATA})), and the calculation of the advective velocities ${\bf v}^{n+0.5}$.

These precision choices already define most of the mixed-precision shallow-water model. Only the precision choice for the elliptic solver cycle---algorithmic steps 2 to 4---still needs to be discussed. Here, three options are explored: 
\begin{itemize}
\item For option one, double precision is employed in the entire elliptic solver. As a result solver convergence is indistinguishable from the double precision shallow-water model (figures \ref{fig:convergence_DPSPMP} a), b) and figures \ref{fig:Lat_Convergence_DPSPMP} a)-b)). 
\item In the second option, single precision is used throughout the entire elliptic solver cycle. The consequence of this choice is that the latitude-wise $L_2$-norms of the final residual error fields show a reduction of the convergence rate near the poles. The error is smaller (about an order of magnitude) when compared to results from figure \ref{fig:Lat_Convergence_DPSPMP} d), but still clearly visible. However, the reduced solver convergence at the poles is not causing an increase of the initial residual error fields near the poles. Thus one can conclude that the rest of the mixed-precision shallow-water model time-step is not affected, possibly also due to the presence of the polar absorber. For the orographic flow test-case, overall solver convergence is reduced by up to an order of magnitude. For the RHW4, using single precision reduces the overall solver convergence by about 2 orders of magnitude, which is mainly due to a reduced solver convergence rate near the poles.
\item The third option is a blend of the first two options. For this option, the elliptic solver cycle employs single precision everywhere except for a three zonal band wide range closest to the North- and Southpole respectively where all calculatuions are performed in double precision. With this choice, solver convergence is indistinguishable from double precision for the orographic flow test-case. For the RHW4, there are small differences found for the fourth solver iteration only. As a result, overall loss in solver convergence is at most 1 order of magnitude when compared to double precision.
\end{itemize}

These three options are all viable and it is ultimately the modeller's choice which of the options to choose. 
Here, we decide to proceed with the single precision elliptic solver cycle (option two). There are strong arguments for this choice. The disadvantage of a slightly reduced overall convergence rate in the $L_2$-norm seems insignificant as it is clearly understood where this reduction stems from, and there is no negative impact on the shallow-water model's or the elliptic solver's behaviour in any way. Especially since the poles are regions where the polar absorbers are acting as a regularizing element anyhow.

The elliptic solver convergence of the resulting mixed-precision shallow-water model is shown in figure \ref{fig:convergence_DPSPMP} e)-f) and figure \ref{fig:Lat_Convergence_DPSPMP} e)-f). The global, as well as latitude-wise, representation of the solver convergence show that much of the behaviour of the double precision model variant is restored. Figure \ref{fig:Lat_Convergence_DPSPMP} e) shows that there is no accumulation of numerical errors at the poles anymore for the orographic wave test-case (RHW4 not shown but qualitatively the same). As discussed earlier, the convergence rate at the poles is slightly reduced, see figure \ref{fig:Lat_Convergence_DPSPMP} f). The solver's overall convergence, figure \ref{fig:convergence_DPSPMP} e)-f), is then completely restored up to the second and third solver iteration for the RHW4 and the orographic flow test-case respectively while differences remain for sub-sequent iterations. Concerning criterion (iii), the difference in deviations from the genuine solution is one order of magnitude smaller when compared to the difference found between single precision and double precision from figures \ref{fig:deviations_DPSP} e)-f) (not shown here).

\subsubsection{Reformulation of the Initial Residual $r_0$ Calculation}\label{SP_r0}
Based on figure \ref{fig:Lat_Error_MP}, a straightforward single precision calculation of $r_0$ was shown insufficient. In the previous section, double precision was instead used for this caclulation. However, since the calculation of $r_0$ is a computationally expensive operation, there is a strong incentive to explore further precision reduction via other means. 

Rather than using higher precision, for some Krylov subspace methods a popular option to manage the accumulation of round-off errors in the residual error field is a complete recalculation of the residual after a set number of solver iterations (\cite{doi:10.1137/S1064827599353865}). However, each recalculation adds another application of the linear operator $\mathcal{L}$ to the total cost of the solver. In our setup we only encounter up to $4$ solver iterations until solver convergence, which renders a complete recalculation of the residual values to be computationally inefficient. Additionally, if the numerical errors in $r_0$ become as large as the value itself near the poles, recalculation of $r_0$ becomes meaningless.

Here, we explore a different direction. The numerical errors in $r_0$ ultimately result from the truncation of the fluid thickness field $\Phi$ to single precision. We present how splitting parts of the residual calculation into the form of a base part plus a correction term can significantly reduce the numerical error in $r_0$. Since the operator $\mathcal{L}$ is linear by design reformulating the calculation of $r_0$ can be done with minimal effort.

As a preparation for splitting the operator calculation, first the fluid thickness variable $\Phi$ is split into two parts. Instead of a double precision variable $\Phi$, the fluid thickness field is now chosen to be described by two single precision fields: $\Phi\approx\hat{\Phi}+\tilde{\Phi}$, where $\tilde{\Phi}$ denotes a correction to some to-be-defined base fluid thickness field $\hat{\Phi}$. 

At the beginning of the simulation, $\hat{\Phi}$ and $\tilde{\Phi}$ are set via the following procedure:
\begin{enumerate}
\item $\hat{\Phi}$ is set to the initial fluid thickness truncated to single precision 
\begin{equation}\hat{\Phi}=trunc(\Phi^{0}).\end{equation}
\item
$\tilde{\Phi}$ is then defined as the difference calculated in double precision arithmetic between $\hat{\Phi}$ and $\Phi^{0}$ truncated to single precision 
\begin{equation}\tilde{\Phi}=trunc(\Phi^{0}-\hat{\Phi}).\end{equation}
\end{enumerate}
The average value of $\hat{\Phi}$ is then on the order or $10^4m$ to $10^5m$, while the average value of $\tilde{\Phi}$ is $10^{-3}m$---a difference of 7 to 8 orders of magnitude which is consistent with the machine epsilon of single precision. The relative error of the sum $\hat{\Phi}+\tilde{\Phi}$ when calculated in double precision arithmetic is on the order of $10^{-15}$ which is consistent with the double precision machine epsilon.

During the simulation, the solver increments to fluid thickness are added to $\tilde{\Phi}$ each time-step. Thus, the average values of $\tilde{\Phi}$ increase in time which increases the relative error in $\tilde{\Phi}$ and by definition also of the sum $\hat{\Phi}+\tilde{\Phi}$. After about 100 time-steps, the average values of $\tilde{\Phi}$ are on the order of $10^{1}m$, and the relative error of the sum $\hat{\Phi}+\tilde{\Phi}$ performed in double precision arithmetic increases to about $10^{-11}$. If this continued further, eventually the relative errors of the sum $\hat{\Phi}+\tilde{\Phi}$ would reach the single precision machine epsilon.

As a result, in our approach we decide to update $\hat{\Phi}$ after every 100 time-steps. This is a good compromise between precision and computational overhead. The procedure to update $\hat{\Phi}$ and $\tilde{\Phi}$ works as follows:
\begin{enumerate}
\item Temporarily store the sum 
\begin{equation}\Phi^{DP}=\hat{\Phi}+\tilde{\Phi}\end{equation} in a double precision variable $\Phi^{DP}$
\item $\hat{\Phi}$ is set to the initial fluid thickness truncated to single precision 
\begin{equation}\hat{\Phi}=trunc(\Phi^{DP}).\end{equation}
\item
$\tilde{\Phi}$ is then defined as the difference calculated in double precision arithmetic between $\hat{\Phi}$ and $\Phi^{0}$ truncated to single precision 
\begin{equation}\tilde{\Phi}=trunc(\Phi^{DP}-\hat{\Phi}).\end{equation}
\end{enumerate}
Since this update happens so rarely throughout the model simulation, the computational overhead is negligible.  

So far, the split of a double precision variable into two single precision variables has not resulted in any compuational gains. The trick to make this approach computationally efficient lies in rewriting the gradient calculation of $\Phi$, i.e. the partial derivatives $\frac{\partial \Phi}{\partial x^{J}}$, in the linear operator $\mathcal{L}$ of definition ~\eqref{L_Operator}. The rewritten linear operator reads as following:
\begin{equation}\label{L_Operator_SP}
\mathcal{L}\left(\Phi\right):=\sum^{M}_{I=1}\frac{\partial}{\partial
x^{I}} \left(\sum^{M}_{J=1}A^{IJ}
\left(\frac{\partial \hat{\Phi}}{\partial x^{J}}+\frac{\partial \tilde{\Phi}}{\partial x^{J}}\right)
+ B^{I}\Phi^{SP}\right)-C\Phi^{SP}~. 
\end{equation} 
In this, $\Phi^{SP}=\hat{\Phi}+\tilde{\Phi}$ is the fluid thickness in single precision. The $\frac{\partial \hat{\Phi}}{\partial x^{J}}$ only change every 100 time-steps and can thus be precomputed and stored. Compared to the original definition ~\eqref{L_Operator}, the operator ~\eqref{L_Operator_SP} consists of two more floating point operations per grid-point---the addition of the partial derivatives---but each of these is now performed in single precision arithmetic.

 \begin{figure}[!htb] 
\vspace{0.3cm}
\minipage{0.48\textwidth}\centering\textbf{RHW4}\par\medskip
 \includegraphics[width=1.0\linewidth, trim=0cm 0cm 0cm 0cm, clip]{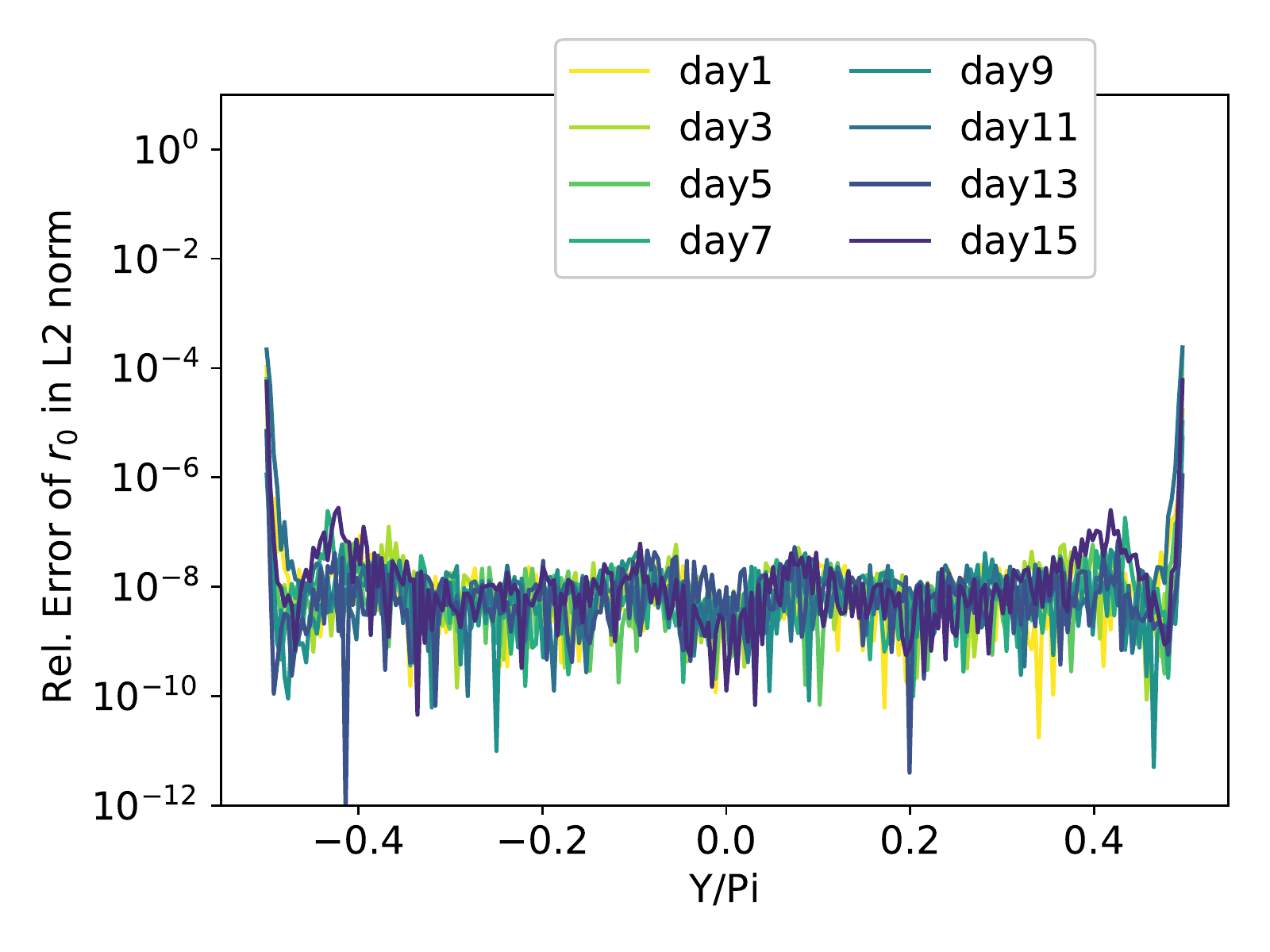} 
\endminipage\hfill
\minipage{0.48\textwidth}\centering\textbf{Orographic Flow}\par\medskip
 \includegraphics[width=1.0\linewidth, trim=0cm 0cm 0 0cm, clip]{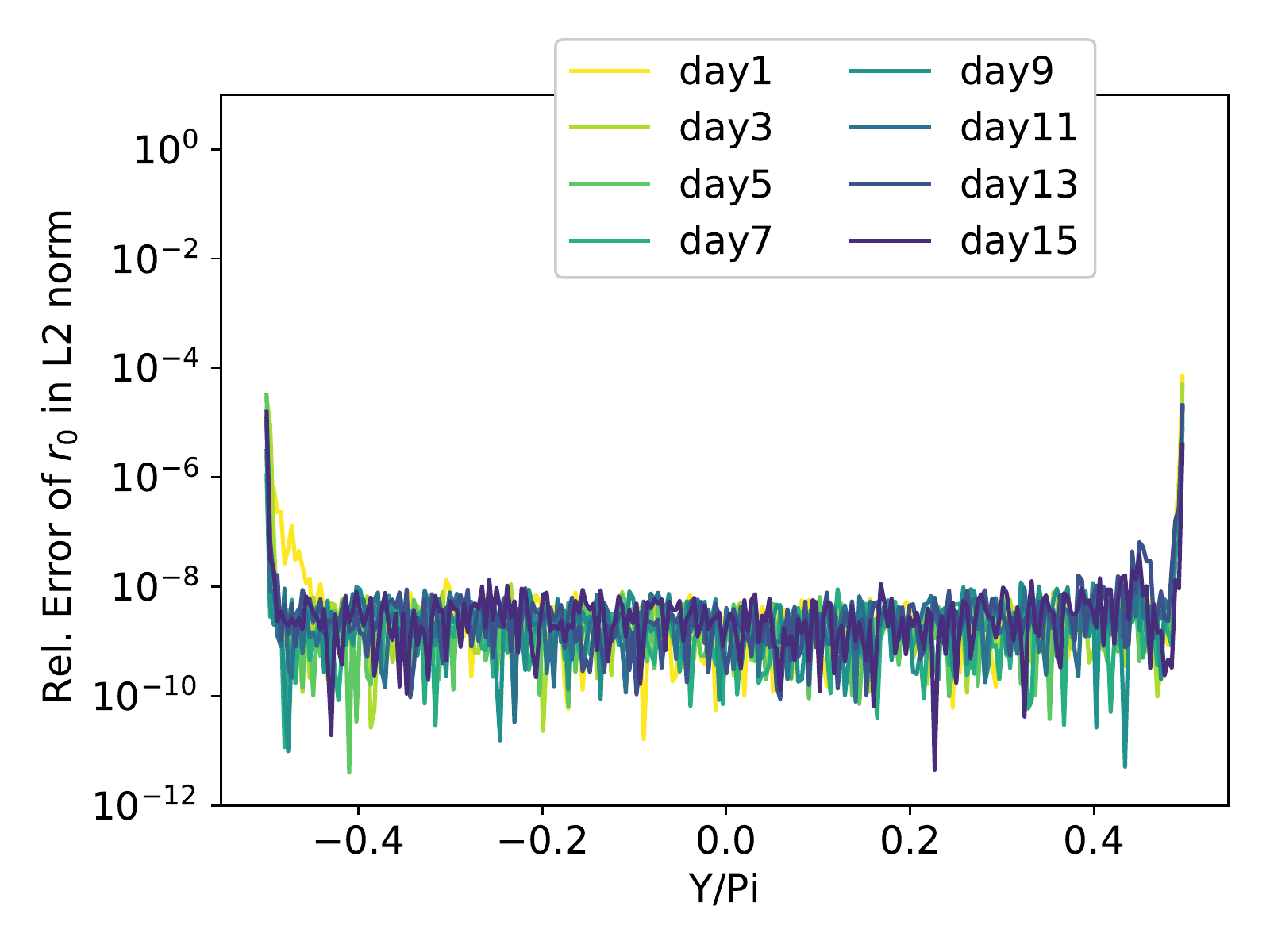}
\endminipage

\caption{Latitude-wise relative numerical error of the inital residual $r_0$ calculated with the operator ~\eqref{L_Operator_SP} with respect to the original operator ~\eqref{L_Operator} in double precision arithmetic in the $L_2$-norm for the RHW4 (left) and the orographic flow test-case (right) in 2-day time interval snapshots; darker colours indicate later simulation time.}
\label{fig:Lat_Error_RP}
\end{figure}
Comparing figures \ref{fig:Lat_Error_MP} and the latitude-wise relative numerical errors when $r_0$ is calculated using the single precision operator ~\eqref{L_Operator_SP}, see figures \ref{fig:Lat_Error_RP},
the relative numerical error is strongly reduced. Near the poles, the reduction in relative errors is 4 orders of magnitude, while for the rest of the computational domain it is about 2 orders of magnitude, reaching the theoretical limit given by the single precision machine epsilon for most of the domain.

Using the new operator ~\eqref{L_Operator_SP} in conjunction with the mixed-precision model of \S\ref{MP_SWM} reveals that using the new operator throughout the entire computational domain is not desirable. This is because, although the relative error in each calculation of the initial residual error field $r_0$ is greatly reduced near the poles, the remaining errors are exacerbated strongly due to the use of single precision in algorithmic steps 2-4. This leads to a vastly reduced convergence rate near the poles, with the latitude-wise $L_2-$norm of the final residual errors being up to $10^{-1}$ for both test-cases, looking very similarly to figure \ref{fig:Lat_Convergence_DPSPMP} d) that was found for the full single precision shallow-water model variant. As a consequence, the latitude-wise $L_2$-norm of the initial residual error field $r_0$ near the poles can be as much as one order of magnitude larger than the values found for the mixed-precision shallow-water model of \S\ref{MP_SWM}, see again figure \ref{fig:Lat_Convergence_DPSPMP} e). 

To avoid this accumulation of numerical errors near the poles, from hereon the standard operator ~\eqref{L_Operator} is instead used for a three zonal band wide range closest to the North- and Southpole respectively, where all calculatuions are performed in double precision. Model behaviour is then completely recovered to that of the mixed-precision shallow-water model of \S\ref{MP_SWM}.

\subsection{Mixed-precision with Half Precision for the Elliptic Solver}\label{MPGCR}

In this section the aim is to take down precision to half precision for as many parts of the elliptic solver as possible while still satisfying a reasonably high solver accuracy. 

The initial residual error $r_0$ calculation in algorithmic step 1 is not considered for half precision arithmetic as the machine epsilon of half precision is $9.77\cdot 10^{-4}$ which means it is far larger than our targeted solver accuracy. The discussion in \S\ref{SP_r0} thus indicates single precision to be the limit within this setting.

The focus is here primarily on the preconditioning and the application of the linear operator ~\eqref{L_Operator} in algorithmic steps 1 and 3. This is in part motivated by the fact that these steps are by far the most computationally intensive parts of the elliptic solver, see the normalized runtimes for GCR(3) next to the flow chart in figure \ref{fig:tempevo15}. It is found that these parts of the solver can indeed be taken down to half precision for the most part. Two exceptions from this rule however need to be made. The first exception concerns the last operation of the linear operator ~\eqref{L_Operator} where $C\Phi$ is subtracted from the previous terms. This calculation is done in single precision because $C\Phi$ and the remaining terms of the linear operator ~\eqref{L_Operator} are found to be of similar size and the subtraction of both terms introduces large numerical errors if done in half precision. The second exception is the area close to the poles. Since the operator ~\eqref{L_Operator} is ill-conditioned near the poles, half precision is found to be crititcal in these regions. Thus, for grid-points within a three zonal-band wide range closest to the North- and Southpole all calculations of the preconditioner and the linear operator are performed in single precision.

In comparison to the preconditioning and the application of the linear operator, algorithmic steps 2 and 4 are computationally cheap in terms of the normalized runtimes, consisting only of global sums and variable updates. The large range in the normalized cost of step 4 is due to the size of the sum changing with $\nu$. Concerning the updates of the residual error field $r_{\nu+1}$
and the accumulation of solver increments to fluid thickness in the variable $\Delta\Phi_{\nu+1}$, both are kept in single precision. Both require precision higher than the machine epsilon of half precision to satisfy the solver accuracy. Concerning $\mathcal{L}(p_{\nu+1})$, experiments where the sum is calculated in half precision arithmetic immediately reduces the attainable solver accuracy to 4 orders of magnitude for both test-cases. Additionally all the summands are only available in single precision, which makes single precision the preferred option. However, the sum to obtain $p_{\nu+1}$ can be performed in half precision. Since all the summands are in half precision as well this also makes sense from a computational point-of-view.

Concerning the global sums for $\beta$ and $\alpha_l$ in algorithmic steps 2 and 4, single precision is the straightforward choice for these caclulations. This is simply because all of the involved input variables are single precision and truncating them to half precision first would likely result in a computational overhead.

 \begin{figure}[!htb] 
\vspace{0.3cm}
\minipage{0.48\textwidth}\centering
(a)
 \includegraphics[width=1.0\linewidth, trim=0cm 0cm 0cm 1.2cm, clip]{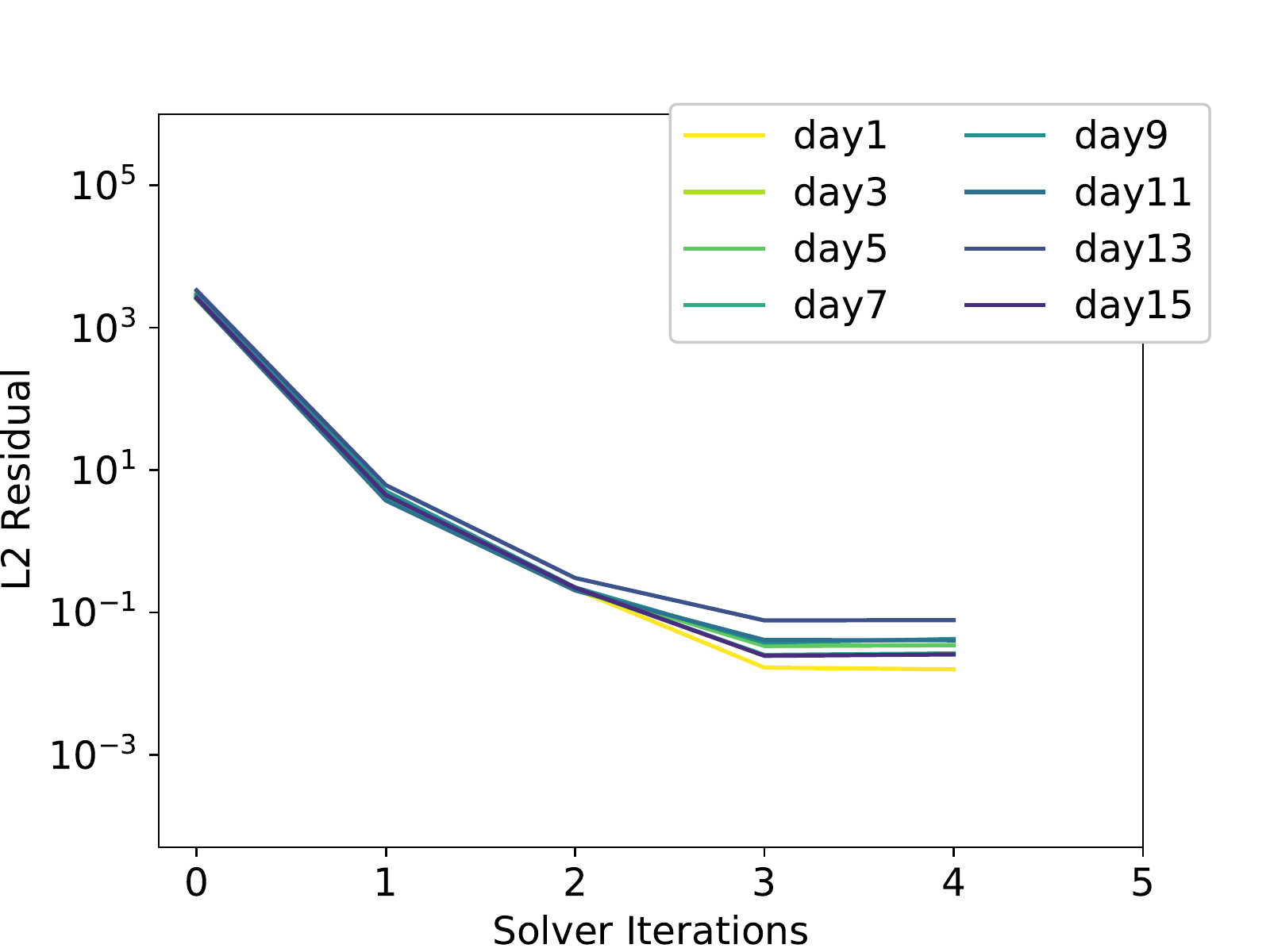} 
\endminipage\hfill
\minipage{0.48\textwidth}\centering
(b)
 \includegraphics[width=1.0\linewidth, trim=0cm 0cm 0 1.2cm, clip]{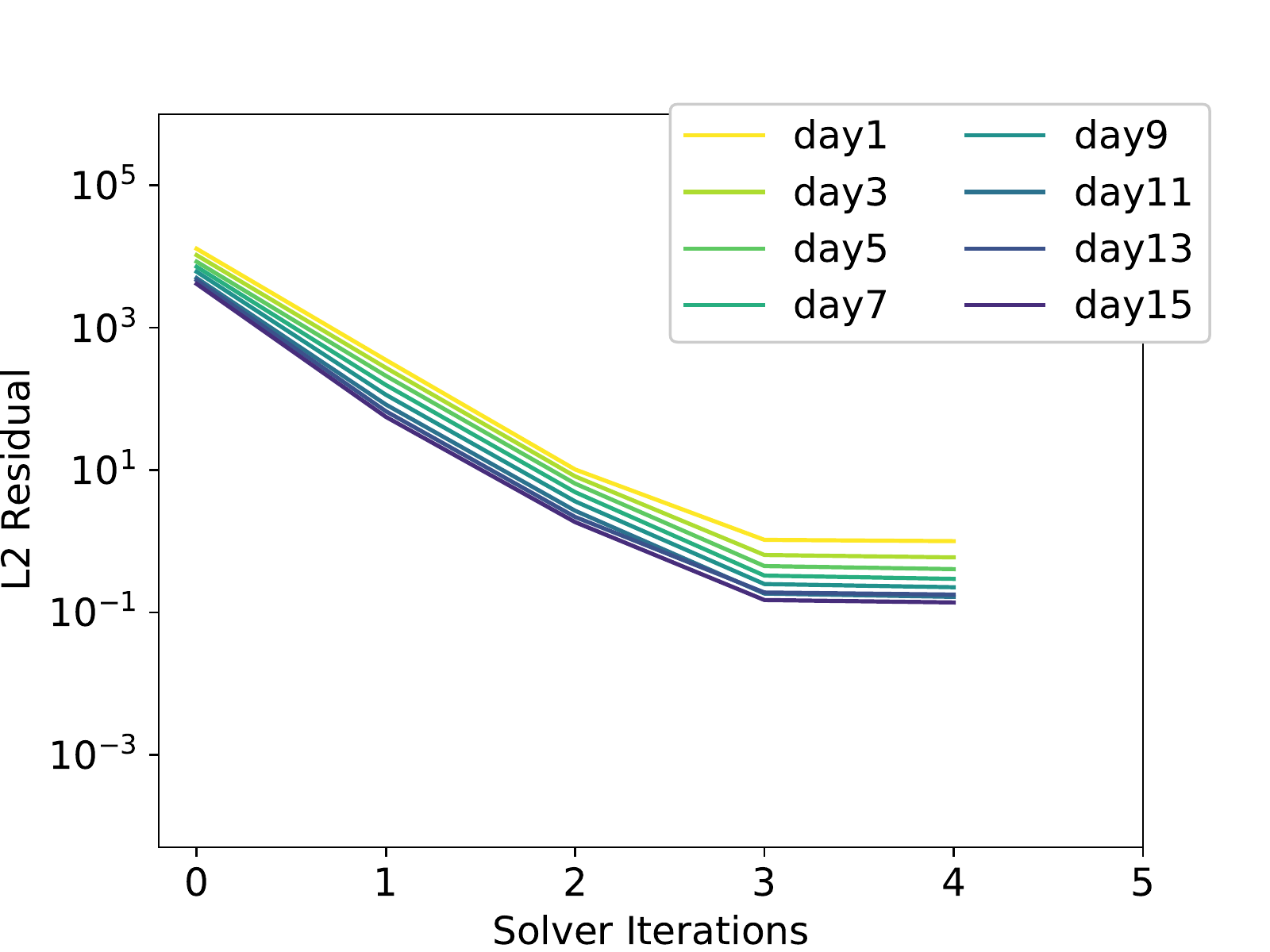}
\endminipage
\vspace{0.3cm}
\minipage{0.48\textwidth}\centering
(c)
 \includegraphics[width=1.0\linewidth, trim=0cm 0cm 0 0.2cm, clip]{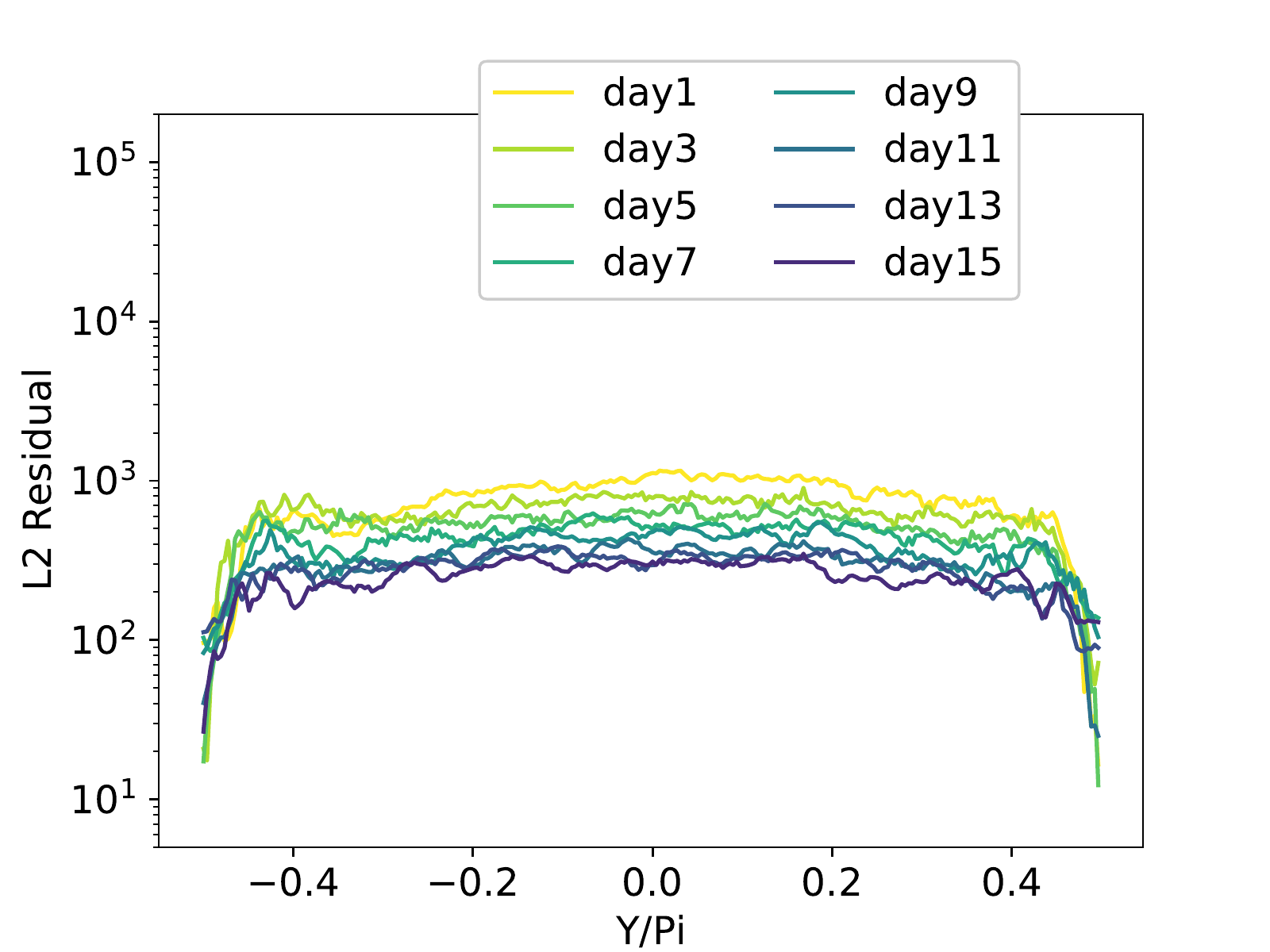} 
\endminipage\hfill
\minipage{0.48\textwidth}\centering
(d)
 \includegraphics[width=1.0\linewidth, trim=0cm 0cm 0 0.2cm, clip]{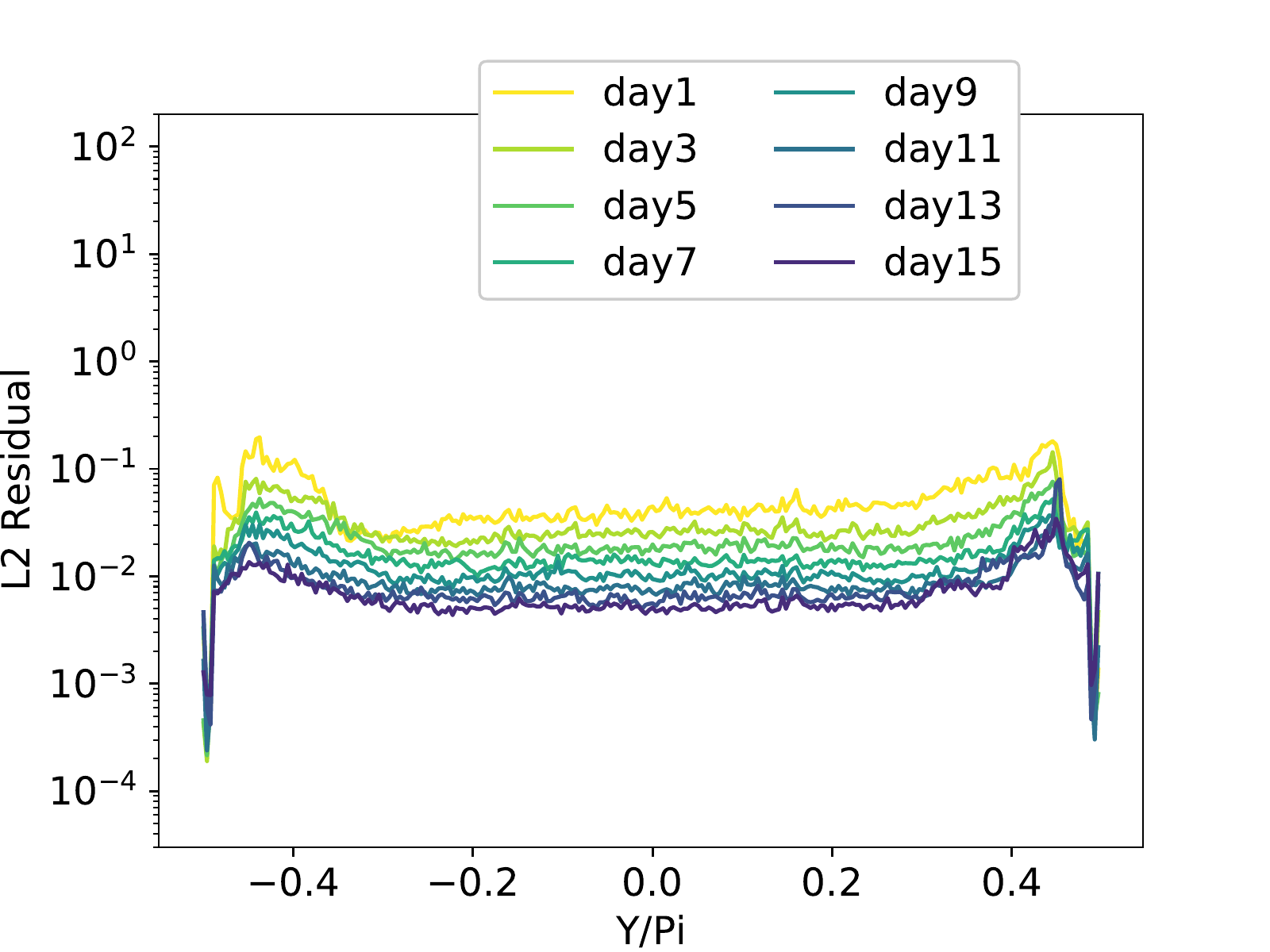}
\endminipage
\vspace{0.3cm}

\caption{Convergence rates of the RHW4 a) and the orographic flow test-case b) for the mixed-precision elliptic solver variant with half precision, and the corresponding latitude-wise inital c) and final d) residual error in the $L_2$-norm for the orographic flow test-case.}
\label{fig:convergence_Latconv_RP}
\end{figure}

The elliptic solver convergence of the resulting mixed-precision shallow-water model is illustrated in figure \ref{fig:convergence_Latconv_RP} a)-d). The global, as well as latitude-wise, representation of the solver convergence show that much of the behaviour of the double precision model variant is preserved. Even though precision is reduced to such a low level, the overall accuracy reached within four solver iterations is still about $1\cdot 10^{-5}$ for the RHW4 and $5\cdot 10^{-5}$ for the orographic flow test-case, see figures \ref{fig:convergence_Latconv_RP} a)-b). For both test-cases, there is a clear degradation of solver convergence for the last solver iteration as there is no further reduction in the $L2$-norm of the residual error fields. For the RHW4, solver convergence is already slightly reduced for the third iteration similarly to what could be found for the mixed-precision shallow-water model in figure \ref{fig:convergence_DPSPMP} e). The plateuing for the last solver iteration indicates that the accuracy limit has been reached for this mixed-precision elliptic solver variant. 

Figure \ref{fig:convergence_Latconv_RP} c) shows that there is no accumulation of numerical errors at the poles for the orographic wave test-case. In Figure \ref{fig:convergence_Latconv_RP} d), the impact of enforcing higher precision in the vicinity of the poles becomes visible. The final values for the latitude-wise residual error $L2$-norm near the poles are smaller in comparison to the rest of the computational domain, where the effect of low precision manifests in overall larger values when compared to previous results, see again figures \ref{fig:Lat_Convergence_DPSPMP} b), d), f). This however means that the convergence rate is now very consistent throughout the computational domain, with similar residual error reductions at all latitudes. The corresponding figures for the RHW4 are not shown but look qualitatively the same. 

Although the final residual error values are generally larger, the difference in deviations from the genuine solution---criterion (iii)---is comparable to the mixed-precision shallow-water model introduced in \S\ref{MP_SWM} for both test-cases, i.e. the difference is one order of magnitude smaller than the difference found between single precision and double precision from figures \ref{fig:deviations_DPSP} e)-f).

For completeness, the two coarser model resolutions $256\times128$ and $128\times64$ were run using the mixed-precision shallow-water model described in this section. As expected, it is found that the results are qualitatively the same (not shown here) and the analysis that has been done for the $512\times256$ applies as well to the coarser model resolutions. The vicinity near the poles for which higher precision levels are used is chosen to be two and one for the $256\times128$ and the $128\times64$ model resolution respectively.

\subsection{Expected Computational Performance of the Mixed-Precision Solver}
Ultimately, we are interested in increasing the solver's, and thus the entire model's, efficiency. For the simulations done in \S\ref{SP_prec_SWM} and \S\ref{MP_SWM}, actual time measurements can be performed to estimate the computational savings. The single precision shallow-water model of \S\ref{SP_prec_SWM} has a shorter time-to-solution than the double precision shallow-water model, the model runtime is reduced by 35 $\%$. The mixed-precision shallow-water model variant of \S\ref{MP_SWM} was 30 $ \%$ faster than the double precision variant. For larger problem sizes, these differences between double and single precision performance are generally expected to become even larger. For instance, for state-of-the-art W\&C models single precision simulations are typically reporting a decrease of computing time by ~40$\%$ in comparison to simulations in double precision \cite{Vana2017a,Maynard2019}.

For the mixed-precision elliptic solver with half-precision arithmetic (\S\ref{MPGCR}), it is, however, non-trivial to perform a reliable and generic estimate of potential savings from the mixed-precision approach for the elliptic solver. This is due to the lack of robust and precise performance models for solvers. Furthermore, the various possible combinations of software and hardware in compute systems, such as different types of computing chips (CPU, GPU, TPUs, FPGAs...), different compilers, the memory layout (cache sizes and hierarchy, shared and/or distributed memory access...), connectivity between computing nodes (network topology, bandwidth...) are simply too abundant to make general statements.

To provide a rough estimate of potential savings, we here evaluate the number of floating point operations and their respective precision level. In accordance with the reduction in computational bits from double precision, we assume a speed-up of factor 2 for a reduction to single precision level and respective factor of 4 for half precision. These assumptions are in accordance with the peak performance of modern processors and is also justified by the increase of cache, memory and bandwidth efficiency. Some hardware accelerators that are optimised for machine learning, such as TensorCores on NVIDIA Volta GPUs, can even get a much higher relative increase in peak performance for half precision arithmetic. However, since modern weather and climate models are typically not able to utilise peak performance, these numbers may turn out to be too optimistic if a model simulation is limited by latency of communication or if the operations are not vectorized.

To estimate the total reduction of runtime we multiply the number of floating point operations that has been performed with the three different precision levels (double, single and half) with their respective estimated cost (1.0, 0.5, 0.25) for the different algorithmic steps. Including the exceptions for precision levels that we made in \S\ref{MPGCR}, the overall runtime can be expected to reduce by a factor of $3.3$, when comparing the mixed-precision elliptic solver with half-precision to the double precision variant. To further put these numbers into perspective, completing the first solver iteration with the double precision elliptic solver would already be as computationally expensive as running the entire four solver iterations with the mixed-precision solver. The result by the UK MetOffice \cite{Maynard2019} that a single precision solver is yielding a speed-up close to a factor of two in comparison to a double precision solver indicates that our cost estimate may be achievable on real hardware.

\section{Conclusion}

Mixed-precision in a representative semi-implicit shallow-water model in general and its preconditioned elliptic solver in particular is found to be possible and overall advantagous when compared to the default double precision option. The described mixed-precision shallow-water model using double and single precision clearly outperforms the double precision model in terms of time-to-solution while maintaining the same solution quality and similar solver convergence behaviour. Concerning the elliptic solver, most of the computationally expensive components of the solver can be performed almost entirely in half precision arithmetic. 

This being said, it is found that the special structure of our problem does not permit a naive reduction of precision. We should expect the exacerbation of artificial modes in the prognostic variable fields of the model when going to too low precision levels in the elliptic solver. In our configuration, precision reduction close to the poles---the two grid-singularities that were present---caused spurious behavior even for the single precision shallow-water model. Overall, as the solver precision is further and further reduced to its limits throughout this paper from double precision to precision levels as low as half precision, a very clear progression is seen in terms of an increase in numerical errors, reduced convergence rates, and the increasing need for assigning higher precision levels for parts of the computational domain.

On the other hand, it is shown that given some knowledge about the specific structure of the application, issues with reduced precision can in fact be mitigated by using higher precision arithmetic in parts of the computational domain. The path towards a computationally efficient mixed-precision elliptic solver can be a challenging one though, as it requires a holistic way of thinking about the interplay between numerical errors from a local precision reduction, to the resulting global model errors, and the solver's overall convergence rate. Understanding this interplay may sometimes require in-depth analysis and in some cases even the reformulation of parts of the existing solver algorithm. The approach we found might also be beneficial in other applications.

All in all, results are similar to \cite{Baboulin2009,Carson2018,Haidar2018,Anzt2019,Goddeke2007,Idomura2018,Amritkar2015,Furuichi2011} in that the preconditioning step is very robust against reducing precision and seems to be the part of the elliptic solver where performance gains can be expected while many of the other parts can be quite sensitive to reducing precision. It is an important finding that this general structure can also be found in our specific problem from geophysical fluid dynamics, using a conjugated residual solver with a preconditioner based on tridiagonal inversion, a key component in 3D atmospheric semi-implicit grid-point models.

A reccurent theme in previous publications on reduced precision in W\&C models is the motivation that there are irreducible uncertainties (such as initial condition uncertainties, stochastic representations of unresolved subgrid-scale processes) present in these models which gives a justification for accepting a certain amount of additional model errors, in the form of reducing precision, in order to increase the overall model's computational efficiency \cite{DuebenMWR2014,Chantry2019,Saffin2019,DuebenJCP2014,Palmer2012,Hatfield2018}. It should be noted that in the elliptic solver this idea of accepting a certain amount of additional model error to increase computational efficiency is at the very core of the approach and is inherently represented and used in the process of setting the solver accuracy.

In summary, given our results, we do not see any major obstacle that would prevent us from using reduced precision arithmetic for an elliptic solver applied to a full 3-dimensional atmospheric model. There is much experience within the atmospheric modeling community concerning the behavior of the height/pressure solve via iterative methods on various computational grids which can be used to circumvent possible pitfalls when reducing precision and instead guide the path towards efficient mixed-precision elliptic solvers for W\&C models. 

\section*{Acknowledgments} Jan Ackmann was funded via the European
Research Council project ITHACA (grant No. 741112). Peter D. Dueben
gratefully acknowledges funding from the Royal Society for his
University Research Fellowship as well as funding from the ESIWACE
2 project. ESIWACE 2 has received funding from the European Union's
Horizon 2020 research and innovation program under grant agreement
823988. NCAR is sponsored by the National Science Foundation.


\begin{thebibliography}{10}

\bibitem{GMDEscape}
A.~M{\"{u}}ller, W.~Deconinck, C.~K{\"{u}}hnlein, G.~Mengaldo, {and co-authors}, 
``{The ESCAPE project: Energy-efficient Scalable Algorithms for Weather
Prediction at Exascale},'' {\em Geoscietific Model Development}, 12 (2019) 4425–4441.

\bibitem{Goddeke2007}
D.~G{\"{o}}ddeke, R.~Strzodka, and S.~Turek, ``{Performance and accuracy of
  hardware-oriented native-, emulated- and mixed-precision solvers in FEM
  simulations},'' {\em International Journal of Parallel, Emergent and
  Distributed Systems}, 22 (2007) 221-256.

\bibitem{Baboulin2009}
M.~Baboulin, A.~Buttari, J.~Dongarra, J.~Kurzak, J.~Langou, J.~Langou,
  P.~Luszczek, and S.~Tomov, ``{Accelerating scientific computations with mixed
  precision algorithms},'' {\em Computer Physics Communications}, 180 (2009) 2526-2533.

\bibitem{Furuichi2011}
M.~Furuichi, D.~A. May, and P.~J. Tackley, ``{Development of a Stokes flow
  solver robust to large viscosity jumps using a Schur complement approach with
  mixed precision arithmetic},'' {\em Journal of Computational Physics}, 230 (2011) 8835-8851.

\bibitem{ChenGJCP2012}
G.~Chen, L.~Chac{\'{o}}n, D.C.~Barnes, ``{An efficient mixed-precision, hybrid CPU-GPU 
implementation of a nonlinearly implicit one-dimensional particle-in-cell algorithm},'' 
{\em Journal of Computational Physics}, 231 (2012) 5374-5388.

\bibitem{Palmer2012}
T.N.~Palmer, ``{Towards the probabilistic Earth-system simulator: a vision
for the future of climate and weather prediction},'' {\em Quarterly Journal
of the Royal Meteorological Society}, 138 (2012) 841-861.

\bibitem{DuebenJCP2014}
P.D.~D{\"{u}}ben, H.~McNamara, T.N.~Palmer, ``{The use of imprecise
  processing to improve accuracy in weather {\&} climate prediction},''
 {\em Journal of Computational Physics}, 271 (2014) 2-18.

\bibitem{Saffin2019}
L.~Saffin, S.~Hatfield, P.~Dueben, T.~Palmer, ``{Reduced‐precision parametrization: lessons from an intermediate‐complexity atmospheric model},'' {\em
Quarterly Journal of Royal Meteorological Society}, 146 (2020) 1590-1607.

\bibitem{DuebenMWR2014}
P.D.~D{\"{u}}ben, T.N.~Palmer, ``{Benchmark tests for numerical weather forecasts 
on inexact hardware},'' {\em Monthly Weather Review}, 142 (2014) 3809-3829.

\bibitem{Chantry2019}
M.~Chantry, T.~Thornes, T.~Palmer, P.~D{\"{u}}ben, ``{Scale-selective precision 
for weather and climate forecasting},'' {\em Monthly Weather Review}, 147 (2019) 
645-655.

\bibitem{Nakano2018}
M.~Nakano, H.~Yashiro, C.~Kodama, H.~Tomita, ``{Single Precision in the Dynamical
Core of a Nonhydrostatic Global Atmospheric Model: Evaluation Using a Baroclinic
Wave Test Case},'' {\em Monthly Weather Review}, 146 (2018) 409-416.

\bibitem{TintoPrims2019}
O.~{Tint{\'{o}} Prims}, M.~C. Acosta, A.~M. Moore, M.~Castrillo, K.~Serradell,
  A.~Cort{\'{e}}s, and F.~J. Doblas-Reyes, ``{How to use mixed precision in
  ocean models: exploring a potential reduction of numerical precision in NEMO
  4.0 and ROMS 3.6},'' {\em Geoscientific Model Development}, 12 (2019) 3135-3148.

\bibitem{Maynard2019}
C.~Maynard and D.~Walters, ``{Mixed-precision arithmetic in the ENDGame
  dynamical core of the Unified Model, a numerical weather prediction and
  climate model code},'' {\em Computer Physics Communications}, 244 (2019) 69-75.

\bibitem{Mengaldo2019}
G.~Mengaldo, A.~Wyszogrodzki, M.~Diamantakis, S.-J. Lock, F.~X. Giraldo, and
  N.~P. Wedi, ``{Current and Emerging Time-Integration Strategies in Global
  Numerical Weather and Climate Prediction},'' {\em Archives of Computational
  Methods in Engineering}, 26 (2019) 663-684.

\bibitem{adcroft2004overview}
A.~Adcroft, C.~Hill, J.-M.~Campin, J.~Marshall, P.~Heimbach, ``{Overview of
the formulation and numerics of the MIT GCM},'' {\em Proceedings of the 
ECMWF seminar series on Numerical Methods, Recent developments in numericali
methods for atmosphere and ocean modelling}, (2004) 139-149.
	
\bibitem{ICON_Korn}
P.~Korn, ``{Formulation of an unstructured grid model for global ocean dynamics},
'' {\em Journal of Computational Physics}, 339 (2017) 525-552.
	
\bibitem{Fesom_Wang}
Q.~Wang, S.~Danilov, D.~Sidorenko, R.~Timmermann, C.~Wekerle, X.~Wang, T.~Jung,
and J.~Schröter, ``{The Finite Element Sea Ice-Ocean Model (FESOM) v.1.4:
formulation of an ocean general circulation model},'' {\em Geoscientific Model 
Development}, 7 (2014) 663–693.

\bibitem{Smolarkiewicz2016}
P.~K. Smolarkiewicz, W.~Deconinck, M.~Hamrud, C.~K{\"{u}}hnlein, G.~Mozdzynski,
  J.~Szmelter, and N.~P. Wedi, ``{A finite-volume module for simulating global
  all-scale atmospheric flows},'' {\em Journal of Computational Physics}, 314 (2016) 287-304.

\bibitem{SKG2017} P.K.~Smolarkiewicz,  C.~K\"uhnlein, W.W.~Grabowski, ``{A
finite-volume module for cloud-resolving simulations global atmospheric
flows},'' {\em Journal of Computational Physics},  341 (2017) 208-229.

\bibitem{Smolarkiewicz2019}
P.~K. Smolarkiewicz, C.~K{\"{u}}hnlein, and N.~P. Wedi, ``{Semi-implicit
  integrations of perturbation equations for all-scale atmospheric dynamics},''
  {\em Journal of Computational Physics}, 376 (2019) 145-159.

\bibitem{Kuhnlein2019}
C.~K{\"{u}}hnlein, W.~Deconinck, R.~Klein, S.~Malardel, Z.~P. Piotrowski, P.~K.
Smolarkiewicz, J.~Szmelter, and N.~P. Wedi, ``{FVM 1.0: a nonhydrostatic
  finite-volume dynamical core for the IFS},'' {\em Geoscientific Model
  Development}, 12 (2019) 651-676.

\bibitem{SmolarSzmelter05} 
P.K.~ Smolarkiewicz, J.~Szmelter, ``{MPDATA: An edge-based unstructured-grid formulation},''
{\em Journal of Computational Physics}, 206 (2005) 624-649.

\bibitem{SzmelterSmolar10}
J.~Szmelter, P.K.~Smolarkiewicz, ``{An edge-based unstructured mesh discretisation in 
geospherical framework},'' {\em Journal of Computational Physics}, 229 (2010) 4980-4995.

\bibitem{KuhnleinSmolar2017} 
C. K\"uhnlein, P.K. Smolarkiewicz, ``{An unstructured-mesh finite-volume MPDATA for 
compressible atmospheric dynamics},'' {\em Journal of Computational Physics}, 334 (2017) 16-30.

\bibitem{Eisenstat_etal1983} S.C.~Eisenstat, H.C.~Elman, M.H. Schultz ``{Variational iterative
methods for nonsymmmetric systems of linear equations},'' {\em SIAM Journal on Numerical Analysis}, 20 (1983) 345-357.

\bibitem{Muller2014} E.~H. M{\"{u}}ller and R.~Scheichl, ``{Massively parallel solvers for elliptic partial differential equations in numerical weather and climate prediction},'' {\em Quarterly Journal of the Royal Meteorological Society}, 140 (2014) 2608-2624.

\bibitem{Yang2017} C.~Yang, W.~Xue, H.~Fu, H.~You, {and co-authors}, ``{10M-Core Scalable Fully-Implicit Solver for Nonhydrostatic Atmospheric Dynamics},'' {\em SC '16: Proceedings of the International Conference for High Performance Computing, Networking, Storage and Analysis},  (2016) 57-68.

\bibitem{Carson2018} E.~Carson and N.~J. Higham, ``{Accelerating the Solution of Linear Systems by
  Iterative Refinement in Three Precisions},'' {\em SIAM Journal on Scientific
  Computing}, 40 (2018) A817-A847.

\bibitem{Haidar2018} A.~Haidar, A.~Abdelfattah, M.~Zounon, P.~Wu, S.~Pranesh, S.~Tomov, and
  J.~Dongarra, ``{The Design of Fast and Energy-Efficient Linear Solvers: On the Potential of Half-Precision Arithmetic and Iterative Refinement Techniques},'' {\em International Conference on Computational Science}, (2018) 586-600.

\bibitem{Anzt2019}
H.~Anzt, G.~Flegar, T.~Gr{\"{u}}tzmacher, and E.~S. Quintana-Ort{\'{i}},
  ``{Toward a modular precision ecosystem for high-performance computing},''
  {\em The International Journal of High Performance Computing Applications},
  33 (2019) 1069-1078.

\bibitem{Idomura2018}
Y.~Idomura, T.~Ina, S.~Yamashita, N.~Onodera, S.~Yamada, and T.~Imamura,
  ``{Communication Avoiding Multigrid Preconditioned Conjugate Gradient Method
  for Extreme Scale Multiphase CFD Simulations},'' {\em 2018 IEEE/ACM 9th
  Workshop on Latest Advances in Scalable Algorithms for Large-Scale Systems
  (scalA)}, (2018) 17-24.

\bibitem{Amritkar2015}
A.~Amritkar and D.~Tafti, ``{Computational Fluid Dynamics Computations Using a
  Preconditioned Krylov Solver on Graphical Processing Units},'' {\em Journal
  of Fluids Engineering}, 138 (2015) 011402.

\bibitem{PRUSA2018331}
J.~M. Prusa, ``{Computation at a coordinate singularity},'' {\em Journal of
  Computational Physics}, 361 (2018) 331-352.

\bibitem{Williamson}
D.~L. Williamson, J.~B. Drake, J.~J. Hack, R.~Jakob, and P.~N. Swarztrauber,
  ``{A standard test set for numerical approximations to the shallow water
  equations in spherical geometry},'' {\em Journal of Computational Physics}, 102 (1992) 211-224.

\bibitem{Prusa2008}
J.~M. Prusa, P.~K. Smolarkiewicz, and A.~A. Wyszogrodzki, ``{EULAG, a
  computational model for multiscale flows},'' {\em Computers {\&} Fluids}, 37 (2008) 1193-1207.

\bibitem{Kurowski2016mwr}
M.~J. Kurowski, D.~K. Wojcik, M.~Z. Ziemianski, B. Rosa, and Z.~P. Piotrowski,
  ``{Convection-permitting regional weather modeling with COSMO-EULAG: Compressible and 
 anelastic solutions for a typical westerly flow over the Alps},'' {\em Monthly Weather 
Review}, 144 (2016) 1961-1982.

\bibitem{Smolarkiewicz1998}
P.~K. Smolarkiewicz and L.~G. Margolin, ``{MPDATA: A Finite-Difference Solver
for Geophysical Flows},'' {\em Journal of Computational Physics}, 140 (1998) 459-480.

\bibitem{smolarkiewicz2000variational} P.~Smolarkiewicz and L.~Margolin,
``{Variational methods for elliptic problems in fluid models},'' 
{\em Proc. ECMWF Workshop on Developments in Numerical Methods for Very High 
Resolution Global Models, 5–7 June 2000, Reading, UK, ECMWF}, (2000) 137–159

\bibitem{smsz11} P.K.~Smolarkiewicz, J.~Szmelter, ``{A nonhydrostatic
unstructured-mesh soundproof model for simulation of internal gravity
waves},'' {\em Acta Geophysica}, 59 (2011) 1109-1134.

\bibitem{Smolarkiewicz1997}
P.~K. Smolarkiewicz, V.~Grubi{\v{s}}i{\'{c}}, L.~G. Margolin, ``{On Forward-in-Time Differencing
  for Fluids: Stopping Criteria for Iterative Solutions of Anelastic Pressure
  Equations},'' {\em Monthly Weather Review}, 125 (1997) 647-654.

\bibitem{Haurwitz} 
B. Haurwitz, ``{The motion of atmospheric disturbances on the spherical
Earth},'' {\em Journal of Marine Research}, 3 (1940) 254-267

\bibitem{Temam2006}
R.~Temam, ``{Suitable initial conditions},'' {\em Journal of Computational
Physics}, 218 (2006) 443-450.

\bibitem{Hoskins73} 
B.~J. Hoskins, ``{Stability of the Rossby-Haurwitz wave},''
{\em Quarterly Journal of the Royal Meteorological Society}, 99 (1973) 723-745.

\bibitem{Thuburn} 
J.~Thuburn and Y.~Li,``{Numerical simulations of
Rossby-Haurwitz waves},''
{\em Tellus A: Dynamic Meteorology and Oceanography}, 52 (2000) 181-189.

\bibitem{grose} 
W.~L. Grose and B.~J. Hoskins, ``{On the influence of
orography on large-scale atmospheric flow},'' {\em Journal of the 
Atmospheric Sciences},'' 36 (1979), 223-234.

\bibitem{smmw01} 
P.~K. Smolarkiewicz, L.~G. Margolin and A.~A. Wyszogrodzki,
``{A Class of Nonhydrostatic Global Models},'' {\em Journal of the 
Atmospheric Sciences},'' 58 (2001), 349-364.

\bibitem{Dawson2017}
A.~Dawson and P.~D. D{\"{u}}ben, ``rpe v5: an emulator for reduced
  floating-point precision in large numerical simulations,'' {\em Geoscientific
  Model Development}, 10 (2017) 2221-2230.

\bibitem{Klower:2019:PAF:3316279.3316281}
M.~Kl{\"{o}}wer, P.~D. D{\"{u}}ben, and T.~N. Palmer, ``{Posits as an
  Alternative to Floats for Weather and Climate Models},'' in {\em Proceedings
  of the Conference for Next Generation Arithmetic 2019}, CoNGA'19, (New York,
  NY, USA), 2 (2019) 1-8.

\bibitem{doi:10.1137/S1064827599353865}
H.~van~der Vorst and Q.~Ye, ``{Residual Replacement Strategies for Krylov
  Subspace Iterative Methods for the Convergence of True Residuals},'' {\em
  SIAM Journal on Scientific Computing}, 22 (2000) 835-852.

\bibitem{Vana2017a} F.~V{\'{a}}{\v{n}}a, P.~D{\"{u}}ben, S.~Lang, T.~Palmer, M.~Leutbecher, 
D.~Salmond, and G.~Carver, ``{Single Precision in Weather Forecasting Models: An Evaluation 
with the IFS},'' {\em Monthly Weather Review}, 145 (2017) 495-502.


\bibitem{Hatfield2018} S.~Hatfield, A.~Subramanian, T.~Palmer,
P.~D{\"{u}}ben, ``{Improving Weather Forecast Skill through
Reduced-Precision Data Assimilation},'' {\em Monthly Weather
  Review}, 146 (2018) 49-62.


\end{thebibliography}
\newpage

\end{document}